\documentclass[%
 reprint,
 amsmath,amssymb,
 aps,
 prb,longbibliography,
 superscriptaddress
]{revtex4-2}

\usepackage{graphicx,xcolor}% Include figure files
\usepackage{dcolumn}% Align table columns on decimal point
\usepackage{bm}% bold math
\usepackage[percent]{overpic}
\usepackage{physics}
\usepackage{mathtools}
\usepackage{hyperref}
\usepackage{booktabs}
\usepackage{longtable}
\usepackage[pdftex]{pict2e}
\usepackage{upgreek}

\hypersetup{colorlinks=true,%
            urlcolor=blue,%
            citecolor=blue,%
            linkcolor=blue}

\begin{document}
\title{{\it Ab initio} derivation of lattice gauge theory dynamics for cold gases in optical lattices}

\author{Federica Maria Surace}%
\email{fsurace@caltech.edu}
\affiliation{Department of Physics and Institute for Quantum Information and Matter,
California Institute of Technology, Pasadena, California 91125, USA}
\author{Pierre Fromholz}%
 \email{pierre.fromholz@unibas.ch}
\affiliation{The Abdus Salam International Centre for Theoretical Physics (ICTP), strada Costiera 11, 34151 Trieste,
Italy}
\affiliation{International School for Advanced Studies (SISSA), via Bonomea 265, 34136 Trieste, Italy} 
\author{Nelson\,\surname{Darkwah\,Oppong}}
\altaffiliation{Current address: JILA, University of Colorado and National Institute of Standards and Technology, and Department of Physics, University of Colorado, Boulder, Colorado 80309, USA}
\affiliation{Faculty of Physics,
Ludwig-Maximilians-Universit\"at M\"unchen, Schellingstr. 4, D-80799 Munich, Germany}
\affiliation{Munich Center for Quantum Science and Technology (MCQST), Schellingstr. 4, D-80799 Munich, Germany}
\author{Marcello Dalmonte}
 \affiliation{The Abdus Salam International Centre for Theoretical Physics (ICTP), strada Costiera 11, 34151 Trieste,
Italy}
\affiliation{International School for Advanced Studies (SISSA), via Bonomea 265, 34136 Trieste, Italy}
\author{Monika Aidelsburger}
\email{monika.aidelsburger@physik.uni-muenchen.de}
\affiliation{Faculty of Physics,
Ludwig-Maximilians-Universit\"at M\"unchen, Schellingstr. 4, D-80799 Munich, Germany}
\affiliation{Munich Center for Quantum Science and Technology (MCQST), Schellingstr. 4, D-80799 Munich, Germany}

\date{\today}

\begin{abstract}
We introduce a method for quantum simulation of U$(1)$ lattice gauge theories coupled to matter, utilizing alkaline-earth(-like) atoms in state-dependent optical lattices. The proposal enables the study of both gauge and fermionic-matter fields without integrating out one of them in one and two dimensions. We focus on a realistic and robust implementation that utilizes the long-lived metastable clock state available in alkaline-earth(-like) atomic species. Starting from an {\it ab initio} modelling of the experimental setting, we systematically carry out a derivation of the target U$(1)$ gauge theory. This approach allows us to identify and address conceptual and practical challenges for the implementation of lattice gauge theories that -- while pivotal for a successful implementation -- have never been rigorously addressed in the literature: those include the specific engineering of lattice potentials to achieve the desired structure of Wannier functions, and the subtleties involved in realizing the proper separation of energy scales to enable gauge-invariant dynamics. We discuss realistic experiments that can be carried out within such a platform using the fermionic isotope $^{173}$Yb, addressing via simulations all key sources of imperfections, and provide concrete parameter estimates for relevant energy scales in both one- and two-dimensional settings.
\end{abstract}

\maketitle
\section{Introduction}

In the last decade, the rapid development of quantum simulators has motivated an increasingly large interest in possible applications to nuclear and particle physics. The study of lattice gauge theories (LGTs)~\cite{Wilson74,KogutReviewLGT,Montvay1994}, one of the most successful theoretical frameworks to regularize strong interacting field theories, could take great advantage of the use of quantum devices. First formulated in the 1970s~\cite{Wilson74}, classical simulations of LGTs based on Monte Carlo sampling soon became a pillar of our understanding of quantum chromodynamics (QCD)~\cite{Montvay1994}, with applications as diverse as low-energy spectra~\cite{RevModPhys.84.449,detmold2019hadrons}, phase diagrams~\cite{detar2009qcd,fukushima2010phase,philipsen2019constraining}, and even precision measurements in the context of the recently puzzling muon magnetic moment results~\cite{borsanyi2021leading}. Quantum simulators promise to extend our understanding of LGTs to regimes that are presently inaccessible to Monte Carlo methods, including real-time dynamics, or the physics of the early universe and neutron stars~\cite{Wiese:2013kk,Dalmonte:2016jk,Zohar2015,banuls2020simulating,aidelsburger_cold_2022,davoudi2022quantum}.

Starting from early theoretical proposals, the field of quantum simulation of LGTs has rapidly evolved driven by two factors: the development of quantum simulation tools and schemes tailored to the specificity of gauge theories (in particular, gauge invariance), and a series of first experimental steps that have been taken to demonstrate the feasibility of the proposed schemes. The former has been pioneered by the trapped ion experiment reported in Ref.~\cite{ExpPaper} (later extended in Ref.~\cite{nguyen_digital_2022}), where the dynamics of a few-site Schwinger model [i.e., quantum electrodynamics in (1+1)-dimension (d)] with up to six sites was demonstrated. Moreover, building blocks of matter-gauge interactions have been successfully demonstrated in cold atom settings, including both $\mathbb{Z}_2$~\cite{schweizer2019floquet} and U$(1)$~\cite{mil2020scalable} gauge theories in 1d, utilizing the quantum link formulation (QLM) of LGTs, where the dimension of the local Hilbert space of the gauge link is truncated and therefore finite. More recently, large-scale quantum simulations of Abelian LGTs have been reported in Rydberg atom arrays (Schwinger model~\cite{Bernien2017,surace2020lattice}, as well as a (2+1)-d Ising-Higgs gauge theory~\cite{semeghini2021probing}) and with ultracold bosonic atoms in tilted optical superlattices~\cite{yang2020observation,zhou_thermalization_2022,wang_interrelated_2022}. In the continuum, quantum simulation of a topological gauge theory was realized in an optically-dressed Bose-Einstein condensate by realising a one-dimensional reduction of the Chern-Simons theory, the so-called chiral BF theory~\cite{frolian2022realising}.

The first generation of experimental realizations has already proven that quantum simulators of LGTs can reach system sizes and timescales at the boundaries of the capabilities of classical numerical simulations~\cite{Bernien2017,surace2020lattice}. Nevertheless, many challenges still have to be overcome before we can utilize quantum devices for making accurate predictions on complex gauge theories like QCD. Despite the recent experimental progress mentioned above, there is no direct path at this point towards quantum simulation of LGTs with fermionic matter in more than one dimension, where both gauge and matter fields are simulated~\cite{zohar_quantum_2022}. In the pioneering experiments implemented with trapped ions~\cite{ExpPaper,nguyen_digital_2022} gauge fields are eliminated by a Jordan-Wigner transformation, which maps the original Schwinger model to a spin model with exotic long-range interactions. A related approach is followed for implementations of QLMs in Rydberg atom arrays, where the matter fields are integrated out~\cite{surace2020lattice}. Keeping both matter and gauge fields is more challenging and most proposals require implementations based on ultracold mixtures~\cite{mil2020scalable}, which significantly increases the experimental complexity. QLMs on the other hand are characterized by a finite-dimensional Hilbert space for the gauge degrees of freedom, offering the possibility of implementing matter and gauge degrees of freedom with a single atomic species. This has been demonstrated with ultracold bosons in tilted optical superlattice potentials~\cite{yang2020observation,zhou_thermalization_2022,wang_interrelated_2022} and a possible extension of this scheme to higher dimensions is currently explored theoretically~\cite{osborne_large-scale_2022}. Moreover, schemes based on Floquet engineering appear challenging due to the presence of higher-order terms that need to be suppressed in order to respect gauge invariance~\cite{schweizer2019floquet}, for instance by implementations of additional stabilizers~\cite{halimeh_stabilizing_2022}. In order to overcome these limitations, we have developed a new scheme for the realization of U(1) QLMs with ultracold alkaline-earth(-like) atoms (AELA) that offers a direct implementation of fermionic matter and gauge fields, as well as a straightforward extension to two dimensions.

To ensure a robust experimental implementation, it is crucial to further bridge the gap between theoretical proposals, which focus on conceptual developments and novel implementation schemes, and experimental realizations, which require microscopic derivations of the gauge-theory dynamics. The last step is vital in order to understand at a qualitative and quantitative level the impact of often neglected challenges or even roadblocks---including particle losses, limited coherence time (e.g., due to spontaneous emission), and practical difficulties (such as challenges in realizing the required optical potentials).

In this work, we introduce a novel scheme and present an {\it ab initio} derivation of a U(1) lattice gauge theory in both one and two spatial dimensions using AELA in optical lattices. The backbone framework is an implementation that relies on protecting gauge invariance utilizing a combination of energy penalty and locality, through a specific design of state-dependent optical potentials, that are particularly well suited for atomic species with a long-lived electronically excited state~\cite{Riegger2018,Heinz2020}.
We carry out a thorough numerical study of the experimental parameters of the optical lattice needed to obtain a regime for our quantum simulation that features the best ratio between coherent and incoherent dynamics. We compute the parameters of the lattice model, and simulate the corresponding microscopic dynamics, showing that a moderate amount of on-site disorder would only mildly affect the observed time evolution.

\begin{figure}
    \centering
    \includegraphics[width=\linewidth]{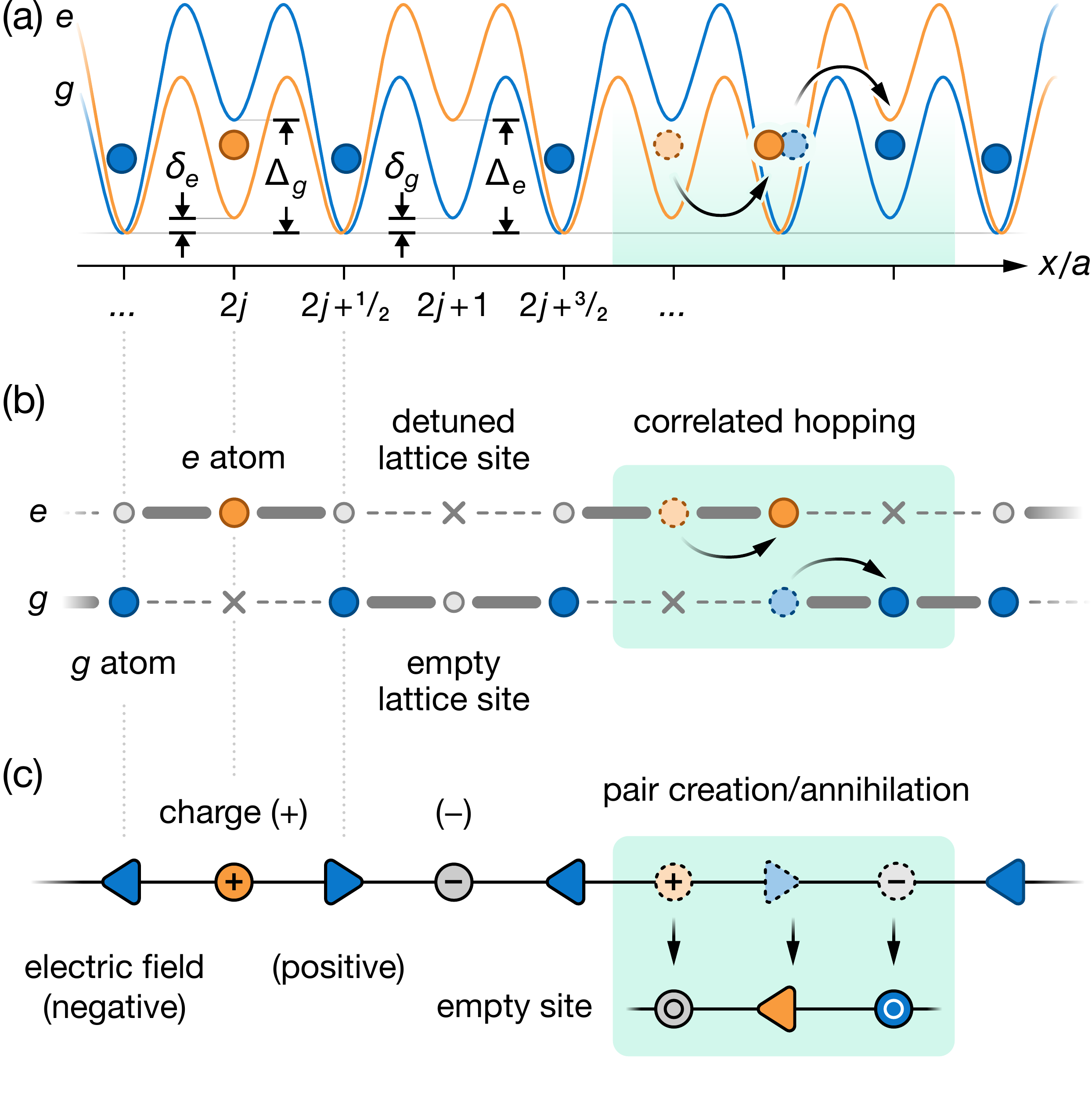}
    \caption{\label{fig:potential}
    \textbf{Illustration of the mapping between atomic states in the optical lattice potential and the U(1) quantum link model considered in our proposal.}
    (a)~Schematic optical lattice potentials for the $g$ (blue line) and $e$ atoms (orange line) together with the relevant energy scales $\delta_{g,e}$ and $\Delta_{g,e}$.
    Circles indicate an exemplary initial state of $g$ (blue circles) and $e$ atoms (orange circles) in the optical lattice.
    The black arrows on the right indicate the correlated hopping of atoms between adjacent lattice sites.
    (b)~Simplified schematic of the optical lattice potential in panel~(a) showing the bonds across which hopping is energetically allowed (thick gray lines) and forbidden (dashed gray lines).
    (c)~State in the quantum link model corresponding to the atomic configuration in panels~(a) and (b). In the mapping, every lattice site that can only be occupied by $e$ or $g$ atoms is interpreted as a matter site, shown as circles with the $+$ ($-$) labels indicating filled sites with charge $+$ ($-$). The lattice sites that can be occupied by both $e$ and $g$ are interpreted as link $(l)$ or gauge-field sites, where triangles pointing right (left) indicate positive (negative) electric field.
    Colors indicate the corresponding internal state of the atoms [see panels~(a) and (b)] in the proposed experimental implementation and gray indicates an empty site.
    Note that the correlated hopping of atoms is mapped to a gauge-invariant pair creation/annihilation process, as shown on the right.}
\end{figure}

\section{Quantum link model}

In this work we focus on the realization of a U$(1)$ lattice gauge theory with fermionic matter, the lattice version of the Schwinger model, i.e., quantum electrodynamics in one spatial dimension~\cite{Schwinger1951}. Despite its simplicity, the Schwinger model displays several salient features in common with more complicated ones, including confinement, chiral symmetry breaking, and non-trivial real-time dynamics~\cite{calzetta_book}.

In one spatial dimension, the Hamiltonian has the form~\cite{KogutReviewLGT}:
\begin{multline}
\label{eq:U1}
    H_\text{LGT}=-w\sum_j \left(\psi_j^\dagger U^{\phantom\dagger}_{j,j+1}\psi^{\phantom\dagger}_{j+1}+ \text{H.c.}\right)\\
    +m\sum_j (-1)^j \psi_j^\dagger \psi^{\phantom\dagger}_j+g \sum_j (E^{\phantom\dagger}_{j,j+1}+\mathcal E_{0})^2,
\end{multline}
where $\psi_j^\dagger, \psi^{\phantom\dagger}_j$ are fermionic creation and annihilation operators on site $j$ of a 1d lattice. The operators $U_{j,j+1}$ and $E_{j,j+1}$ are respectively the parallel transporter and the electric field operators, with commutation relation $[E_{i,i+1},U_{j,j+1}]=\delta_{ij}U_{j,j+1}$: these operators represent a U$(1)$ gauge field  on the link connecting the sites $j$ and $j+1$. The nearest-neighbor hopping term, of amplitude $w$, is made gauge invariant by the parallel transporter $U_{j,j+1}$. The fermionic mass is staggered, according to the Kogut-Susskind formulation~\cite{KogutSusskindFormulation}: on even sites, an occupied fermionic site represents a ``positron" with charge $+1$, while ``electrons", of charge $-1$, are represented by holes on odd sites [Fig.~\ref{fig:potential}(c)]. We can therefore define the local charge as
\begin{equation}
    q_j = \psi_j^\dagger \psi^{\phantom\dagger}_j-\frac{1-(-1)^j}{2}.
\end{equation}
The mass term in Eq.~(\ref{eq:U1}) assigns the mass $m$ to both electrons and positrons. Finally, the term proportional to $g$ is the energy of the electric field, and $\mathcal E_0$ represents a static background electric field.
The Hamiltonian in Eq.~(\ref{eq:U1}) has a gauge symmetry generated by the local operators $G_j$, defined as

\begin{equation}
    G_j=E_{j,j+1}-E_{j-1,j}-q_j.
\end{equation}

The physical states for the LGT are the ones that satisfy the local constraint (\textit{Gauss' law}) $G_j\ket{\Psi}=0$ for every site $j$.

In the following, we will consider the quantum link formulation of the model~\cite{QLink1,QLink2,QLink3,QLink4}: in this formulation, all the gauge fields are represented by a finite $d$-dimensional Hilbert space (we choose $d=2$), and the operators $E_{j,j+1}$ and $U_{j,j+1}$ have the form of $S^z$ and $S^+$ operators respectively. Compared to the usual Wilsonian lattice gauge theories~\cite{Wilson74,KogutReviewLGT}, this formulation is particularly suitable for quantum simulations, because it exploits discrete quantum degrees of freedom.

 Here we focus on the spin-$1/2$ representation~\cite{Banerjee2012}. In this case, the electric field has the two possible values $E_{j,j+1}=\pm 1/2$ [Fig.~\ref{fig:potential}(c)], that have the same energy for $\mathcal E_0=0$. This choice, with half-integer values of the electric field, is generally denoted as having a {\it topological angle} $\theta=\pi$ (in contrast to the case of integer electric field values, having $\theta=0$)~\cite{surace2020lattice}. The topological angle can be tuned by changing the static background field $\mathcal E_0$, whose effect is to split the degeneracy between the two electric field states~\cite{halimeh_tuning_2022,cheng_tunable_2022}. For the spin-$1/2$ representation it is useful to define $\tau=2g\mathcal E_0$. Then, the Hamiltonian (\ref{eq:U1}) becomes (up to an additive constant)

\begin{multline}
\label{eq:QLM}
    H_{\text{QLM}}=-w\sum_j \left(\psi_j^\dagger U^{\phantom\dagger}_{j,j+1}\psi^{\phantom\dagger}_{j+1}+ \text{H.c.}\right)\\
    +m\sum_j (-1)^j \psi_j^\dagger \psi^{\phantom\dagger}_j+\tau \sum_j E_{j,j+1}.
\end{multline}

With this notation, choosing $\tau\neq 0$ effectively changes the topological angle to $\theta\neq \pi$. We note that the model above can be exactly mapped to a spin chain via direct integration of Gauss' law~\cite{surace2020lattice}.

\section{Quantum simulation}
\label{sec:onedim}
\subsection{Optical lattice}
\label{sec:opt-lattice}

In the proposed experimental setup, we consider cold fermionic atoms in two different electronic states $\alpha=\{g,e\}$, realized by the ground and meta-stable excited clock states $g\equiv{^1\mathrm{S}}_0$ and $e\equiv{^3\mathrm{P}}_0$ of AELA. The atoms are considered to be spin polarized in a given nuclear Zeeman state $m_F$, so that the corresponding Hamiltonian is given by $H=H_\text{non-int}+H_\text{int}$, with~\cite{Gorshkov2010}
\begin{equation}\label{eq:opt}
    \begin{split}
        H_{\text{non-int}} = & \sum_{\alpha} \int \mathrm{d}^3\mathbf{r} \Psi^\dagger_{\alpha}\left( \mathbf{r} \right)\left(- \frac{\hbar^2}{2M}\nabla^2+V_{\alpha} \left(\mathbf{r}\right)\right)\Psi^{\phantom\dagger}_{\alpha }\left( \mathbf{r} \right), \\
        H_{\text{int}}=& g_{eg}^-\int\mathrm{d}^3\mathbf{r} \rho_e\left(\mathbf{r}\right)\rho_g\left(\mathbf{r}\right).
    \end{split}
\end{equation}
Here $\Psi_{\alpha}(\mathbf{r})$ denotes the fermion field operator for atoms in the internal state $\lvert \alpha\, m_F \rangle$. The density operators are defined as $\rho_{\alpha }\left(\mathbf{r} \right) =\Psi^\dagger_{\alpha }(\mathbf{r})\Psi^{\phantom\dagger}_{\alpha }(\mathbf{r})$. Since the atoms are polarized in the same nuclear Zeeman state, the interaction strength $g_{eg}^-=4\pi \hbar^2 a_{eg}^-/M$ (atomic mass $M$) is associated with the scattering length $a_{eg}^-$ of the antisymmetric electronic state~\cite{Scazza14,zhang_spectroscopic_2014}. The term $V_{\alpha } \left(\mathbf{r}\right)$ denotes a 3d lattice potential $V_\alpha(\mathbf{r})=
V^x_\alpha(x)+V^y_\alpha(y)+V^z_\alpha(z)$, where $V^x_\alpha(x)$ is the state-dependent potential depicted in Fig.~\ref{fig:potential}(a) and $V_{\alpha}^y(y)$ and $V_{\alpha}^z(z)$ are deep state-independent optical lattices with amplitude $F_g=F_e$ and lattice spacing $d_y=d_z$ that isolate individual 1d chains and provide strong radial confinement. For simplicity, we choose equal amplitudes for the transverse lattices along $y$ and $z$. The state-dependent lattice along $x$ is defined as

\begin{align}
    V^x_\alpha(x)=&-A_\alpha\sin^2\left(\frac{\pi}{2a}x+\varphi\right)-B_\alpha\sin^2\left(\frac{\pi}{a}x\right) \nonumber\\
    &-C_\alpha\sin^2\left(\frac{2\pi}{a}x+\frac{\pi}{2}\right).
    \label{eq:opt-lat-pot}
\end{align}

It has a unit cell of length $2a$ with three ``low"-energy lattice sites and one ``high"-energy site, which suppresses tunneling to that site as shown schematically in Fig.~\ref{fig:potential}(b). The triple wells of the $g$ and $e$ lattices are shifted relative to each other by a distance~$a$.

\begin{figure}
    \centering
    \includegraphics[width=\linewidth]{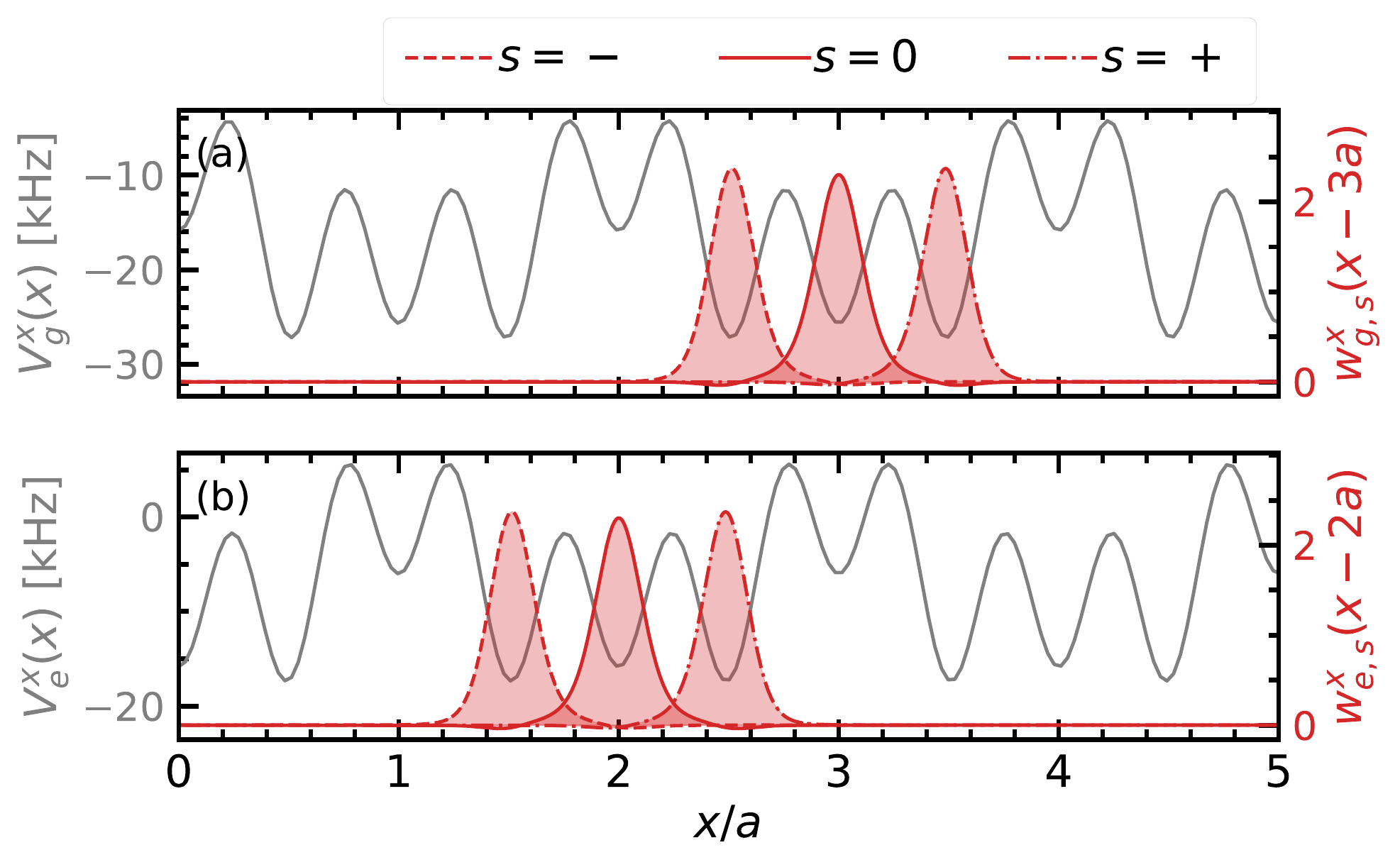}
    \caption{\textbf{ Optical lattice potential and Wannier functions.} The $x$-component of the optical lattice potential defined in Eq.~(\ref{eq:opt-lat-pot}) for the (a) $g$ and (b) $e$ atoms is plotted in gray for $\varphi=0$. The parameters $A_\alpha,B_\alpha,C_\alpha$ are reported in Table~\ref{tab:params}. In red, the $x$-components of the Wannier functions centered on site $x=3a$ and $x=2a$ are shown. The parameter $s=\{-,0,+\}$ labels the three different orbitals in the unit cell.}
    \label{fig:wannier}
\end{figure}

The optical potential along $x$ can be realized by superimposing three different optical lattices. 
Each of them is generated from a pair of monochromatic laser beams at either the magic wavelength $\lambda_m$~\cite{takamoto_optical_2005,ludlow_systematic_2006}, which corresponds to a state-independent potential, or the anti-magic wavelength $\lambda_{am}$~\cite{Yi08}, where the potentials for atoms in the $g$ and $e$ state are equal in magnitude but have opposite signs. Moreover, the lattice spacing can be set by tuning the intersection angle $\theta$ between the interfering pair of laser beams according to $\lambda / \left[ 2 \sin(\theta / 2) \right]$. The two shorter-spacing lattices in Eq.~(\ref{eq:opt-lat-pot}) are operated at the magic wavelength $\lambda_m$ ($B_e = B_g$ and $C_e = C_g$) and intersection angles $\theta_C = 180^\circ$ and $\theta_B = 60^\circ$. The corresponding lattice spacings are $a=\lambda_m/2$ and $2a$, such that their combination yields a symmetric double-well potential~\cite{sebby-strabley_lattice_2006,folling_direct_2007}. The third long-lattice at lattice spacing $4a$, which can be generated at a smaller intersection angle, is operated at~$\lambda_{am}$ with $A_g = -A_e$ generating a triple well potential that is shifted for $g$ and $e$ atoms as shown in Fig.~\ref{fig:potential}(a) for $\varphi=0$. Note that the required optical potentials could also be generated using a hybrid approach using a combination of optical lattices and tweezers, which have recently been employed for Hubbard-type physics~\cite{Spar21,Young22b}.

\subsection{Lattice Hamiltonian}
To obtain a lattice Hamiltonian for the model, we assume that only the three lowest Bloch bands are occupied both for the $g$ and $e$ states, and we express the field operator $\Psi_{\alpha}(\mathbf{r})$ in terms of the Wannier functions $w_{\alpha, s}$, where $s=\{-,0,+\}$ labels the three Wannier centers in a unit cell (Fig.~\ref{fig:wannier}):

\begin{multline}
    \Psi_g(\mathbf{r})=\sum_{j \text{ odd}} \left[w_{g,+}(\mathbf{r}-\mathbf{r}_{j})c_{j+1/2}\right.\\
    \left.+w_{g,0}(\mathbf{r}-\mathbf{r}_j)c_j +w_{g,-}(\mathbf{r}-\mathbf{r}_{j})c_{j-1/2}\right],
\end{multline}

\begin{multline}
    \Psi_e(\mathbf{r})=\sum_{j \text{ even}} \left[w_{e,+}(\mathbf{r}-\mathbf{r}_{j})d_{j+1/2}\right.\\
    \left.+w_{e,0}(\mathbf{r}-\mathbf{r}_j)d_j +w_{e,-}(\mathbf{r}-\mathbf{r}_{j})d_{j-1/2}\right],
\end{multline}
where $\mathbf{r}_j =ja\hat x$, $\hat x$ is the unit vector and $c_j$ ($d_j$) is the lattice fermionic annihilation operator of a $g$ atom ($e$ atom) on lattice site $j$.

Substituting the expressions for the field operators in Eq.~(\ref{eq:opt}), we obtain the lattice Hamiltonian (see Appendix~\ref{app:wannier})
\begin{equation}
\label{eq:Hlatt}
    H_{\text{latt}}=H_g+H_e+H_U+H_D+H_\text{lr}+\text{const},
\end{equation}
where $H_g$ and $H_e$ denote the terms containing hopping and chemical potentials of the $g$ and $e$ atoms within a single triple well respectively
\begin{equation}
    H_g=\sum_{j \text{ odd}}\left[-t_g \left(c_{j}^\dagger c^{\phantom\dagger}_{j+1/2}+c_{j}^\dagger c^{\phantom\dagger}_{j-1/2}+\text{H.c.}\right)+\delta_g c_j^\dagger c^{\phantom\dagger}_j\right],
\end{equation}

\begin{equation}
H_e=\sum_{j \text{ even}}\left[-t_e \left(d_{j}^\dagger d^{\phantom\dagger}_{j+1/2}+d_{j}^\dagger d^{\phantom\dagger}_{j-1/2}+\text{H.c.}\right)+\delta_e d_j^\dagger d^{\phantom\dagger}_j\right].
\end{equation}
We assumed, for the moment, that $\varphi=0$, so the model is symmetric under reflections centered on the matter sites: this implies that the chemical potentials of the sites $s=+$ and $s=-$ are the same (and can be chosen as a reference level and set to zero). 

\begin{figure}
    \centering
    \includegraphics[width=\linewidth]{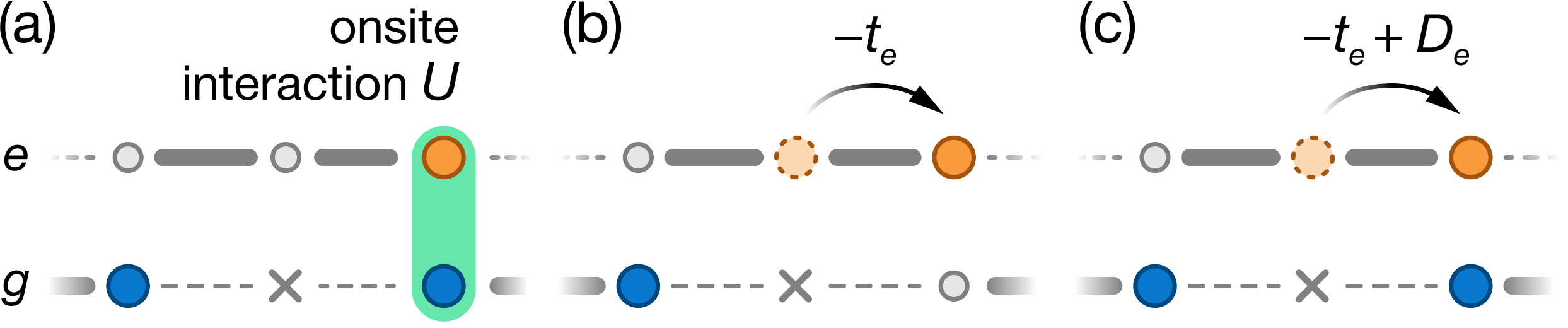}
    \caption{\label{fig:UD}
    \textbf{Illustration of the interacting terms in $H_U$ and~$H_D$.}
    (a) On-site interaction $U$ between a single $g$ (blue circle) and $e$ atom (orange circle).
    (b) Hopping of a single $e$ atom to an empty lattice site. (c) In the presence of interactions, tunneling is additionally modified by a density-assisted tunneling with amplitude $D_e$.
    }
\end{figure}

The terms $H_U$ and $H_D$ are obtained from the interacting term in Eq.~(\ref{eq:opt}), and read (see Fig.~\ref{fig:UD})
\begin{equation}
    H_U= U\sum_j d^\dagger_{j+1/2}d^{\phantom\dagger}_{j+1/2}c^\dagger_{j+1/2}c^{\phantom\dagger}_{j+1/2},
\end{equation}

\begin{align}
\label{eq:HD}
    H_{D}&=D_g\sum_{j \text{ odd}}\left(d^\dagger_{j+1/2}d^{\phantom\dagger}_{j+1/2}c_{j}^\dagger c^{\phantom\dagger}_{j+1/2}\right.\\
    \nonumber
    &\qquad \left.+d^\dagger_{j-1/2}d^{\phantom\dagger}_{j-1/2}c_{j}^\dagger c^{\phantom\dagger}_{j-1/2}+\text{H.c.}\right)\\
    \nonumber
    &+D_e\sum_{j \text{ even}}\left(c^\dagger_{j+1/2}c^{\phantom\dagger}_{j+1/2}d_{j}^\dagger d^{\phantom\dagger}_{j+1/2}\right.\\
    \nonumber
    & \qquad \left.+c^\dagger_{j-1/2}c^{\phantom\dagger}_{j-1/2}d_{j}^\dagger d^{\phantom\dagger}_{j-1/2}+\text{H.c.}\right).
\end{align}
Finally, $H_\text{lr}$ contains all the additional terms, of the form of longer-range hoppings and interactions, that have very small amplitudes and can be neglected (we explicitly verify that these terms are negligible for the parameters reported in Section \ref{sec:param}).

It is useful to define $\epsilon=(\delta_g-\delta_e)/2$, $\delta=(\delta_g+\delta_e)/2$, and the total number of atoms on each site $j+1/2$, which corresponds to a link in the QLM (Fig.~\ref{fig:potential}),
\begin{equation}
    n_{j+1/2}^{(l)}=d_{j+1/2}^\dagger d^{\phantom\dagger}_{j+1/2}+c_{j+1/2}^\dagger c^{\phantom\dagger}_{j+1/2}.
\end{equation}
Here $(l)$ is a redundant superscript to indicate we are on a link. We now assume $\epsilon, t_\alpha, D_\alpha \ll \delta, U-\delta$: in this regime it is convenient to split the Hamiltonian into three parts $H_{\text{latt}}=H_0+H_1+H_\text{lr}$ with different energy scales, i.e.,
\begin{equation}
\label{eq:H0}
    H_0=  (N_g+N_e-N_l)\delta+ \sum_j \left(-\delta+\frac{U}{2}n_{j+1/2}^{(l)}\right)(n_{j+1/2}^{(l)}-1),
\end{equation}
where $N_g$ and $N_e$ are the total numbers of atoms in the $g$ and $e$ states respectively and $N_l$ is the total number of links.
From Eq.~(\ref{eq:H0}) it is immediate to see that for $\delta, U-\delta>0$ the lowest energy states of $H_0$ have exactly one atom (either $g$ or $e$) on each half-integer site, i.e., $n_{j+1/2}^{(l)}=1$ for every $j$: a double occupancy $n_{j+1/2}^{(l)}=2$ costs energy $U-\delta$, while having a hole $n_{j+1/2}^{(l)}=0$ costs energy $\delta$. 
The term $H_1$ has the form
\begin{multline}\label{eq:H1}
    H_1=\sum_{j \text{ odd}}\left[-t_g \left(c_{j}^\dagger c^{\phantom\dagger}_{j+1/2}+c_{j}^\dagger c^{\phantom\dagger}_{j-1/2}+\text{H.c.}\right)+\epsilon\, c_j^\dagger c^{\phantom\dagger}_j\right]\\
    +\sum_{j \text{ even}}\left[-t_e \left(d_{j}^\dagger d^{\phantom\dagger}_{j+1/2}+d_{j}^\dagger d^{\phantom\dagger}_{j-1/2}+\text{H.c.}\right)-\epsilon\, d_j^\dagger d^{\phantom\dagger}_j\right]\\
    +H_D,
\end{multline}
where $H_D$ is the Hamiltonian in Eq.~(\ref{eq:HD}).

We initialize the system with two $g$ atoms for every $g$ triple well and one $e$ atom for every $e$ triple well (all these quantities are locally conserved, if we neglect $H_\text{lr}$). The effective Hamiltonian describing the resonant dynamics is obtained using perturbation theory: we neglect $H_\text{lr}$, and we treat $H_1$ as a perturbation to $H_0$. To second order, the effective Hamiltonian has the form (see Appendix \ref{app:A})

\begin{multline}\label{eq:correction}
    H^{(2)}_{\text{eff}}=-w\sum_{j \text{ odd}} (c_j^\dagger c^{\phantom\dagger}_{j+1/2}d^\dagger_{j+1/2}d^{\phantom\dagger}_{j+1} +\text{H.c})\\
    -w\sum_{j \text{ even}} (d_j^\dagger d^{\phantom\dagger}_{j+1/2}c^\dagger_{j+1/2}c^{\phantom\dagger}_{j+1} +\text{H.c})\\
    +m\sum_{j \text{ odd}} c_j^\dagger c^{\phantom\dagger}_{j}-m\sum_{j \text{ even}} d_j^\dagger d^{\phantom\dagger}_{j},
\end{multline}
with

\begin{equation}
\label{eq:w}
    w=\frac{t_g t_eU}{\delta(\delta-U)} +\frac{-D_e t_g-D_g t_e+D_e D_g}{\delta-U},
\end{equation}

\begin{equation}
\label{eq:m}
    m=\epsilon+\frac{2t_g^2-t_e^2}{2\delta}-\frac{(t_g-D_g)^2-2(t_e-D_e)^2}{2(U-\delta)}.
\end{equation}

\subsection{Mapping to the quantum link model}
We now prove that there is an exact mapping between the effective Hamiltonian $H^{(2)}_\text{eff}$ and the quantum link Hamiltonian in Eq.~(\ref{eq:QLM}).

The fermionic operator $\psi_j$ for the matter is defined as

\begin{equation}
\label{eq:mapmat}
\psi_j =    \left\{ \begin{matrix}
c_j \quad j\text{ odd},\\
d_j \quad j\text{ even},
\end{matrix}\right.
\end{equation}
(and an analogous definition is used for $\psi_j^\dagger$). 
The electric field $E_{j,j+1}$ and the parallel transporter $U_{j,j+1}$ on the link are represented by
\begin{equation}
\label{eq:mape}
E_{j,j+1}=\frac{(-1)^j}{2} (c_{j+1/2}^\dagger c^{\phantom\dagger}_{j+1/2}-d_{j+1/2}^\dagger d^{\phantom\dagger}_{j+1/2}),
\end{equation}
\begin{equation}
U_{j,j+1}= \left\{ \begin{matrix}
 c^{\phantom\dagger}_{j+1/2} d_{j+1/2}^\dagger \quad j\text{ odd},\\
 d^{\phantom\dagger}_{j+1/2}c_{j+1/2}^\dagger\quad j\text{ even},
\end{matrix}\right.
\end{equation}
and satisfy the desired commutation relation $[E_{i,i+1},U_{j,j+1}]=\delta_{i,j}U_{j,j+1}$.
With the definitions in Eqs. (\ref{eq:mapmat}) and (\ref{eq:mape}), the operator $G_j$ takes the form
\begin{equation}
    G_j=\frac{1}{2}(n_{j+1/2}^{(l)}+n_{j-1/2}^{(l)})-n_{j}^{(b)}+\frac{(1-(-1)^j)}{2},
\end{equation}
with $n_j^{(b)}$ being the number of atoms in the $j$-th triple-well (or ``block"), i.e.,
\begin{equation}
    n_{j}^{(b)}=\left\{ \begin{matrix}
c_j^\dagger c^{\phantom\dagger}_j+c_{j+1/2}^\dagger c^{\phantom\dagger}_{j+1/2}+c_{j-1/2}^\dagger c^{\phantom\dagger}_{j-1/2} \; j\text{ odd},\\
d_j^\dagger d^{\phantom\dagger}_j+d_{j+1/2}^\dagger d^{\phantom\dagger}_{j+1/2}+d_{j-1/2}^\dagger d^{\phantom\dagger}_{j-1/2} \; j\text{ even}.
\end{matrix}\right.
\end{equation}

\begin{figure}
    \centering
    \includegraphics[width=\linewidth]{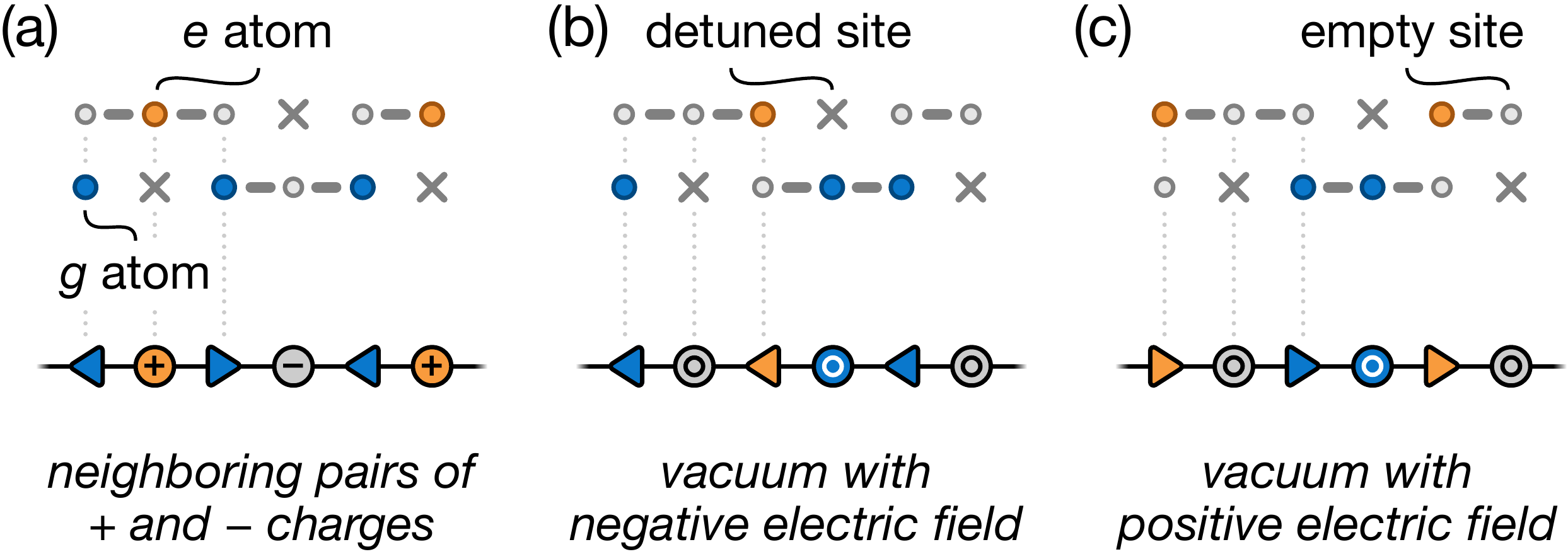}
    \caption{\label{fig:mapping}
    \textbf{Examples of the mapping between atomic configurations in the optical lattice and gauge-invariant states in the $\text{U(1)}$ QLM.}
    In each panel, the atomic states in the optical lattice are shown in the top row and the states in the QLM are shown in the bottom row.
    The panels show (a)~the state with neighboring pairs of $+$ and $-$ charges, (b)~the vacuum state with homogeneous negative electric field $E=-1/2$, and (c)~the vacuum state with homogeneous positive electric field $E=+1/2$.}
\end{figure}

With this mapping, which is schematically shown in Fig.~\ref{fig:potential}, we obtain that $H_\text{eff}^{(2)}$ is equivalent to the Hamiltonian $H_\text{QLM}$ with $\tau=0$.
The gauge-invariant subspace corresponds to the sector with $n_{j+1/2}=1$ and $n_j^{(b)}=[3-(-1)^j]/2$ for every $j$. Some examples of gauge-invariant states are shown in Fig.~\ref{fig:mapping}.

In Fig.~\ref{fig:mapping}(a), all $g$ atoms sit on the links, and all $e$ atoms are on the matter sites: the corresponding electric field takes values $E_{j,j+1}=(-1)^j/2$, while $\psi_j^\dagger \psi_j=1,0$ for even and odd sites respectively, leading to alternating positive and negative charges on matter sites. This state is the ground state of the model in the limit $m\rightarrow -\infty$. Similarly, it is easy to show that the states represented in Fig.~\ref{fig:mapping}(b) and \ref{fig:mapping}(c) have no charges on matter sites, and have uniform (negative or positive) electric field. These states (\textit{vacua}) are degenerate ground states in the limit $m\gg |w|$ with $\tau=0$, while the degeneracy is split for $\tau\neq 0$.

\subsection{Theta term}
We now show how to tune the parameters of the optical lattice to obtain $\tau\neq 0$.
The lattice Hamiltonian Eq.~(\ref{eq:Hlatt}) was derived with the assumption that $\varphi=0$. We now slightly perturb this model, by introducing a small shift $\varphi \ll 1$. To first order in $\varphi$, the shift produces an additional potential along $x$
\begin{equation}
   V_{\alpha}^x\rightarrow V_{\alpha}^x-A_{\alpha}\sin\left(\frac{\pi}{a}x\right)\varphi.
\end{equation}
The main effect of this additional term is to change the chemical potential at half-integer positions $x=(j+1/2)a$ by a quantity $-A_{\alpha}\varphi(-1)^j$. We obtain
\begin{multline}
    H_{\text{latt}}\rightarrow H_{\text{latt}}-\sum_{j}(-1)^j\varphi(A_g c_{j+1/2}^\dagger c^{\phantom\dagger}_{j+1/2}\\
    \qquad+A_ed_{j+1/2}^\dagger d^{\phantom\dagger}_{j+1/2})\\
    =H_{\text{latt}}-\sum_j (-1)^j\varphi \left[\frac{A_g+A_e}{2}n^{(l)}_{j+1/2}\right.\\
    \left.+\frac{A_g-A_e}{2}(c_{j+1/2}^\dagger c^{\phantom\dagger}_{j+1/2}-d_{j+1/2}^\dagger d^{\phantom\dagger}_{j+1/2})\right].
\end{multline}
The term $\sum_j (-1)^j n^{(l)}_{j+1/2}$ cancels in the resonant sector, and the remaining term is mapped to $\sum_j \tau E_{j,j+1}$, with
\begin{equation}
    \tau=(A_e-A_g)\varphi.
\end{equation}

\section{Experimental implementation}

The ab-initio calculation of the band structure and Wannier functions allows us to estimate the energy scales involved in the quantum simulation. These estimations explicitly verify that the desired parameter range is achievable in present-day experiments. The quantitative estimation of the parameters is also crucial to understand the limitations of our proposal such as the amplitude and duration of the signal and to identify the main sources of error such as the population of higher bands, higher-order perturbative processes, longer-range terms, and dissipation.

To set values, we choose the fermionic isotope ${{}^\text{173}\text{Yb}}$ with mass \mbox{$M \approx 173 u$} and the interorbital scattering length \mbox{$a_{eg}^-=219.7\, a_0$}~\cite{Hoefer15}; here $u$ denotes the atomic mass unit and $a_0$ the Bohr radius.
However, we note that our proposal can be similarly applied to other fermionic AELA species such as ${{}^\text{171}\text{Yb}}$ and ${{}^\text{87}\text{Sr}}$.
The experimental parameters require scaling to account for the modified atomic mass and scattering length~\cite{Goban18,Ono19} with no conceptual change in the design of the experiment.

\subsection{Realistic parameters}
\label{sec:param}
We define $\Delta_{g/e}$ as the gap between the third and fourth energy band in the lattice for the $g/e$ atoms (see Section \ref{app:wannier}). The parameters used here are chosen to satisfy the hierarchy of energy scales
\begin{equation}
    \Delta_{g/e}\gg \delta_{g/e}, U-\delta_{g/e} \gg t_{g/e}, D_{g/e}\gg \text{terms in }H_\text{lr}.
\end{equation}

We note that for the 3d lattice potential $V_\alpha(\mathbf{r})$ the Wannier functions obtained by solving the non-interacting Hamiltonian $H_\text{non-int}$ can be factorized in the three directions. The hoppings $t_\alpha$, the chemical potentials $\delta_\alpha$, and the gaps $\Delta_\alpha$  do not depend on the $y$ and $z$ components $\phi_\alpha^y(y)$ and $\phi_\alpha^z(z)$ of the Wannier functions (see Appendix \ref{app:wannier}). The interactions $U$ and $D_\alpha$, on the other hand, are proportional to the quantity
\begin{equation}
    J_{yz}=\int \mathrm d y \, \mathrm d z \, |\phi_g^y(y)|^2 |\phi_e^y(y)|^2 |\phi_g^z(z)|^2 |\phi_e^z(z)|^2.
\end{equation}
We can therefore tune $F_\alpha$ and $d_{y/z}$ to change the value of $J_{yz}$ and thus enhance or suppress the interaction terms $U$ and $D_{g/e}$ independently from the other parameters.

In Table~\ref{tab:params}, we report a possible choice for the parameters of the optical lattice, and the corresponding parameters of the lattice Hamiltonian. We choose $a=\lambda_{\text{m}}=0.7594\,\mathrm{\upmu m}$~\cite{Lemke09} and $J_{yz}=48.566\;\mathrm{\upmu m}^{-2}$.

\begin{table}[t]
\caption{\textbf{Experimental parameters.} All values are given in units of $h\cdot$kHz.}\vspace{0.5em}
\begin{ruledtabular}
\begin{tabular}{@{}ccc@{}} 
$\;A_g=-A_e\;$ & $\;B_g=B_e\;$ & $\;C_g=C_e\;$ \\ \midrule
$
9.827$ & $
6.343$ & $
15.832$ \\ 
\end{tabular}
\end{ruledtabular}

\vspace{0.4cm}
\begin{ruledtabular}
\begin{tabular}{@{}ccccc@{}} 
$\;\Delta_g=\Delta_e\;$ & $\;\delta_g=\delta_e\;$ & $\;U\;$ & $\;t_g=t_e\;$ & $\;D_g=D_e\;$\\ \midrule
$\;7.22
\;$ & $\;
1.02\;$ & $\;
2.03\;$ & $\;0.085
\;$ & $\;
0.023\;$\\ 
\end{tabular}
\end{ruledtabular}
\label{tab:params}
\end{table}

With these parameters, from Eqs.~(\ref{eq:w}) and (\ref{eq:m}) we obtain $m/h= 9\,$Hz and $w/h= -18\,$Hz; where $h$ denotes Planck's constant.
The value of $J_{yz}$ reported here can be obtained with a transverse confinement $\;F_g/h=F_e/h=48.9 \;
$kHz and $d_{y/z}=\lambda_\text{m}/2$. For this choice of transverse potential, the hopping in the $y$ and $z$ direction is $t^{y/z}_\alpha/h=2\,$Hz, much smaller than the relevant scales $m$ and $w$.
The transverse hopping can be made even smaller by using two beams intersecting at a shallow angle instead of using retro-reflected beams: the lattice spacing $d_{y/z}$ is increased, and the same value of $J_{yz}$ is obtained for a larger $F_\alpha$, thus suppressing the transverse hopping.

We explicitly check that all the terms included in $H_\text{lr}$, which were neglected in the derivation of $H_\text{QLM}$, are small with respect to $w$. The nearest-neighbor density-density interaction is of the order of $\sim 0.6 \; h
\cdot$Hz, while the hopping between the sites $s=+$ and $s=-$ in a triple well is $\sim 2\;h
\cdot$Hz, and is negligible because it is suppressed by the on-site interaction $U$.

The parameters in Table~\ref{tab:params} can be readily generalized to other atomic species: we can define the adimensional lattice constant $\tilde a=a/\lambda_\text{m}$ and the adimensional parameter $\tilde A_\alpha = A_\alpha 2M/\hbar^2\lambda_\text{m}^2$. Similarly we can define dimensionless parameters for the other energy scales $B_\alpha$, $C_\alpha$, $\Delta_\alpha$, $\delta_\alpha$, $U$, $t_\alpha$, $D_\alpha$, $m$, $w$, and for the quantity $J_{yz}$. Implementations with different atomic species but same adimensional parameters of the optical lattice yield the same adimensional values for the terms in the lattice Hamiltonian.

\subsection{Initial state preparation}
We propose two distinct ways of a two-step preparation of the system in gauge-invariant initial states as shown in Fig.~\ref{fig:mapping}, which require specific atom configurations.
First, the correct atom-number distribution needs to be prepared (independent of the internal state). One option is to prepare it starting from a sample with one $g$ atom per lattice site followed by the removal of atoms on selected lattice sites yielding the desired occupation. This is a standard technique in quantum gas microscopes~\cite{weitenberg_single-spin_2011,gross_quantum_2021}.
Alternatively, the atoms could directly be placed at their desired location with moving optical tweezer potentials~\cite{Young22b}.
In the second step, $g$ atoms can be converted to $e$ atoms on selected lattice sites using a global clock laser excitation pulse exploiting the local differential light shifts $\delta_\alpha$ and $\Delta_\alpha$ [Fig.~\ref{fig:potential}(a)].
The conversion could also be performed locally by using clock laser light focused onto single lattice sites.

\section{Real-time dynamics}
\label{sec:dyn}
Using numerical simulations, we study here the real-time dynamics of the model of Hamiltonian (\ref{eq:Hlatt}) that we compare with the dynamics of the quantum link model in Eq.~(\ref{eq:QLM}). In both cases we start with the initial state shown in Fig.~\ref{fig:mapping}(a). The time evolution of the model is simulated exactly for a system of length $4a$ with periodic boundary conditions, with the parameters reported in Section \ref{sec:param}. Longer-range terms from $H_{\text{lr}}$ in Eq.~(\ref{eq:Hlatt}) are included in the numerical simulation: the dynamics is exact as long as higher bands are not occupied.

\begin{figure}[t!]
    \centering
    \includegraphics[width=\linewidth]{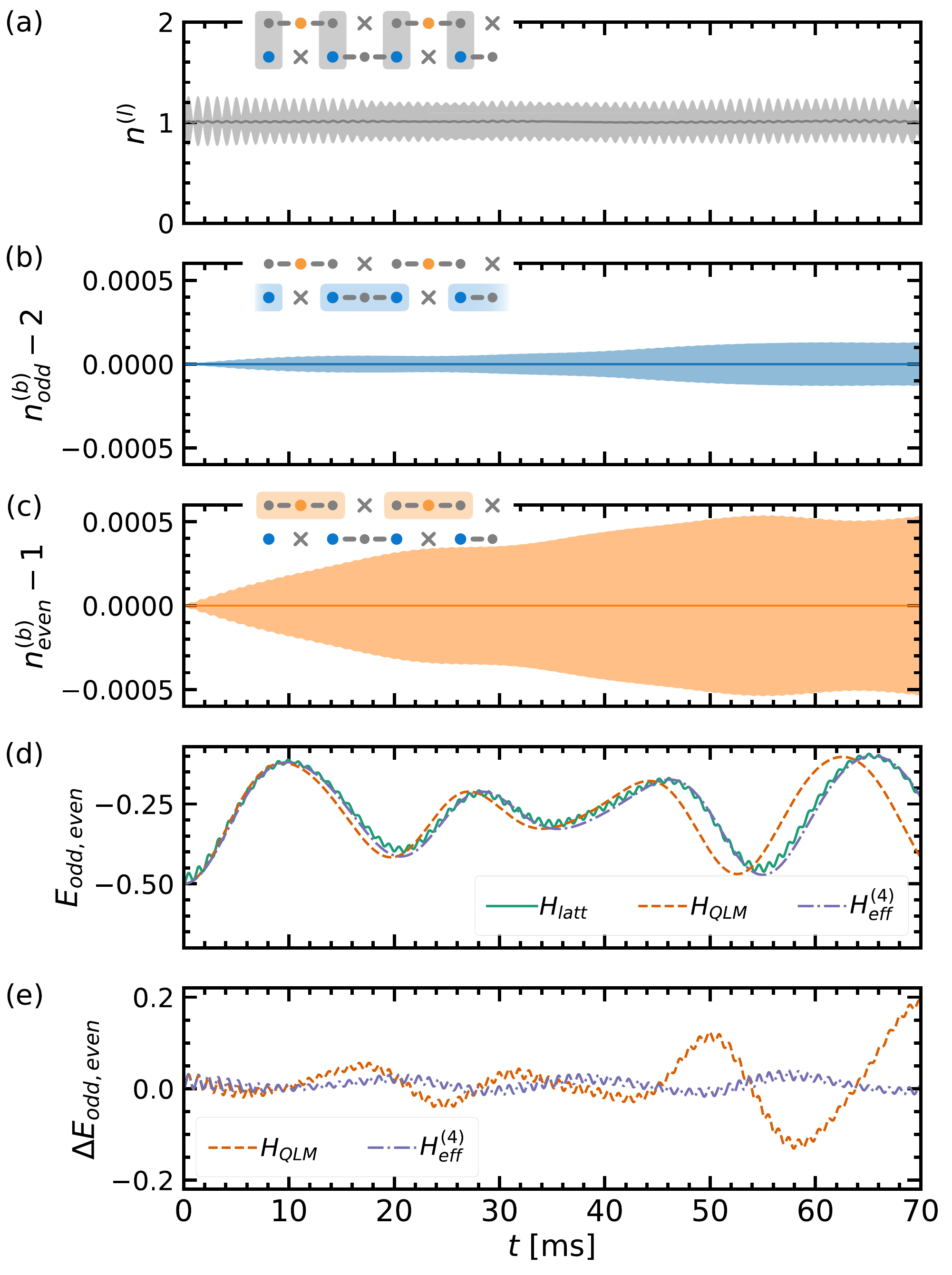}
    \caption{
    \textbf{Dynamics in the $\text{U(1)}$ quantum link model.}
    Time evolution of (a) the number of atoms per link, (b,c) the number of atoms on each odd/even block. For each observable $O$, the shaded area indicates the interval $\langle O\rangle\pm \sqrt{\text{Var}(O)}$. The initial state is depicted in Fig.~\ref{fig:mapping}(a) and is evolved under $H_\text{latt}$. The system has periodic boundary conditions and finite size $4a$.
    The top-left schematic in each panel illustrates the observable.
    (d) Time evolution of the electric field on an odd-even link. The exact dynamics given by $H_\text{latt}$ are compared with the second-order effective Hamiltonian [using Eqs.~(\ref{eq:w},\ref{eq:m})] and with the fourth-order effective Hamiltonian $H_\text{eff}^{(4)}$. The latter two are both equivalent to $H_\text{QLM}$. (e) Difference $\Delta E_\text{odd,even}= \langle E_\text{odd,even}\rangle_{H_\text{latt}}-\langle E_\text{odd,even}\rangle_{H}$ (the subscript denotes the Hamiltonian that generates the time evolution) for the two cases $H=H_\text{QLM}$ and $H=H_\text{eff}^{(4)}.$}
    \label{fig:dyn1}
\end{figure}

\subsection{Gauge invariance and gauge field}\label{sec:gauge}
The system is effectively gauge invariant as long as Gauss' law applies. To quantify the violation of Gauss' law in Eq.~(\ref{eq:Hlatt}), we examine the time evolution of $n_{j+1/2}^{(l)}$, and $n_j^{(b)}$. The simulation results are shown in Fig.~\ref{fig:dyn1}(a)-(c). We find that the conditions  $n_{j}^{(b)}=[3+(-1)^j]/2$ and $n_{j+1/2}^{(l)}=1$ are preserved up to $5\cdot 10^{-4}$ and $10^{-1}$ respectively within the first $70\,$ms after initialization, which should be compared to the characteristic interaction timescale $\hbar/|w|\simeq 8.8\,$ms.

In Fig.~\ref{fig:dyn1}(d) we examine the evolution of the electric field with (i) the Hamiltonian $H_{\text{latt}}$, (ii) the second-order effective Hamiltonian $H_{\text{eff}}^{(2)}$, equivalent to $H_{\text{QLM}}$, and (iii) the fourth-order effective Hamiltonian $H_{\text{eff}}^{(4)}$ (from the Schrieffer Wolff procedure). The differences between the electric field value obtained in case (i) and with the approximate Hamiltonians are plotted in Fig.~\ref{fig:dyn1}(e). For times up to $\sim 40$ ms, the results obtained in the three cases are in good agreement: this shows that the second-order effective Hamiltonian $H_{\text{QLM}}$ captures the main features of the time evolution, at least at short and intermediate time scales, and that longer-range terms and higher-order perturbative processes are minor sources of error. For times of the order of $\gtrsim 50\,$ms, we find that fourth-order corrections have to be considered in order to obtain a good prediction of the time evolution in the experiment. We remark that the fourth-order corrections do not violate Gauss' law, but correspond to additional gauge-invariant terms. A feature of the evolution induced by $H_{\text{latt}}$ that is not observed in the effective Hamiltonians is the presence of fast oscillations with small amplitude. These oscillations have a frequency compatible with the energy scale of $H_0$ and are averaged out in the perturbative approach. 

\begin{figure}
    \centering
    \begin{overpic}[width=0.9\linewidth]{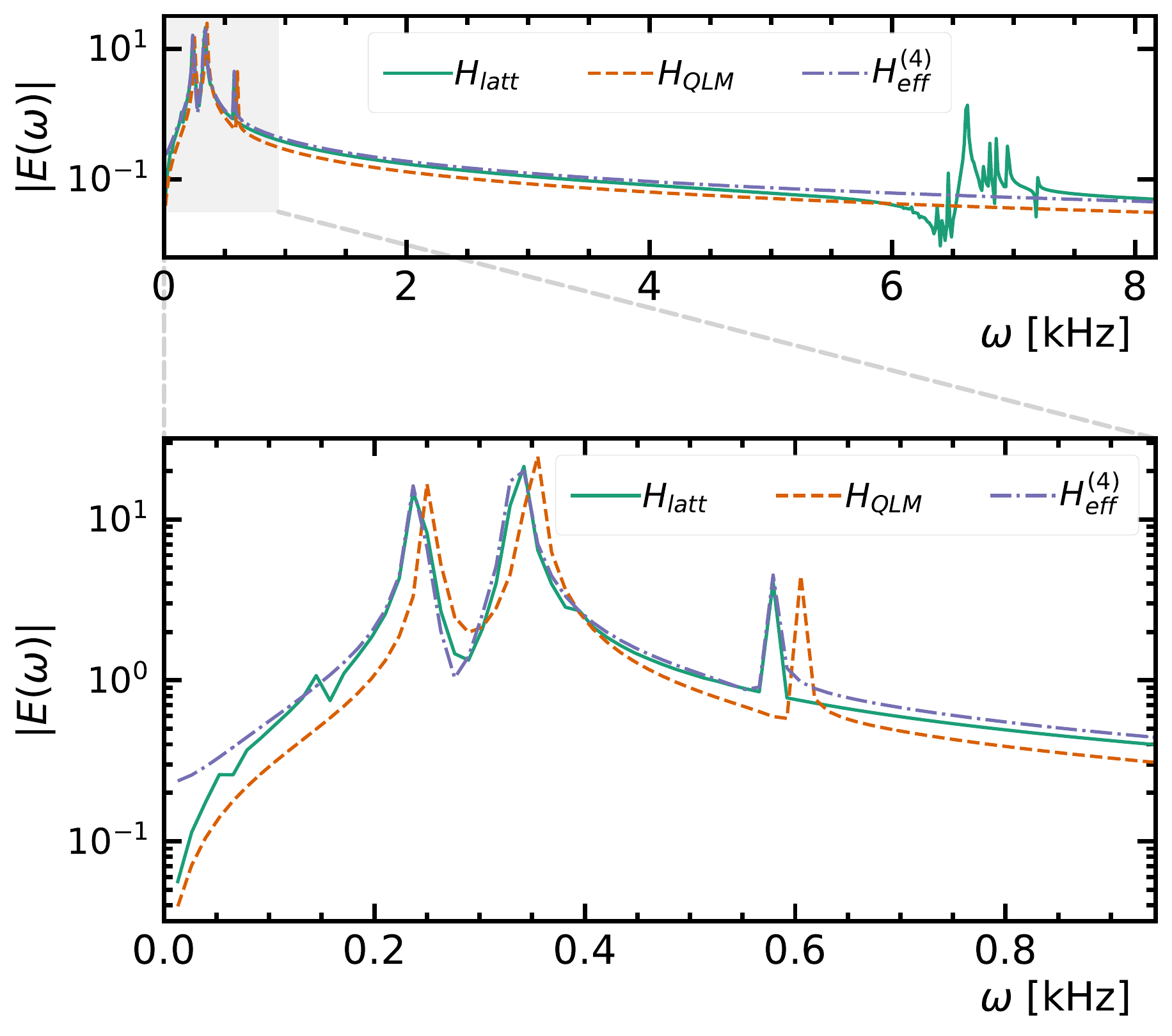}
    \put(0,87){(a)}
    \put(0,48){(b)}
    \end{overpic}
    \caption{
    \textbf{Fourier transform of the time evolution of the electric field.}
    (a) Spectrum corresponding to the time evolution shown in Fig.~\ref{fig:dyn1}(d).
    Panel~(b) shows a zoom-in of panel~(a) revealing details at small frequencies.}
    \label{fig:fourier1}
\end{figure}

To make this separation of energy scales more evident, we examine the Fourier transform of the signal $E(\omega)=\int_0^{t_{\max}}E(t)e^{-i\omega t}\mathrm{d}t$ in Fig.~\ref{fig:fourier1}. As expected, the evolution under $H_\text{latt}$ shows peaks at frequency $\omega\sim \delta/\hbar=(U-\delta)/\hbar=6.4$ kHz [Fig.~\ref{fig:fourier1}(a)], that are not observed for the effective Hamiltonians. Zooming in at smaller frequencies [Fig.~\ref{fig:fourier1}(b)], we see that the agreement in the Fourier transforms is good between $H_\text{latt}$ and $H_\text{QLM}$ and is excellent between $H_\text{latt}$ and $H_{\text{eff}}^{(4)}$. 

\subsection{Dissipation}
Finite dissipation can become a crucial issue in the experiment when it occurs on time scales comparable to the relevant dynamic evolution of the system.
Therefore, we focus on identifying the fastest dissipation process.
This allows us to estimate for how long the experimental system closely follows the coherent dynamics of our model.
For the AELA in an optical lattice considered here, the typically dominant  dissipation channels are lossy collisions between pairs of atoms and off-resonant scattering of optical lattice photons.
In the following, we evaluate the relevance of both for our proposed experimental implementation.

Lossy collisions between pairs of atoms with one or both of the two atoms in the $e$~state can lead to a particularly fast atom loss~\cite{Gorshkov2010,Scazza14}.
For our proposed implementation, however, double occupancies of lattice sites are (purposefully) strongly suppressed, either by fermionic quantum statistics ($ee$-pairs) or a large on-site interaction energy ($eg$-pairs).
As a consequence, lossy collisions between pairs are not expected to be a limiting factor.

In contrast, off-resonant photon scattering from lattice photons was identified as the dominant limiting factor in our proposed implementation.
While off-resonant photon scattering eventually leads to heating and atom loss, another effect could become relevant at earlier times.
When an atom in the $e$~state scatters an optical lattice photon, short-lived intermediate states can be populated and decay back to the $g$~state with a finite probability.
Conversion of $e$~atoms to $g$~atoms due to this optical pumping process has already been observed and characterized in optical lattice experiments with AELA~\cite{Riegger18,darkwahoppong:2022}.
Here, we employ these results to estimate the expected time scale for our parameters (see Table~\ref{tab:params}).
Focusing on the proposed implementation in Section~\ref{sec:opt-lattice} for ${{}^\text{173}}\text{Yb}$ and the off-resonant scattering of magic-wavelength lattice light, we estimate a repumping rate $\Gamma \approx 111\,\text{mHz}$.
Comparison of this estimate to the quantity $|w|/\hbar = 113\,\text{Hz}$ suggests that the dynamics of our model can be faithfully observed for many characteristic time scales.

\subsection{Disorder}
\label{sec:disorder}
An experimental implementation could also exhibit finite disorder.
Based on previous experimental work~\cite{Endres16}, disorder occurs in particular when utilizing hybrid potentials generated with both optical lattices and optical tweezers.
To estimate how much disorder affects the dynamic evolution of the system, we consider a quenched disorder in the chemical potentials of the atoms of the form

\begin{align}
    H_W=& H_{\text{latt}}+\sum_{j \text{ odd}}\left[W^{\phantom\dagger}_{g,j+1/2}c^\dagger_{j+1/2}c^{\phantom\dagger}_{j+1/2}\right.\\ \nonumber
    &\qquad\left.+W^{\phantom\dagger}_{g,j-1/2}c^\dagger_{j-1/2}c^{\phantom\dagger}_{j-1/2}+W^{\phantom\dagger}_{j}c^\dagger_{j}c^{\phantom\dagger}_{j}\right]\\ \nonumber
    &+\sum_{j \text{ even}}\left[W^{\phantom\dagger}_{e,j+1/2}d^\dagger_{j+1/2}d^{\phantom\dagger}_{j+1/2}\right.\\ \nonumber
    &\qquad\left.+W^{\phantom\dagger}_{e,j-1/2}d^\dagger_{j-1/2}d^{\phantom\dagger}_{j-1/2}+W^{\phantom\dagger}_{j}d^\dagger_{j}d^{\phantom\dagger}_{j}\right],
\end{align}
where $W_{\alpha, j+1/2}$, $W_j$ are taken randomly from a uniform distribution in the interval $[0,W)$.

\begin{figure}
    \centering
    \includegraphics[width=\linewidth]{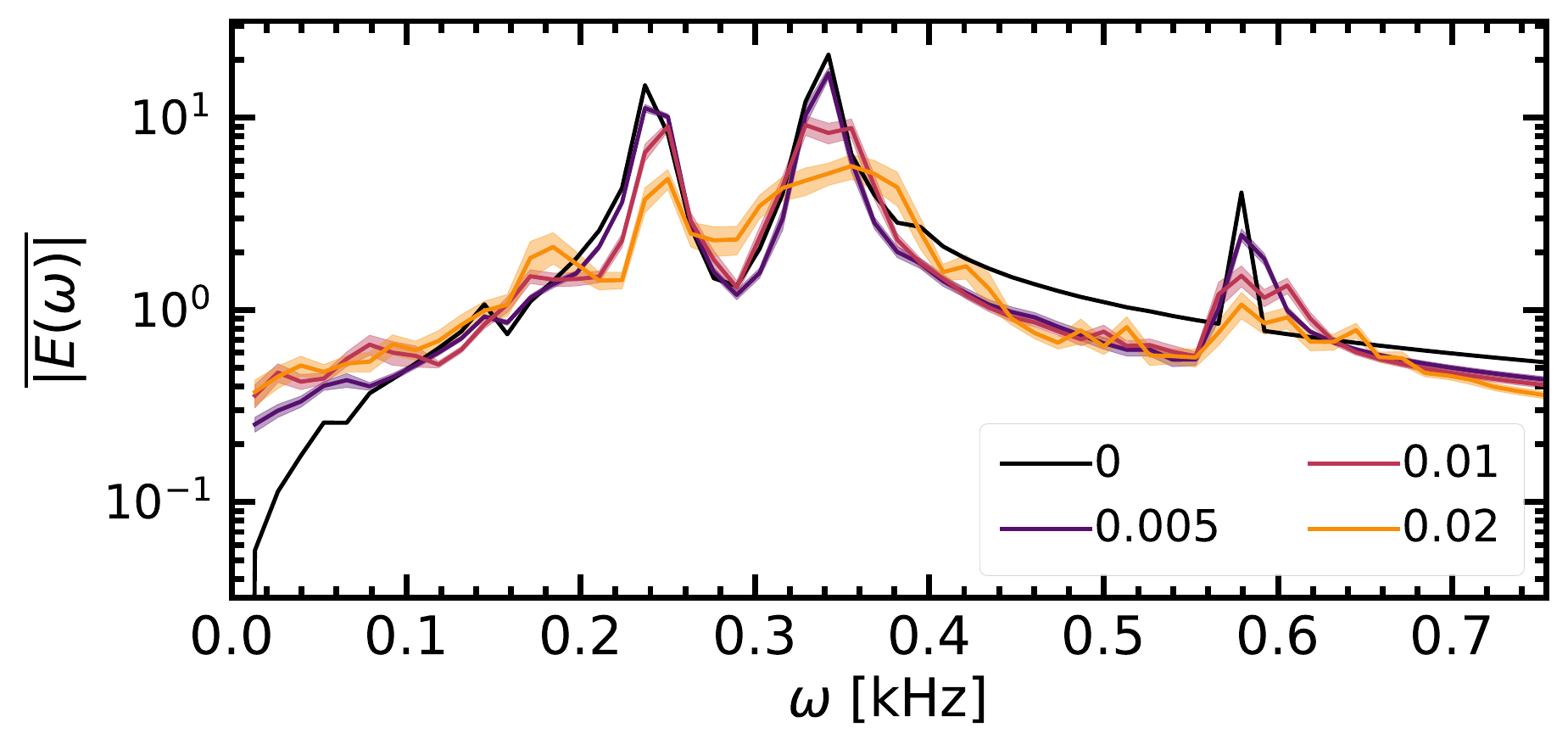}
    \caption{
    \textbf{Influence of disorder on the dynamics of the electric field.}
    Fourier transform of the time evolution of the electric field under $H_W$ for different values of $W$. The legend indicates the values of $W/h$ in kHz. The results are averaged over 100 disorder realizations.}
    \label{fig:dis1}
\end{figure}

In Fig.~\ref{fig:dis1} we plot the Fourier transform $E(\omega)$ averaged over $100$ disorder realizations for different values of $W$. We find that disorder has the effect of smearing out the peaks, which nevertheless remain visible up to $W/h\sim 0.01$ kHz.

\section{The 2D model}
We now generalize the implementation examined in the previous sections to the quantum link model in two spatial dimensions, and we show how this can be simulated with realistic experimental setups.

\subsection{Quantum link model}
The two-dimensional quantum link model is described by the Hamiltonian~\footnote{Note that here the matter gauge coupling has the same sign on every link, so this model differs from the quantum link model studied, for example, in~\cite{Banerjee2013}. With this choice, the model has a simpler implementation in the experimental setup that we propose. To obtain couplings with sign $\tilde s_{r,r+\hat x}=1$, $\tilde s_{r,r+\hat y}=(-1)^i$, an approach similar to Ref.~\cite{Jaksch_2003} could be applied.}
\begin{equation}
\begin{split}
\label{eq:QLM2D}
    H_{QLM}=&-w\sum_r\sum_{k=\hat x, \hat y} \left(\psi_r^\dagger U^{\phantom\dagger}_{r, r+k}\psi^{\phantom\dagger}_{r+k}+\text{H.c.}\right)\\
    &+m\sum_r s_r \psi_r^\dagger \psi^{\phantom\dagger}_r+\tau\sum_r \sum_{k=\hat x, \hat y} E_{r,r+k},
    \end{split}
\end{equation}
where the sums run over the points $r=(i,j)$ of a two-dimensional square lattice ($i,j$ are integers), and $s_r=(-1)^{i+j}$; here $\hat{x}$ and $\hat{y}$ denote unit vectors along the respective direction. The gauge fields sit on the links and, similarly to the one-dimensional case, are represented by spin variables with finite $d$-dimensional Hilbert spaces (here, $d=2$). The generators of the gauge symmetry read
\begin{equation}
G_r=\sum_{k=\hat x, \hat y} (E_{r, r+k}-E_{r, r-k})-\psi_r^\dagger \psi^{\phantom\dagger}_r+\frac{1-s_r}{2}.
\end{equation}

\begin{figure}
    \centering
    \includegraphics[width=\linewidth]{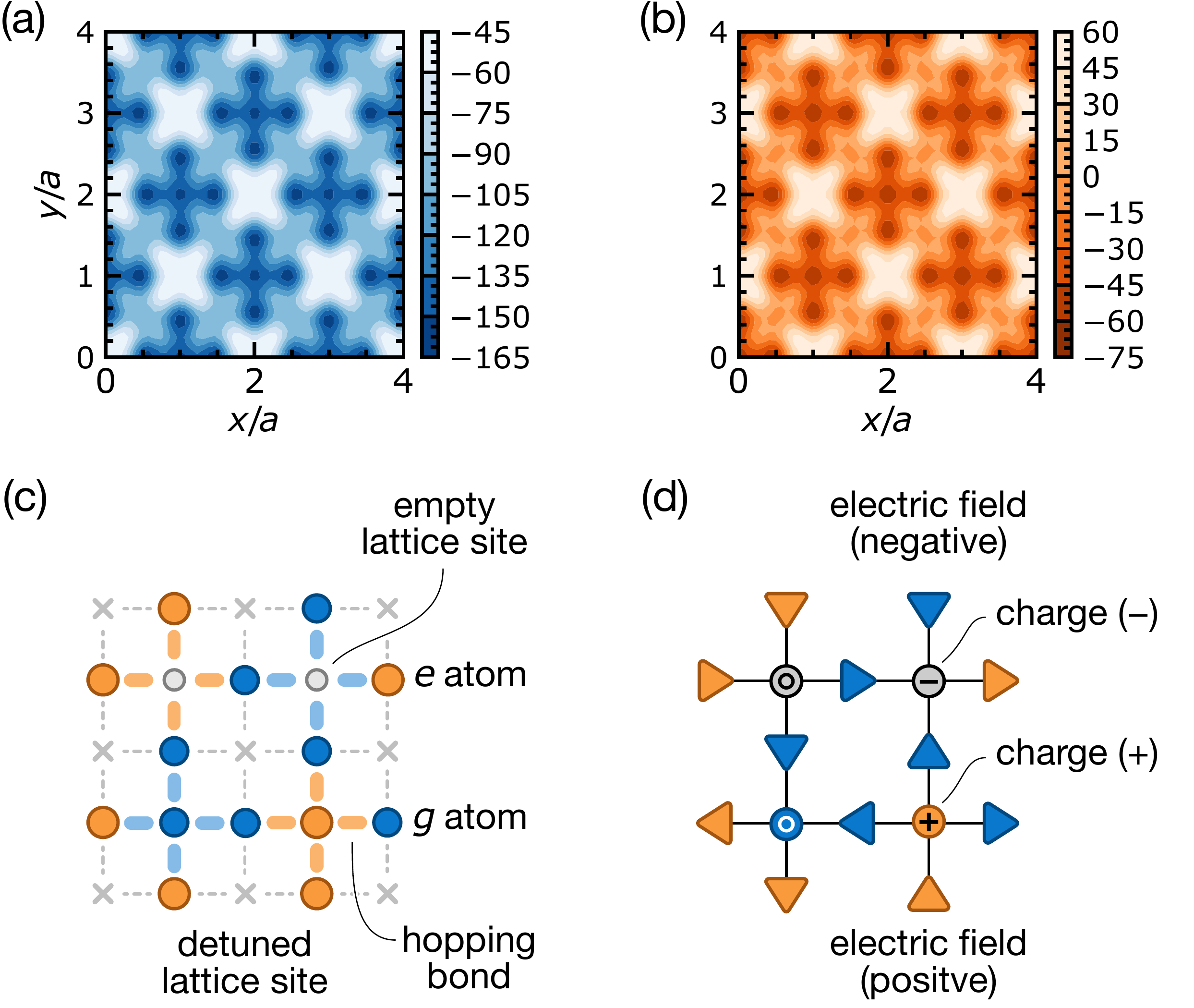}%
    \caption{
    \textbf{Implementation of the quantum link model in two dimensions.}
    Optical lattice for the (a) $g$ and (b) $e$ atoms.
    (c)~Example of a gauge-invariant state belonging to the resonant subspace.
    Blue and orange circles represent $g$ and $e$ atoms, respectively.
    (d)~Corresponding gauge-invariant state in the QLM. Dark/light gray circles indicate the occupied/empty matter sites (with charge $+$, $-$, or no charge).
    Red (blue) arrows represent a link with electric field $E_{r,r+k}=+1/2$ ($E_{r,r+k}=-1/2$).}
    \label{fig:2D}
\end{figure}

A state $\ket{\Psi}$ is gauge-invariant if it satisfies Gauss' law $G_r\ket{\Psi}=0$ for every lattice site $r$. An example of a gauge-invariant state is shown in Fig.~\ref{fig:2D}(d).

\subsection{Quantum simulation}
Our desired optical lattice in two dimensions consists of cross-shaped ``blocks" of $g$ and $e$ sites [see Fig.~\ref{fig:2D}(a-c)]. While in the one-dimensional case each block consisted of a triple well, here a block contains five sites: a central matter site at $r=(x/a,y/a)=(i,j)$, with $i,j$ integers, and four gauge sites around it at positions $r\pm (1/2,0)$ and $r\pm (0,1/2)$. Blocks of $g$ and $e$ sites alternate in a checkerboard pattern, as shown in Fig.~\ref{fig:2D}(c), with overlapping $g$ and $e$ gauge sites. A lattice of this type can be realized with the potential
\begin{equation}
\begin{split}
    &V^{x,y}_\alpha(x,y)=\\ &-A_\alpha\sin^2\left[\frac{\pi}{2a}(x+y)+\varphi\right]-A_\alpha\sin^2\left[\frac{\pi}{2a}(x-y)\right]\\ &\quad -B_\alpha\sin^2\left[\frac{\pi}{a}(x+y)\right]
    -B_\alpha\sin^2\left[\frac{\pi}{a}(x-y)\right]\\ &\quad -C_\alpha\sin^2\left(\frac{2\pi}{a}x+\frac{\pi}{2}\right)-C_\alpha\sin^2\left(\frac{2\pi}{a}y+\frac{\pi}{2}\right).
    \end{split}
\end{equation}
Fig.~\ref{fig:2D}(a,b) depicts the profiles of $V_g^{x,y}$ and $V_e^{x,y}$ for the values of $A_g, A_e, B_g, B_e,C_g, C_e$ reported in Table~\ref{tab:params2D}. 

The steps for deriving the lattice Hamiltonian and mapping it to the QLM are analogous to the one-dimensional case. We report here the mapping of the operators:
\begin{align}
E_{r,r+k}&=\frac{s_r}{2}(c_{r+k/2}^\dagger c^{\phantom\dagger}_{r+k/2}-d_{r+k/2}^\dagger d^{\phantom\dagger}_{r+k/2}),\\
\label{eq:mapmat2}
\psi_r &=    \left\{ \begin{matrix}
c_r \; \text{ if } s_r=-1,\\
d_r \; \text{ if } s_r=+1,
\end{matrix}\right.\\
U_{r,r+k}&= \left\{ \begin{matrix}
 c^{\phantom\dagger}_{r+k/2} d_{r+k/2}^\dagger \hspace{0.8cm} \text{if } s_r=+1,\\
 d^{\phantom\dagger}_{r+k/2}c_{r+k/2}^\dagger \hspace{0.8cm} \text{if } s_r=-1.
\end{matrix}\right.
\end{align}
The system is initialized with three $g$ atoms on each odd block, and two $e$ atoms on each even block.
To illustrate the mapping, we show in Fig.~\ref{fig:2D}(c) and (d) an example of a resonant state in the local occupation basis and the corresponding gauge-invariant state in the quantum link model.

Using the same derivation as the one-dimensional case, we obtain the effective Hamiltonian Eq.~(\ref{eq:QLM2D}). In Table \ref{tab:params2D} we report the parameters obtained for the lattice potential depicted in Fig.~\ref{fig:2D}(a,b). We set $a=\lambda_{m}=0.7594\,\mathrm{\upmu m}$ and $J_z\equiv \int \mathrm d z |\phi_g^z(z)|^2|\phi_e^z(z)|^2=19.235\; \mathrm{\upmu m}^{-1}$. For this choice of parameters we obtain $m=
10\, h\cdot$Hz and $w=
20\, h\cdot$Hz for the QLM Hamiltonian Eq.~(\ref{eq:QLM2D}).

\begin{table}[b]
\caption{\textbf{Experimental parameters for the two-dimensional quantum link model}. All values are given in units of $h\cdot$kHz.}\vspace{1em}
\begin{ruledtabular}
\begin{tabular}{@{}ccc@{}} 
$\;A_g=-A_e\;$ & $\;B_g=B_e\;$ & $\;C_g=C_e\;$ \\ \midrule
$
50.029$ & $
25.915$ & $
27.516$ \\ 
\end{tabular}
\end{ruledtabular}

\vspace{0.4cm}
\begin{ruledtabular}
\begin{tabular}{@{}ccccc@{}} 
$\;\Delta\;$ & $\;\delta_g=\delta_e\;$ & $\;U\;$ & $\;t_g=t_e\;$ & $\;D_g=D_e\;$\\ \midrule
$\;
9.63\;$ & $\;
1.08\;$ & $\;
2.16\;$ & $\;
0.087\;$ & $\;
0.033\;$\\ 
\end{tabular}
\end{ruledtabular}
\label{tab:params2D}
\end{table}

\section{Conclusions}

We have presented a proposal for the scalable quantum simulation of lattice gauge theories coupling (staggered) fermions to U(1) gauge fields utilizing a mixture of alkaline-earth(-like) atoms in both a ground and a metastable state in optical potentials. The key element of our proposal is a careful treatment of the full system dynamics, that are derived {\it ab initio} from microscopic interactions between atoms and light, and atoms themselves. While the proposal can be applied to a variety of atomic species, we have drawn a complete blueprint utilizing concrete estimates based on $^{173}$Yb atoms.

Our treatment highlights concrete challenges in the quantum simulation of lattice gauge theories that have so far mostly been overlooked. In particular, it makes clear that the superposition of lattice potentials required for such simulations, while certainly realistic experiment-wise, gives rise to complicated band structures that must be quantitatively understood to access the reliability and feasibility of any quantum simulation. The reason for this is twofold: band separation can become much smaller than what is naively expected, making protection of gauge invariance very challenging; in parallel, intrinsic energy scales of desired processes can be considerably reduced with respect to simplistic deep lattice estimates based on highly localized Wannier functions. These limitations are particularly pernicious for single-body terms in the lattice potential, whose estimate crucially requires a quantitative approach as the one carried out here. 

Within the context of our proposal, we have shown that optimal parameter regimes can still be found for observing the correct and expected LGT dynamics. This can be achieved thanks to the detailed microscopic understanding our treatment leads to. We have demonstrated this conclusion by comparing numerical simulations of both ideal and effective dynamics of string relaxation, including also effects of inhomogeneities. 

Based on our findings, we believe that the {\it ab initio} approach we propose will be, in the long term, the one needed to fully determine the capabilities of quantum simulators of lattice gauge theories. We have taken the first step beyond Abelian 1D models, by extending them to 2D geometries. Future works will be fundamental to address the experimental capabilities to realize non-Abelian lattice gauge theories, which to date have been proposed only in very few settings~\cite{Banerjee2013,Zohar_2013,Tagliacozzo_2013,Stannigel:2014bf,Mezzacapo2015,Rico_2018,Kasper2020,Davoudi20,gonzalez2022hardware}.

\begin{acknowledgements}
We thank M. Burrello, G. Pagano and E. Rico for insightful discussions, and F. Scazza for collaboration on a related work.
The work of M.~D., P.~F. and F.~S. was partly supported by the ERC under grant number 758329 (AGEnTh), and by the MIUR Programme FARE (MEPH). M.~A. and N.~D.~O. acknowledge funding from the Deutsche Forschungsgemeinschaft (DFG, German Research Foundation) under Germany’s Excellence Strategy – EXC-2111 –
390814868, from the European Research Council (ERC) under the European Union’s Horizon 2020 research and innovation program
(grant agreement No. 803047) and from the German Federal Ministry of Education and Research via the funding program quantum technologies -- from basic research to market (contract number 13N15895 FermiQP). M.~A. and M.~D. further acknowledge funding within the QuantERA II Programme that has received funding from the European Union's Horizon 2020 research and innovation programme under Grant Agreement No 101017733.
\end{acknowledgements}

\bibliography{bib}

%apsrev4-2.bst 2019-01-14 (MD) hand-edited version of apsrev4-1.bst
%Control: key (0)
%Control: author (8) initials jnrlst
%Control: editor formatted (1) identically to author
%Control: production of article title (0) allowed
%Control: page (0) single
%Control: year (1) truncated
%Control: production of eprint (0) enabled
\begin{thebibliography}{75}%
\makeatletter
\providecommand \@ifxundefined [1]{%
 \@ifx{#1\undefined}
}%
\providecommand \@ifnum [1]{%
 \ifnum #1\expandafter \@firstoftwo
 \else \expandafter \@secondoftwo
 \fi
}%
\providecommand \@ifx [1]{%
 \ifx #1\expandafter \@firstoftwo
 \else \expandafter \@secondoftwo
 \fi
}%
\providecommand \natexlab [1]{#1}%
\providecommand \enquote  [1]{``#1''}%
\providecommand \bibnamefont  [1]{#1}%
\providecommand \bibfnamefont [1]{#1}%
\providecommand \citenamefont [1]{#1}%
\providecommand \href@noop [0]{\@secondoftwo}%
\providecommand \href [0]{\begingroup \@sanitize@url \@href}%
\providecommand \@href[1]{\@@startlink{#1}\@@href}%
\providecommand \@@href[1]{\endgroup#1\@@endlink}%
\providecommand \@sanitize@url [0]{\catcode `\\12\catcode `\$12\catcode
  `\&12\catcode `\#12\catcode `\^12\catcode `\_12\catcode `\%12\relax}%
\providecommand \@@startlink[1]{}%
\providecommand \@@endlink[0]{}%
\providecommand \url  [0]{\begingroup\@sanitize@url \@url }%
\providecommand \@url [1]{\endgroup\@href {#1}{\urlprefix }}%
\providecommand \urlprefix  [0]{URL }%
\providecommand \Eprint [0]{\href }%
\providecommand \doibase [0]{https://doi.org/}%
\providecommand \selectlanguage [0]{\@gobble}%
\providecommand \bibinfo  [0]{\@secondoftwo}%
\providecommand \bibfield  [0]{\@secondoftwo}%
\providecommand \translation [1]{[#1]}%
\providecommand \BibitemOpen [0]{}%
\providecommand \bibitemStop [0]{}%
\providecommand \bibitemNoStop [0]{.\EOS\space}%
\providecommand \EOS [0]{\spacefactor3000\relax}%
\providecommand \BibitemShut  [1]{\csname bibitem#1\endcsname}%
\let\auto@bib@innerbib\@empty
%</preamble>
\bibitem [{\citenamefont {Wilson}(1974)}]{Wilson74}%
  \BibitemOpen
  \bibfield  {author} {\bibinfo {author} {\bibfnamefont {K.~G.}\ \bibnamefont
  {Wilson}},\ }\bibfield  {title} {\bibinfo {title} {Confinement of quarks},\
  }\href {https://doi.org/10.1103/PhysRevD.10.2445} {\bibfield  {journal}
  {\bibinfo  {journal} {Phys. Rev. D}\ }\textbf {\bibinfo {volume} {10}},\
  \bibinfo {pages} {2445} (\bibinfo {year} {1974})}\BibitemShut {NoStop}%
\bibitem [{\citenamefont {Kogut}(1979)}]{KogutReviewLGT}%
  \BibitemOpen
  \bibfield  {author} {\bibinfo {author} {\bibfnamefont {J.~B.}\ \bibnamefont
  {Kogut}},\ }\bibfield  {title} {\bibinfo {title} {An introduction to lattice
  gauge theory and spin systems},\ }\href
  {https://doi.org/10.1103/RevModPhys.51.659} {\bibfield  {journal} {\bibinfo
  {journal} {Rev. Mod. Phys.}\ }\textbf {\bibinfo {volume} {51}},\ \bibinfo
  {pages} {659} (\bibinfo {year} {1979})}\BibitemShut {NoStop}%
\bibitem [{\citenamefont {Montvay}\ and\ \citenamefont
  {Muenster}(1994)}]{Montvay1994}%
  \BibitemOpen
  \bibfield  {author} {\bibinfo {author} {\bibfnamefont {I.}~\bibnamefont
  {Montvay}}\ and\ \bibinfo {author} {\bibfnamefont {G.}~\bibnamefont
  {Muenster}},\ }\href@noop {} {\emph {\bibinfo {title} {{Quantum Fields on a
  lattice}}}}\ (\bibinfo  {publisher} {Cambridge Univ. Press, Cambridge},\
  \bibinfo {year} {1994})\BibitemShut {NoStop}%
\bibitem [{\citenamefont {Fodor}\ and\ \citenamefont
  {Hoelbling}(2012)}]{RevModPhys.84.449}%
  \BibitemOpen
  \bibfield  {author} {\bibinfo {author} {\bibfnamefont {Z.}~\bibnamefont
  {Fodor}}\ and\ \bibinfo {author} {\bibfnamefont {C.}~\bibnamefont
  {Hoelbling}},\ }\bibfield  {title} {\bibinfo {title} {Light hadron masses
  from lattice qcd},\ }\href {https://doi.org/10.1103/RevModPhys.84.449}
  {\bibfield  {journal} {\bibinfo  {journal} {Rev. Mod. Phys.}\ }\textbf
  {\bibinfo {volume} {84}},\ \bibinfo {pages} {449} (\bibinfo {year}
  {2012})}\BibitemShut {NoStop}%
\bibitem [{\citenamefont {Detmold}\ \emph {et~al.}(2019)\citenamefont
  {Detmold}, \citenamefont {Edwards}, \citenamefont {Dudek}, \citenamefont
  {Engelhardt}, \citenamefont {Lin}, \citenamefont {Meinel}, \citenamefont
  {Orginos},\ and\ \citenamefont {Shanahan}}]{detmold2019hadrons}%
  \BibitemOpen
  \bibfield  {author} {\bibinfo {author} {\bibfnamefont {W.}~\bibnamefont
  {Detmold}}, \bibinfo {author} {\bibfnamefont {R.~G.}\ \bibnamefont
  {Edwards}}, \bibinfo {author} {\bibfnamefont {J.~J.}\ \bibnamefont {Dudek}},
  \bibinfo {author} {\bibfnamefont {M.}~\bibnamefont {Engelhardt}}, \bibinfo
  {author} {\bibfnamefont {H.-W.}\ \bibnamefont {Lin}}, \bibinfo {author}
  {\bibfnamefont {S.}~\bibnamefont {Meinel}}, \bibinfo {author} {\bibfnamefont
  {K.}~\bibnamefont {Orginos}},\ and\ \bibinfo {author} {\bibfnamefont
  {P.}~\bibnamefont {Shanahan}},\ }\bibfield  {title} {\bibinfo {title}
  {Hadrons and nuclei},\ }\href
  {https://doi.org/https://doi.org/10.1140/epja/i2019-12902-4} {\bibfield
  {journal} {\bibinfo  {journal} {Eur. Phys. J. A}\ }\textbf {\bibinfo {volume}
  {55}},\ \bibinfo {pages} {1} (\bibinfo {year} {2019})}\BibitemShut {NoStop}%
\bibitem [{\citenamefont {DeTar}\ and\ \citenamefont
  {Heller}(2009)}]{detar2009qcd}%
  \BibitemOpen
  \bibfield  {author} {\bibinfo {author} {\bibfnamefont {C.}~\bibnamefont
  {DeTar}}\ and\ \bibinfo {author} {\bibfnamefont {U.}~\bibnamefont {Heller}},\
  }\bibfield  {title} {\bibinfo {title} {{QCD} thermodynamics from the
  lattice},\ }\href
  {https://doi.org/https://doi.org/10.1140/epja/i2009-10825-3} {\bibfield
  {journal} {\bibinfo  {journal} {Eur. Phys. J. A}\ }\textbf {\bibinfo {volume}
  {41}},\ \bibinfo {pages} {405} (\bibinfo {year} {2009})}\BibitemShut
  {NoStop}%
\bibitem [{\citenamefont {Fukushima}\ and\ \citenamefont
  {Hatsuda}(2010)}]{fukushima2010phase}%
  \BibitemOpen
  \bibfield  {author} {\bibinfo {author} {\bibfnamefont {K.}~\bibnamefont
  {Fukushima}}\ and\ \bibinfo {author} {\bibfnamefont {T.}~\bibnamefont
  {Hatsuda}},\ }\bibfield  {title} {\bibinfo {title} {The phase diagram of
  dense {QCD}},\ }\href {https://doi.org/10.1088/0034-4885/74/1/014001}
  {\bibfield  {journal} {\bibinfo  {journal} {Rep. Prog. Phys.}\ }\textbf
  {\bibinfo {volume} {74}},\ \bibinfo {pages} {014001} (\bibinfo {year}
  {2010})}\BibitemShut {NoStop}%
\bibitem [{\citenamefont {Philipsen}(2019)}]{philipsen2019constraining}%
  \BibitemOpen
  \bibfield  {author} {\bibinfo {author} {\bibfnamefont {O.}~\bibnamefont
  {Philipsen}},\ }\bibfield  {title} {\bibinfo {title} {Constraining the qcd
  phase diagram at finite temperature and density},\ }\href
  {https://arxiv.org/abs/1912.04827} {\bibfield  {journal} {\bibinfo  {journal}
  {arxiv:1912.04827}\ } (\bibinfo {year} {2019})}\BibitemShut {NoStop}%
\bibitem [{\citenamefont {Borsanyi}\ \emph {et~al.}(2021)\citenamefont
  {Borsanyi}, \citenamefont {Fodor}, \citenamefont {Guenther}, \citenamefont
  {Hoelbling}, \citenamefont {Katz}, \citenamefont {Lellouch}, \citenamefont
  {Lippert}, \citenamefont {Miura}, \citenamefont {Parato}, \citenamefont
  {Szabo} \emph {et~al.}}]{borsanyi2021leading}%
  \BibitemOpen
  \bibfield  {author} {\bibinfo {author} {\bibfnamefont {S.}~\bibnamefont
  {Borsanyi}}, \bibinfo {author} {\bibfnamefont {Z.}~\bibnamefont {Fodor}},
  \bibinfo {author} {\bibfnamefont {J.}~\bibnamefont {Guenther}}, \bibinfo
  {author} {\bibfnamefont {C.}~\bibnamefont {Hoelbling}}, \bibinfo {author}
  {\bibfnamefont {S.}~\bibnamefont {Katz}}, \bibinfo {author} {\bibfnamefont
  {L.}~\bibnamefont {Lellouch}}, \bibinfo {author} {\bibfnamefont
  {T.}~\bibnamefont {Lippert}}, \bibinfo {author} {\bibfnamefont
  {K.}~\bibnamefont {Miura}}, \bibinfo {author} {\bibfnamefont
  {L.}~\bibnamefont {Parato}}, \bibinfo {author} {\bibfnamefont
  {K.}~\bibnamefont {Szabo}}, \emph {et~al.},\ }\bibfield  {title} {\bibinfo
  {title} {Leading hadronic contribution to the muon magnetic moment from
  lattice {QCD}},\ }\href {https://doi.org/10.1038/s41586-021-03418-1}
  {\bibfield  {journal} {\bibinfo  {journal} {Nature}\ }\textbf {\bibinfo
  {volume} {593}},\ \bibinfo {pages} {51} (\bibinfo {year} {2021})}\BibitemShut
  {NoStop}%
\bibitem [{\citenamefont {Wiese}(2013)}]{Wiese:2013kk}%
  \BibitemOpen
  \bibfield  {author} {\bibinfo {author} {\bibfnamefont {U.~J.}\ \bibnamefont
  {Wiese}},\ }\bibfield  {title} {\bibinfo {title} {{Ultracold quantum gases
  and lattice systems: quantum simulation of lattice gauge theories}},\ }\href
  {https://doi.org/10.1002/andp.201300104} {\bibfield  {journal} {\bibinfo
  {journal} {Ann. Phys.}\ }\textbf {\bibinfo {volume} {525}},\ \bibinfo {pages}
  {777} (\bibinfo {year} {2013})}\BibitemShut {NoStop}%
\bibitem [{\citenamefont {Dalmonte}\ and\ \citenamefont
  {Montangero}(2016)}]{Dalmonte:2016jk}%
  \BibitemOpen
  \bibfield  {author} {\bibinfo {author} {\bibfnamefont {M.}~\bibnamefont
  {Dalmonte}}\ and\ \bibinfo {author} {\bibfnamefont {S.}~\bibnamefont
  {Montangero}},\ }\bibfield  {title} {\bibinfo {title} {{Lattice gauge
  theories simulations in the quantum information era}},\ }\href
  {https://doi.org/10.1080/00107514.2016.1151199} {\bibfield  {journal}
  {\bibinfo  {journal} {Contemp. Phys.}\ }\textbf {\bibinfo {volume} {57}},\
  \bibinfo {pages} {388} (\bibinfo {year} {2016})}\BibitemShut {NoStop}%
\bibitem [{\citenamefont {Zohar}\ \emph {et~al.}(2016)\citenamefont {Zohar},
  \citenamefont {Cirac},\ and\ \citenamefont {Reznik}}]{Zohar2015}%
  \BibitemOpen
  \bibfield  {author} {\bibinfo {author} {\bibfnamefont {E.}~\bibnamefont
  {Zohar}}, \bibinfo {author} {\bibfnamefont {I.}~\bibnamefont {Cirac}},\ and\
  \bibinfo {author} {\bibfnamefont {B.}~\bibnamefont {Reznik}},\ }\bibfield
  {title} {\bibinfo {title} {{Quantum Simulations of Lattice Gauge Theories
  using Ultracold Atoms in Optical Lattices}},\ }\href
  {https://doi.org/10.1088/0034-4885/79/1/014401} {\bibfield  {journal}
  {\bibinfo  {journal} {Rep. Prog. Phys.}\ }\textbf {\bibinfo {volume} {79}},\
  \bibinfo {pages} {014401} (\bibinfo {year} {2016})}\BibitemShut {NoStop}%
\bibitem [{\citenamefont {Ba$\tilde{\text{n}}$uls}\ \emph
  {et~al.}(2020)\citenamefont {Ba$\tilde{\text{n}}$uls}, \citenamefont {Blatt},
  \citenamefont {Catani}, \citenamefont {Celi}, \citenamefont {Cirac},
  \citenamefont {Dalmonte}, \citenamefont {Fallani}, \citenamefont {Jansen},
  \citenamefont {Lewenstein}, \citenamefont {Montangero} \emph
  {et~al.}}]{banuls2020simulating}%
  \BibitemOpen
  \bibfield  {author} {\bibinfo {author} {\bibfnamefont {M.~C.}\ \bibnamefont
  {Ba$\tilde{\text{n}}$uls}}, \bibinfo {author} {\bibfnamefont
  {R.}~\bibnamefont {Blatt}}, \bibinfo {author} {\bibfnamefont
  {J.}~\bibnamefont {Catani}}, \bibinfo {author} {\bibfnamefont
  {A.}~\bibnamefont {Celi}}, \bibinfo {author} {\bibfnamefont {J.~I.}\
  \bibnamefont {Cirac}}, \bibinfo {author} {\bibfnamefont {M.}~\bibnamefont
  {Dalmonte}}, \bibinfo {author} {\bibfnamefont {L.}~\bibnamefont {Fallani}},
  \bibinfo {author} {\bibfnamefont {K.}~\bibnamefont {Jansen}}, \bibinfo
  {author} {\bibfnamefont {M.}~\bibnamefont {Lewenstein}}, \bibinfo {author}
  {\bibfnamefont {S.}~\bibnamefont {Montangero}}, \emph {et~al.},\ }\bibfield
  {title} {\bibinfo {title} {Simulating lattice gauge theories within quantum
  technologies},\ }\href {https://doi.org/10.1140/epjd/e2020-100571-8}
  {\bibfield  {journal} {\bibinfo  {journal} {Eur. Phys. J. D}\ }\textbf
  {\bibinfo {volume} {74}},\ \bibinfo {pages} {1} (\bibinfo {year}
  {2020})}\BibitemShut {NoStop}%
\bibitem [{\citenamefont {Aidelsburger}\ \emph {et~al.}(2022)\citenamefont
  {Aidelsburger}, \citenamefont {Barbiero}, \citenamefont {Bermudez},
  \citenamefont {Chanda}, \citenamefont {Dauphin}, \citenamefont
  {González-Cuadra}, \citenamefont {Grzybowski}, \citenamefont {Hands},
  \citenamefont {Jendrzejewski}, \citenamefont {Jünemann} \emph
  {et~al.}}]{aidelsburger_cold_2022}%
  \BibitemOpen
  \bibfield  {author} {\bibinfo {author} {\bibfnamefont {M.}~\bibnamefont
  {Aidelsburger}}, \bibinfo {author} {\bibfnamefont {L.}~\bibnamefont
  {Barbiero}}, \bibinfo {author} {\bibfnamefont {A.}~\bibnamefont {Bermudez}},
  \bibinfo {author} {\bibfnamefont {T.}~\bibnamefont {Chanda}}, \bibinfo
  {author} {\bibfnamefont {A.}~\bibnamefont {Dauphin}}, \bibinfo {author}
  {\bibfnamefont {D.}~\bibnamefont {González-Cuadra}}, \bibinfo {author}
  {\bibfnamefont {P.~R.}\ \bibnamefont {Grzybowski}}, \bibinfo {author}
  {\bibfnamefont {S.}~\bibnamefont {Hands}}, \bibinfo {author} {\bibfnamefont
  {F.}~\bibnamefont {Jendrzejewski}}, \bibinfo {author} {\bibfnamefont
  {J.}~\bibnamefont {Jünemann}}, \emph {et~al.},\ }\bibfield  {title}
  {\bibinfo {title} {Cold atoms meet lattice gauge theory},\ }\href
  {https://doi.org/10.1098/rsta.2021.0064} {\bibfield  {journal} {\bibinfo
  {journal} {Phil. Trans. R. Soc. A.}\ }\textbf {\bibinfo {volume} {380}},\
  \bibinfo {pages} {20210064} (\bibinfo {year} {2022})}\BibitemShut {NoStop}%
\bibitem [{\citenamefont {Davoudi}\ \emph {et~al.}(2022)\citenamefont
  {Davoudi}, \citenamefont {Bauer}, \citenamefont {Balantekin}, \citenamefont
  {Bhattacharya}, \citenamefont {Carena}, \citenamefont {de~Jong},
  \citenamefont {Draper}, \citenamefont {El-Khadra}, \citenamefont {Gemelke},
  \citenamefont {Hanada} \emph {et~al.}}]{davoudi2022quantum}%
  \BibitemOpen
  \bibfield  {author} {\bibinfo {author} {\bibfnamefont {Z.}~\bibnamefont
  {Davoudi}}, \bibinfo {author} {\bibfnamefont {C.}~\bibnamefont {Bauer}},
  \bibinfo {author} {\bibfnamefont {A.}~\bibnamefont {Balantekin}}, \bibinfo
  {author} {\bibfnamefont {T.}~\bibnamefont {Bhattacharya}}, \bibinfo {author}
  {\bibfnamefont {M.}~\bibnamefont {Carena}}, \bibinfo {author} {\bibfnamefont
  {W.~A.}\ \bibnamefont {de~Jong}}, \bibinfo {author} {\bibfnamefont
  {P.}~\bibnamefont {Draper}}, \bibinfo {author} {\bibfnamefont
  {A.}~\bibnamefont {El-Khadra}}, \bibinfo {author} {\bibfnamefont
  {N.}~\bibnamefont {Gemelke}}, \bibinfo {author} {\bibfnamefont
  {M.}~\bibnamefont {Hanada}}, \emph {et~al.},\ }\bibfield  {title} {\bibinfo
  {title} {Quantum simulation for high energy physics},\ }\href
  {https://arxiv.org/abs/2204.03381} {\bibfield  {journal} {\bibinfo  {journal}
  {arXiv:2204.03381}\ } (\bibinfo {year} {2022})}\BibitemShut {NoStop}%
\bibitem [{\citenamefont {Martinez}\ \emph {et~al.}(2016)\citenamefont
  {Martinez}, \citenamefont {Muschik}, \citenamefont {Schindler}, \citenamefont
  {Nigg}, \citenamefont {Erhard}, \citenamefont {Heyl}, \citenamefont {Hauke},
  \citenamefont {Dalmonte}, \citenamefont {Monz}, \citenamefont {Zoller},\ and\
  \citenamefont {Blatt}}]{ExpPaper}%
  \BibitemOpen
  \bibfield  {author} {\bibinfo {author} {\bibfnamefont {E.~A.}\ \bibnamefont
  {Martinez}}, \bibinfo {author} {\bibfnamefont {C.~A.}\ \bibnamefont
  {Muschik}}, \bibinfo {author} {\bibfnamefont {P.}~\bibnamefont {Schindler}},
  \bibinfo {author} {\bibfnamefont {D.}~\bibnamefont {Nigg}}, \bibinfo {author}
  {\bibfnamefont {A.}~\bibnamefont {Erhard}}, \bibinfo {author} {\bibfnamefont
  {M.}~\bibnamefont {Heyl}}, \bibinfo {author} {\bibfnamefont {P.}~\bibnamefont
  {Hauke}}, \bibinfo {author} {\bibfnamefont {M.}~\bibnamefont {Dalmonte}},
  \bibinfo {author} {\bibfnamefont {T.}~\bibnamefont {Monz}}, \bibinfo {author}
  {\bibfnamefont {P.}~\bibnamefont {Zoller}},\ and\ \bibinfo {author}
  {\bibfnamefont {R.}~\bibnamefont {Blatt}},\ }\bibfield  {title} {\bibinfo
  {title} {{Real-time dynamics of lattice gauge theories with a few-qubit
  quantum computer}},\ }\href {https://doi.org/10.1038/nature18318} {\bibfield
  {journal} {\bibinfo  {journal} {Nature}\ }\textbf {\bibinfo {volume} {534}},\
  \bibinfo {pages} {516} (\bibinfo {year} {2016})}\BibitemShut {NoStop}%
\bibitem [{\citenamefont {Nguyen}\ \emph {et~al.}(2021)\citenamefont {Nguyen},
  \citenamefont {Tran}, \citenamefont {Zhu}, \citenamefont {Green},
  \citenamefont {Alderete}, \citenamefont {Davoudi},\ and\ \citenamefont
  {Linke}}]{nguyen_digital_2022}%
  \BibitemOpen
  \bibfield  {author} {\bibinfo {author} {\bibfnamefont {N.~H.}\ \bibnamefont
  {Nguyen}}, \bibinfo {author} {\bibfnamefont {M.~C.}\ \bibnamefont {Tran}},
  \bibinfo {author} {\bibfnamefont {Y.}~\bibnamefont {Zhu}}, \bibinfo {author}
  {\bibfnamefont {A.~M.}\ \bibnamefont {Green}}, \bibinfo {author}
  {\bibfnamefont {C.~H.}\ \bibnamefont {Alderete}}, \bibinfo {author}
  {\bibfnamefont {Z.}~\bibnamefont {Davoudi}},\ and\ \bibinfo {author}
  {\bibfnamefont {N.~M.}\ \bibnamefont {Linke}},\ }\bibfield  {title} {\bibinfo
  {title} {Digital {Quantum} {Simulation} of the {Schwinger} {Model} and
  {Symmetry} {Protection} with {Trapped} {Ions}},\ }\href
  {https://arxiv.org/abs/2204.03381} {\bibfield  {journal} {\bibinfo  {journal}
  {arXiv:2112.14262}\ } (\bibinfo {year} {2021})}\BibitemShut {NoStop}%
\bibitem [{\citenamefont {Schweizer}\ \emph {et~al.}(2019)\citenamefont
  {Schweizer}, \citenamefont {Grusdt}, \citenamefont {Berngruber},
  \citenamefont {Barbiero}, \citenamefont {Demler}, \citenamefont {Goldman},
  \citenamefont {Bloch},\ and\ \citenamefont
  {Aidelsburger}}]{schweizer2019floquet}%
  \BibitemOpen
  \bibfield  {author} {\bibinfo {author} {\bibfnamefont {C.}~\bibnamefont
  {Schweizer}}, \bibinfo {author} {\bibfnamefont {F.}~\bibnamefont {Grusdt}},
  \bibinfo {author} {\bibfnamefont {M.}~\bibnamefont {Berngruber}}, \bibinfo
  {author} {\bibfnamefont {L.}~\bibnamefont {Barbiero}}, \bibinfo {author}
  {\bibfnamefont {E.}~\bibnamefont {Demler}}, \bibinfo {author} {\bibfnamefont
  {N.}~\bibnamefont {Goldman}}, \bibinfo {author} {\bibfnamefont
  {I.}~\bibnamefont {Bloch}},\ and\ \bibinfo {author} {\bibfnamefont
  {M.}~\bibnamefont {Aidelsburger}},\ }\bibfield  {title} {\bibinfo {title}
  {Floquet approach to $\mathbb{Z}_2$ lattice gauge theories with ultracold
  atoms in optical lattices},\ }\href
  {https://doi.org/10.1038/s41567-019-0649-7} {\bibfield  {journal} {\bibinfo
  {journal} {Nature Physics}\ }\textbf {\bibinfo {volume} {15}},\ \bibinfo
  {pages} {1168} (\bibinfo {year} {2019})}\BibitemShut {NoStop}%
\bibitem [{\citenamefont {Mil}\ \emph {et~al.}(2020)\citenamefont {Mil},
  \citenamefont {Zache}, \citenamefont {Hegde}, \citenamefont {Xia},
  \citenamefont {Bhatt}, \citenamefont {Oberthaler}, \citenamefont {Hauke},
  \citenamefont {Berges},\ and\ \citenamefont
  {Jendrzejewski}}]{mil2020scalable}%
  \BibitemOpen
  \bibfield  {author} {\bibinfo {author} {\bibfnamefont {A.}~\bibnamefont
  {Mil}}, \bibinfo {author} {\bibfnamefont {T.~V.}\ \bibnamefont {Zache}},
  \bibinfo {author} {\bibfnamefont {A.}~\bibnamefont {Hegde}}, \bibinfo
  {author} {\bibfnamefont {A.}~\bibnamefont {Xia}}, \bibinfo {author}
  {\bibfnamefont {R.~P.}\ \bibnamefont {Bhatt}}, \bibinfo {author}
  {\bibfnamefont {M.~K.}\ \bibnamefont {Oberthaler}}, \bibinfo {author}
  {\bibfnamefont {P.}~\bibnamefont {Hauke}}, \bibinfo {author} {\bibfnamefont
  {J.}~\bibnamefont {Berges}},\ and\ \bibinfo {author} {\bibfnamefont
  {F.}~\bibnamefont {Jendrzejewski}},\ }\bibfield  {title} {\bibinfo {title} {A
  scalable realization of local {U(1)} gauge invariance in cold atomic
  mixtures},\ }\href {https://doi.org/10.1126/science.aaz5312} {\bibfield
  {journal} {\bibinfo  {journal} {Science}\ }\textbf {\bibinfo {volume}
  {367}},\ \bibinfo {pages} {1128} (\bibinfo {year} {2020})}\BibitemShut
  {NoStop}%
\bibitem [{\citenamefont {Bernien}\ \emph {et~al.}(2017)\citenamefont
  {Bernien}, \citenamefont {Schwartz}, \citenamefont {Keesling}, \citenamefont
  {Levine}, \citenamefont {Omran}, \citenamefont {Pichler}, \citenamefont
  {Choi}, \citenamefont {Zibrov}, \citenamefont {Endres}, \citenamefont
  {Greiner} \emph {et~al.}}]{Bernien2017}%
  \BibitemOpen
  \bibfield  {author} {\bibinfo {author} {\bibfnamefont {H.}~\bibnamefont
  {Bernien}}, \bibinfo {author} {\bibfnamefont {S.}~\bibnamefont {Schwartz}},
  \bibinfo {author} {\bibfnamefont {A.}~\bibnamefont {Keesling}}, \bibinfo
  {author} {\bibfnamefont {H.}~\bibnamefont {Levine}}, \bibinfo {author}
  {\bibfnamefont {A.}~\bibnamefont {Omran}}, \bibinfo {author} {\bibfnamefont
  {H.}~\bibnamefont {Pichler}}, \bibinfo {author} {\bibfnamefont
  {S.}~\bibnamefont {Choi}}, \bibinfo {author} {\bibfnamefont {A.~S.}\
  \bibnamefont {Zibrov}}, \bibinfo {author} {\bibfnamefont {M.}~\bibnamefont
  {Endres}}, \bibinfo {author} {\bibfnamefont {M.}~\bibnamefont {Greiner}},
  \emph {et~al.},\ }\bibfield  {title} {\bibinfo {title} {Probing many-body
  dynamics on a 51-atom quantum simulator},\ }\href
  {https://doi.org/10.1038/nature24622} {\bibfield  {journal} {\bibinfo
  {journal} {Nature}\ }\textbf {\bibinfo {volume} {551}},\ \bibinfo {pages}
  {579} (\bibinfo {year} {2017})}\BibitemShut {NoStop}%
\bibitem [{\citenamefont {Surace}\ \emph {et~al.}(2020)\citenamefont {Surace},
  \citenamefont {Mazza}, \citenamefont {Giudici}, \citenamefont {Lerose},
  \citenamefont {Gambassi},\ and\ \citenamefont
  {Dalmonte}}]{surace2020lattice}%
  \BibitemOpen
  \bibfield  {author} {\bibinfo {author} {\bibfnamefont {F.~M.}\ \bibnamefont
  {Surace}}, \bibinfo {author} {\bibfnamefont {P.~P.}\ \bibnamefont {Mazza}},
  \bibinfo {author} {\bibfnamefont {G.}~\bibnamefont {Giudici}}, \bibinfo
  {author} {\bibfnamefont {A.}~\bibnamefont {Lerose}}, \bibinfo {author}
  {\bibfnamefont {A.}~\bibnamefont {Gambassi}},\ and\ \bibinfo {author}
  {\bibfnamefont {M.}~\bibnamefont {Dalmonte}},\ }\bibfield  {title} {\bibinfo
  {title} {Lattice gauge theories and string dynamics in rydberg atom quantum
  simulators},\ }\href {https://doi.org/10.1103/PhysRevX.10.021041} {\bibfield
  {journal} {\bibinfo  {journal} {Phys. Rev. X}\ }\textbf {\bibinfo {volume}
  {10}},\ \bibinfo {pages} {021041} (\bibinfo {year} {2020})}\BibitemShut
  {NoStop}%
\bibitem [{\citenamefont {Semeghini}\ \emph {et~al.}(2021)\citenamefont
  {Semeghini}, \citenamefont {Levine}, \citenamefont {Keesling}, \citenamefont
  {Ebadi}, \citenamefont {Wang}, \citenamefont {Bluvstein}, \citenamefont
  {Verresen}, \citenamefont {Pichler}, \citenamefont {Kalinowski},
  \citenamefont {Samajdar} \emph {et~al.}}]{semeghini2021probing}%
  \BibitemOpen
  \bibfield  {author} {\bibinfo {author} {\bibfnamefont {G.}~\bibnamefont
  {Semeghini}}, \bibinfo {author} {\bibfnamefont {H.}~\bibnamefont {Levine}},
  \bibinfo {author} {\bibfnamefont {A.}~\bibnamefont {Keesling}}, \bibinfo
  {author} {\bibfnamefont {S.}~\bibnamefont {Ebadi}}, \bibinfo {author}
  {\bibfnamefont {T.~T.}\ \bibnamefont {Wang}}, \bibinfo {author}
  {\bibfnamefont {D.}~\bibnamefont {Bluvstein}}, \bibinfo {author}
  {\bibfnamefont {R.}~\bibnamefont {Verresen}}, \bibinfo {author}
  {\bibfnamefont {H.}~\bibnamefont {Pichler}}, \bibinfo {author} {\bibfnamefont
  {M.}~\bibnamefont {Kalinowski}}, \bibinfo {author} {\bibfnamefont
  {R.}~\bibnamefont {Samajdar}}, \emph {et~al.},\ }\bibfield  {title} {\bibinfo
  {title} {Probing topological spin liquids on a programmable quantum
  simulator},\ }\href {https://doi.org/10.1126/science.abi8794} {\bibfield
  {journal} {\bibinfo  {journal} {Science}\ }\textbf {\bibinfo {volume}
  {374}},\ \bibinfo {pages} {1242} (\bibinfo {year} {2021})}\BibitemShut
  {NoStop}%
\bibitem [{\citenamefont {Yang}\ \emph {et~al.}(2020)\citenamefont {Yang},
  \citenamefont {Sun}, \citenamefont {Ott}, \citenamefont {Wang}, \citenamefont
  {Zache}, \citenamefont {Halimeh}, \citenamefont {Yuan}, \citenamefont
  {Hauke},\ and\ \citenamefont {Pan}}]{yang2020observation}%
  \BibitemOpen
  \bibfield  {author} {\bibinfo {author} {\bibfnamefont {B.}~\bibnamefont
  {Yang}}, \bibinfo {author} {\bibfnamefont {H.}~\bibnamefont {Sun}}, \bibinfo
  {author} {\bibfnamefont {R.}~\bibnamefont {Ott}}, \bibinfo {author}
  {\bibfnamefont {H.-Y.}\ \bibnamefont {Wang}}, \bibinfo {author}
  {\bibfnamefont {T.~V.}\ \bibnamefont {Zache}}, \bibinfo {author}
  {\bibfnamefont {J.~C.}\ \bibnamefont {Halimeh}}, \bibinfo {author}
  {\bibfnamefont {Z.-S.}\ \bibnamefont {Yuan}}, \bibinfo {author}
  {\bibfnamefont {P.}~\bibnamefont {Hauke}},\ and\ \bibinfo {author}
  {\bibfnamefont {J.-W.}\ \bibnamefont {Pan}},\ }\bibfield  {title} {\bibinfo
  {title} {Observation of gauge invariance in a 71-site {Bose-Hubbard} quantum
  simulator},\ }\href {https://doi.org/10.1038/s41586-020-2910-8} {\bibfield
  {journal} {\bibinfo  {journal} {Nature}\ }\textbf {\bibinfo {volume} {587}},\
  \bibinfo {pages} {392} (\bibinfo {year} {2020})}\BibitemShut {NoStop}%
\bibitem [{\citenamefont {Zhou}\ \emph {et~al.}(2022)\citenamefont {Zhou},
  \citenamefont {Su}, \citenamefont {Halimeh}, \citenamefont {Ott},
  \citenamefont {Sun}, \citenamefont {Hauke}, \citenamefont {Yang},
  \citenamefont {Yuan}, \citenamefont {Berges},\ and\ \citenamefont
  {Pan}}]{zhou_thermalization_2022}%
  \BibitemOpen
  \bibfield  {author} {\bibinfo {author} {\bibfnamefont {Z.-Y.}\ \bibnamefont
  {Zhou}}, \bibinfo {author} {\bibfnamefont {G.-X.}\ \bibnamefont {Su}},
  \bibinfo {author} {\bibfnamefont {J.~C.}\ \bibnamefont {Halimeh}}, \bibinfo
  {author} {\bibfnamefont {R.}~\bibnamefont {Ott}}, \bibinfo {author}
  {\bibfnamefont {H.}~\bibnamefont {Sun}}, \bibinfo {author} {\bibfnamefont
  {P.}~\bibnamefont {Hauke}}, \bibinfo {author} {\bibfnamefont
  {B.}~\bibnamefont {Yang}}, \bibinfo {author} {\bibfnamefont {Z.-S.}\
  \bibnamefont {Yuan}}, \bibinfo {author} {\bibfnamefont {J.}~\bibnamefont
  {Berges}},\ and\ \bibinfo {author} {\bibfnamefont {J.-W.}\ \bibnamefont
  {Pan}},\ }\bibfield  {title} {\bibinfo {title} {Thermalization dynamics of a
  gauge theory on a quantum simulator},\ }\href
  {https://doi.org/10.1126/science.abl6277} {\bibfield  {journal} {\bibinfo
  {journal} {Science}\ }\textbf {\bibinfo {volume} {377}},\ \bibinfo {pages}
  {311} (\bibinfo {year} {2022})}\BibitemShut {NoStop}%
\bibitem [{\citenamefont {Wang}\ \emph {et~al.}(2022)\citenamefont {Wang},
  \citenamefont {Zhang}, \citenamefont {Yao}, \citenamefont {Liu},
  \citenamefont {Zhu}, \citenamefont {Zheng}, \citenamefont {Wang},
  \citenamefont {Zhai}, \citenamefont {Yuan},\ and\ \citenamefont
  {Pan}}]{wang_interrelated_2022}%
  \BibitemOpen
  \bibfield  {author} {\bibinfo {author} {\bibfnamefont {H.-Y.}\ \bibnamefont
  {Wang}}, \bibinfo {author} {\bibfnamefont {W.-Y.}\ \bibnamefont {Zhang}},
  \bibinfo {author} {\bibfnamefont {Z.-Y.}\ \bibnamefont {Yao}}, \bibinfo
  {author} {\bibfnamefont {Y.}~\bibnamefont {Liu}}, \bibinfo {author}
  {\bibfnamefont {Z.-H.}\ \bibnamefont {Zhu}}, \bibinfo {author} {\bibfnamefont
  {Y.-G.}\ \bibnamefont {Zheng}}, \bibinfo {author} {\bibfnamefont {X.-K.}\
  \bibnamefont {Wang}}, \bibinfo {author} {\bibfnamefont {H.}~\bibnamefont
  {Zhai}}, \bibinfo {author} {\bibfnamefont {Z.-S.}\ \bibnamefont {Yuan}},\
  and\ \bibinfo {author} {\bibfnamefont {J.-W.}\ \bibnamefont {Pan}},\
  }\bibfield  {title} {\bibinfo {title} {Interrelated {Thermalization} and
  {Quantum} {Criticality} in a {Lattice} {Gauge} {Simulator}},\ }\href
  {http://arxiv.org/abs/2210.17032} {\bibfield  {journal} {\bibinfo  {journal}
  {arXiv:2210.17032}\ } (\bibinfo {year} {2022})}\BibitemShut {NoStop}%
\bibitem [{\citenamefont {Fr{\"o}lian}\ \emph {et~al.}(2022)\citenamefont
  {Fr{\"o}lian}, \citenamefont {Chisholm}, \citenamefont {Neri}, \citenamefont
  {Cabrera}, \citenamefont {Ramos}, \citenamefont {Celi},\ and\ \citenamefont
  {Tarruell}}]{frolian2022realising}%
  \BibitemOpen
  \bibfield  {author} {\bibinfo {author} {\bibfnamefont {A.}~\bibnamefont
  {Fr{\"o}lian}}, \bibinfo {author} {\bibfnamefont {C.~S.}\ \bibnamefont
  {Chisholm}}, \bibinfo {author} {\bibfnamefont {E.}~\bibnamefont {Neri}},
  \bibinfo {author} {\bibfnamefont {C.~R.}\ \bibnamefont {Cabrera}}, \bibinfo
  {author} {\bibfnamefont {R.}~\bibnamefont {Ramos}}, \bibinfo {author}
  {\bibfnamefont {A.}~\bibnamefont {Celi}},\ and\ \bibinfo {author}
  {\bibfnamefont {L.}~\bibnamefont {Tarruell}},\ }\bibfield  {title} {\bibinfo
  {title} {Realizing a {1D} topological gauge theory in an optically dressed
  {BEC}},\ }\href {https://doi.org/https://doi.org/10.1038/s41586-022-04943-3}
  {\bibfield  {journal} {\bibinfo  {journal} {Nature}\ }\textbf {\bibinfo
  {volume} {608}},\ \bibinfo {pages} {293} (\bibinfo {year}
  {2022})}\BibitemShut {NoStop}%
\bibitem [{\citenamefont {Zohar}(2022)}]{zohar_quantum_2022}%
  \BibitemOpen
  \bibfield  {author} {\bibinfo {author} {\bibfnamefont {E.}~\bibnamefont
  {Zohar}},\ }\bibfield  {title} {\bibinfo {title} {Quantum simulation of
  lattice gauge theories in more than one space dimension -- requirements,
  challenges and methods},\ }\href {https://doi.org/10.1098/rsta.2021.0069}
  {\bibfield  {journal} {\bibinfo  {journal} {Phil. Trans. R. Soc. A.}\
  }\textbf {\bibinfo {volume} {380}},\ \bibinfo {pages} {20210069} (\bibinfo
  {year} {2022})}\BibitemShut {NoStop}%
\bibitem [{\citenamefont {Osborne}\ \emph {et~al.}(2022)\citenamefont
  {Osborne}, \citenamefont {McCulloch}, \citenamefont {Yang}, \citenamefont
  {Hauke},\ and\ \citenamefont {Halimeh}}]{osborne_large-scale_2022}%
  \BibitemOpen
  \bibfield  {author} {\bibinfo {author} {\bibfnamefont {J.}~\bibnamefont
  {Osborne}}, \bibinfo {author} {\bibfnamefont {I.~P.}\ \bibnamefont
  {McCulloch}}, \bibinfo {author} {\bibfnamefont {B.}~\bibnamefont {Yang}},
  \bibinfo {author} {\bibfnamefont {P.}~\bibnamefont {Hauke}},\ and\ \bibinfo
  {author} {\bibfnamefont {J.~C.}\ \bibnamefont {Halimeh}},\ }\bibfield
  {title} {\bibinfo {title} {Large-{Scale} $2+1${D} {U}$(1)$ {Gauge} {Theory}
  with {Dynamical} {Matter} in a {Cold}-{Atom} {Quantum} {Simulator}},\ }\href
  {http://arxiv.org/abs/2211.01380} {\bibfield  {journal} {\bibinfo  {journal}
  {arXiv:2211.01380}\ } (\bibinfo {year} {2022})}\BibitemShut {NoStop}%
\bibitem [{\citenamefont {Halimeh}\ \emph
  {et~al.}(2022{\natexlab{a}})\citenamefont {Halimeh}, \citenamefont {Homeier},
  \citenamefont {Schweizer}, \citenamefont {Aidelsburger}, \citenamefont
  {Hauke},\ and\ \citenamefont {Grusdt}}]{halimeh_stabilizing_2022}%
  \BibitemOpen
  \bibfield  {author} {\bibinfo {author} {\bibfnamefont {J.~C.}\ \bibnamefont
  {Halimeh}}, \bibinfo {author} {\bibfnamefont {L.}~\bibnamefont {Homeier}},
  \bibinfo {author} {\bibfnamefont {C.}~\bibnamefont {Schweizer}}, \bibinfo
  {author} {\bibfnamefont {M.}~\bibnamefont {Aidelsburger}}, \bibinfo {author}
  {\bibfnamefont {P.}~\bibnamefont {Hauke}},\ and\ \bibinfo {author}
  {\bibfnamefont {F.}~\bibnamefont {Grusdt}},\ }\bibfield  {title} {\bibinfo
  {title} {Stabilizing lattice gauge theories through simplified local
  pseudogenerators},\ }\href {https://doi.org/10.1103/PhysRevResearch.4.033120}
  {\bibfield  {journal} {\bibinfo  {journal} {Phys. Rev. Research}\ }\textbf
  {\bibinfo {volume} {4}},\ \bibinfo {pages} {033120} (\bibinfo {year}
  {2022}{\natexlab{a}})}\BibitemShut {NoStop}%
\bibitem [{\citenamefont {Riegger}\ \emph
  {et~al.}(2018{\natexlab{a}})\citenamefont {Riegger}, \citenamefont {{Darkwah
  Oppong}}, \citenamefont {H\"ofer}, \citenamefont {Fernandes}, \citenamefont
  {Bloch},\ and\ \citenamefont {F\"olling}}]{Riegger2018}%
  \BibitemOpen
  \bibfield  {author} {\bibinfo {author} {\bibfnamefont {L.}~\bibnamefont
  {Riegger}}, \bibinfo {author} {\bibfnamefont {N.}~\bibnamefont {{Darkwah
  Oppong}}}, \bibinfo {author} {\bibfnamefont {M.}~\bibnamefont {H\"ofer}},
  \bibinfo {author} {\bibfnamefont {D.~R.}\ \bibnamefont {Fernandes}}, \bibinfo
  {author} {\bibfnamefont {I.}~\bibnamefont {Bloch}},\ and\ \bibinfo {author}
  {\bibfnamefont {S.}~\bibnamefont {F\"olling}},\ }\bibfield  {title} {\bibinfo
  {title} {Localized magnetic moments with tunable spin exchange in a gas of
  ultracold fermions},\ }\href {https://doi.org/10.1103/PhysRevLett.120.143601}
  {\bibfield  {journal} {\bibinfo  {journal} {Phys. Rev. Lett.}\ }\textbf
  {\bibinfo {volume} {120}},\ \bibinfo {pages} {143601} (\bibinfo {year}
  {2018}{\natexlab{a}})}\BibitemShut {NoStop}%
\bibitem [{\citenamefont {Heinz}\ \emph {et~al.}(2020)\citenamefont {Heinz},
  \citenamefont {Park}, \citenamefont {{\v S}anti{\'c}}, \citenamefont
  {Trautmann}, \citenamefont {Porsev}, \citenamefont {Safronova}, \citenamefont
  {Bloch},\ and\ \citenamefont {Blatt}}]{Heinz2020}%
  \BibitemOpen
  \bibfield  {author} {\bibinfo {author} {\bibfnamefont {A.}~\bibnamefont
  {Heinz}}, \bibinfo {author} {\bibfnamefont {A.~J.}\ \bibnamefont {Park}},
  \bibinfo {author} {\bibfnamefont {N.}~\bibnamefont {{\v S}anti{\'c}}},
  \bibinfo {author} {\bibfnamefont {J.}~\bibnamefont {Trautmann}}, \bibinfo
  {author} {\bibfnamefont {S.~G.}\ \bibnamefont {Porsev}}, \bibinfo {author}
  {\bibfnamefont {M.~S.}\ \bibnamefont {Safronova}}, \bibinfo {author}
  {\bibfnamefont {I.}~\bibnamefont {Bloch}},\ and\ \bibinfo {author}
  {\bibfnamefont {S.}~\bibnamefont {Blatt}},\ }\bibfield  {title} {\bibinfo
  {title} {State-dependent optical lattices for the strontium optical qubit},\
  }\href {https://doi.org/10.1103/PhysRevLett.124.203201} {\bibfield  {journal}
  {\bibinfo  {journal} {Phys. Rev. Lett.}\ }\textbf {\bibinfo {volume} {124}},\
  \bibinfo {pages} {203201} (\bibinfo {year} {2020})}\BibitemShut {NoStop}%
\bibitem [{\citenamefont {Schwinger}(1951)}]{Schwinger1951}%
  \BibitemOpen
  \bibfield  {author} {\bibinfo {author} {\bibfnamefont {J.}~\bibnamefont
  {Schwinger}},\ }\bibfield  {title} {\bibinfo {title} {On gauge invariance and
  vacuum polarization},\ }\href {https://doi.org/10.1103/PhysRev.82.664}
  {\bibfield  {journal} {\bibinfo  {journal} {Phys. Rev.}\ }\textbf {\bibinfo
  {volume} {82}},\ \bibinfo {pages} {664} (\bibinfo {year} {1951})}\BibitemShut
  {NoStop}%
\bibitem [{\citenamefont {Calzetta}\ and\ \citenamefont
  {Hu}(2008)}]{calzetta_book}%
  \BibitemOpen
  \bibfield  {author} {\bibinfo {author} {\bibfnamefont {E.~A.}\ \bibnamefont
  {Calzetta}}\ and\ \bibinfo {author} {\bibfnamefont {B.~L.}\ \bibnamefont
  {Hu}},\ }\href@noop {} {\emph {\bibinfo {title} {{Nonequilibrium Quantum
  Field Theory}}}}\ (\bibinfo  {publisher} {Cambridge Univ. Press, Cambridge},\
  \bibinfo {year} {2008})\BibitemShut {NoStop}%
\bibitem [{\citenamefont {Kogut}\ and\ \citenamefont
  {Susskind}(1975)}]{KogutSusskindFormulation}%
  \BibitemOpen
  \bibfield  {author} {\bibinfo {author} {\bibfnamefont {J.}~\bibnamefont
  {Kogut}}\ and\ \bibinfo {author} {\bibfnamefont {L.}~\bibnamefont
  {Susskind}},\ }\bibfield  {title} {\bibinfo {title} {Hamiltonian formulation
  of {W}ilson's lattice gauge theories},\ }\href
  {https://doi.org/10.1103/PhysRevD.11.395} {\bibfield  {journal} {\bibinfo
  {journal} {Phys. Rev. D}\ }\textbf {\bibinfo {volume} {11}},\ \bibinfo
  {pages} {395} (\bibinfo {year} {1975})}\BibitemShut {NoStop}%
\bibitem [{\citenamefont {Horn}(1981)}]{QLink1}%
  \BibitemOpen
  \bibfield  {author} {\bibinfo {author} {\bibfnamefont {D.}~\bibnamefont
  {Horn}},\ }\href {https://doi.org/10.1016/0370-2693(81)90763-2} {\bibfield
  {journal} {\bibinfo  {journal} {Phys. Lett. B}\ }\textbf {\bibinfo {volume}
  {100}},\ \bibinfo {pages} {149} (\bibinfo {year} {1981})}\BibitemShut
  {NoStop}%
\bibitem [{\citenamefont {Orland}\ and\ \citenamefont
  {Rohrlich}(1990)}]{QLink2}%
  \BibitemOpen
  \bibfield  {author} {\bibinfo {author} {\bibfnamefont {P.}~\bibnamefont
  {Orland}}\ and\ \bibinfo {author} {\bibfnamefont {D.}~\bibnamefont
  {Rohrlich}},\ }\href {https://doi.org/10.1016/0550-3213(90)90646-U}
  {\bibfield  {journal} {\bibinfo  {journal} {Nucl. Phys. B}\ }\textbf
  {\bibinfo {volume} {338}},\ \bibinfo {pages} {647} (\bibinfo {year}
  {1990})}\BibitemShut {NoStop}%
\bibitem [{\citenamefont {Chandrasekharan}\ and\ \citenamefont
  {Wiese}(1997)}]{QLink3}%
  \BibitemOpen
  \bibfield  {author} {\bibinfo {author} {\bibfnamefont {S.}~\bibnamefont
  {Chandrasekharan}}\ and\ \bibinfo {author} {\bibfnamefont {U.-J.}\
  \bibnamefont {Wiese}},\ }\bibfield  {title} {\bibinfo {title} {{Quantum link
  models: A discrete approach to gauge theories}},\ }\href
  {https://doi.org/https://doi.org/10.1016/S0550-3213(97)80041-7} {\bibfield
  {journal} {\bibinfo  {journal} {Nuclear Physics B}\ }\textbf {\bibinfo
  {volume} {492}},\ \bibinfo {pages} {455} (\bibinfo {year}
  {1997})}\BibitemShut {NoStop}%
\bibitem [{\citenamefont {Brower}\ \emph {et~al.}(1999)\citenamefont {Brower},
  \citenamefont {Chandrasekharan},\ and\ \citenamefont {Wiese}}]{QLink4}%
  \BibitemOpen
  \bibfield  {author} {\bibinfo {author} {\bibfnamefont {R.}~\bibnamefont
  {Brower}}, \bibinfo {author} {\bibfnamefont {S.}~\bibnamefont
  {Chandrasekharan}},\ and\ \bibinfo {author} {\bibfnamefont {U.-J.}\
  \bibnamefont {Wiese}},\ }\bibfield  {title} {\bibinfo {title} {{QCD as a
  quantum link model}},\ }\href {https://doi.org/10.1103/PhysRevD.60.094502}
  {\bibfield  {journal} {\bibinfo  {journal} {Phys. Rev. D}\ }\textbf {\bibinfo
  {volume} {60}},\ \bibinfo {pages} {094502} (\bibinfo {year}
  {1999})}\BibitemShut {NoStop}%
\bibitem [{\citenamefont {Banerjee}\ \emph {et~al.}(2012)\citenamefont
  {Banerjee}, \citenamefont {Dalmonte}, \citenamefont {M\"uller}, \citenamefont
  {Rico}, \citenamefont {Stebler}, \citenamefont {Wiese},\ and\ \citenamefont
  {Zoller}}]{Banerjee2012}%
  \BibitemOpen
  \bibfield  {author} {\bibinfo {author} {\bibfnamefont {D.}~\bibnamefont
  {Banerjee}}, \bibinfo {author} {\bibfnamefont {M.}~\bibnamefont {Dalmonte}},
  \bibinfo {author} {\bibfnamefont {M.}~\bibnamefont {M\"uller}}, \bibinfo
  {author} {\bibfnamefont {E.}~\bibnamefont {Rico}}, \bibinfo {author}
  {\bibfnamefont {P.}~\bibnamefont {Stebler}}, \bibinfo {author} {\bibfnamefont
  {U.-J.}\ \bibnamefont {Wiese}},\ and\ \bibinfo {author} {\bibfnamefont
  {P.}~\bibnamefont {Zoller}},\ }\bibfield  {title} {\bibinfo {title} {Atomic
  quantum simulation of dynamical gauge fields coupled to fermionic matter:
  From string breaking to evolution after a quench},\ }\href
  {https://doi.org/10.1103/PhysRevLett.109.175302} {\bibfield  {journal}
  {\bibinfo  {journal} {Phys. Rev. Lett.}\ }\textbf {\bibinfo {volume} {109}},\
  \bibinfo {pages} {175302} (\bibinfo {year} {2012})}\BibitemShut {NoStop}%
\bibitem [{\citenamefont {Halimeh}\ \emph
  {et~al.}(2022{\natexlab{b}})\citenamefont {Halimeh}, \citenamefont
  {McCulloch}, \citenamefont {Yang},\ and\ \citenamefont
  {Hauke}}]{halimeh_tuning_2022}%
  \BibitemOpen
  \bibfield  {author} {\bibinfo {author} {\bibfnamefont {J.~C.}\ \bibnamefont
  {Halimeh}}, \bibinfo {author} {\bibfnamefont {I.~P.}\ \bibnamefont
  {McCulloch}}, \bibinfo {author} {\bibfnamefont {B.}~\bibnamefont {Yang}},\
  and\ \bibinfo {author} {\bibfnamefont {P.}~\bibnamefont {Hauke}},\ }\bibfield
   {title} {\bibinfo {title} {Tuning the topological
  $\ensuremath{\theta}$-angle in cold-atom quantum simulators of gauge
  theories},\ }\href {https://doi.org/10.1103/PRXQuantum.3.040316} {\bibfield
  {journal} {\bibinfo  {journal} {PRX Quantum}\ }\textbf {\bibinfo {volume}
  {3}},\ \bibinfo {pages} {040316} (\bibinfo {year}
  {2022}{\natexlab{b}})}\BibitemShut {NoStop}%
\bibitem [{\citenamefont {Cheng}\ \emph {et~al.}(2022)\citenamefont {Cheng},
  \citenamefont {Liu}, \citenamefont {Zheng}, \citenamefont {Zhang},\ and\
  \citenamefont {Zhai}}]{cheng_tunable_2022}%
  \BibitemOpen
  \bibfield  {author} {\bibinfo {author} {\bibfnamefont {Y.}~\bibnamefont
  {Cheng}}, \bibinfo {author} {\bibfnamefont {S.}~\bibnamefont {Liu}}, \bibinfo
  {author} {\bibfnamefont {W.}~\bibnamefont {Zheng}}, \bibinfo {author}
  {\bibfnamefont {P.}~\bibnamefont {Zhang}},\ and\ \bibinfo {author}
  {\bibfnamefont {H.}~\bibnamefont {Zhai}},\ }\bibfield  {title} {\bibinfo
  {title} {Tunable confinement-deconfinement transition in an ultracold-atom
  quantum simulator},\ }\href {https://doi.org/10.1103/PRXQuantum.3.040317}
  {\bibfield  {journal} {\bibinfo  {journal} {PRX Quantum}\ }\textbf {\bibinfo
  {volume} {3}},\ \bibinfo {pages} {040317} (\bibinfo {year}
  {2022})}\BibitemShut {NoStop}%
\bibitem [{\citenamefont {Gorshkov}\ \emph {et~al.}(2010)\citenamefont
  {Gorshkov}, \citenamefont {Hermele}, \citenamefont {Gurarie}, \citenamefont
  {Xu}, \citenamefont {Julienne}, \citenamefont {Ye}, \citenamefont {Zoller},
  \citenamefont {Demler}, \citenamefont {Lukin},\ and\ \citenamefont
  {Rey}}]{Gorshkov2010}%
  \BibitemOpen
  \bibfield  {author} {\bibinfo {author} {\bibfnamefont {A.~V.}\ \bibnamefont
  {Gorshkov}}, \bibinfo {author} {\bibfnamefont {M.}~\bibnamefont {Hermele}},
  \bibinfo {author} {\bibfnamefont {V.}~\bibnamefont {Gurarie}}, \bibinfo
  {author} {\bibfnamefont {C.}~\bibnamefont {Xu}}, \bibinfo {author}
  {\bibfnamefont {P.~S.}\ \bibnamefont {Julienne}}, \bibinfo {author}
  {\bibfnamefont {J.}~\bibnamefont {Ye}}, \bibinfo {author} {\bibfnamefont
  {P.}~\bibnamefont {Zoller}}, \bibinfo {author} {\bibfnamefont
  {E.}~\bibnamefont {Demler}}, \bibinfo {author} {\bibfnamefont {M.~D.}\
  \bibnamefont {Lukin}},\ and\ \bibinfo {author} {\bibfnamefont {A.~M.}\
  \bibnamefont {Rey}},\ }\bibfield  {title} {\bibinfo {title} {{Two-orbital
  SU(N) magnetism with ultracold alkaline-earth atoms}},\ }\href
  {https://doi.org/10.1038/nphys1535} {\bibfield  {journal} {\bibinfo
  {journal} {Nature Physics}\ }\textbf {\bibinfo {volume} {6}},\ \bibinfo
  {pages} {289} (\bibinfo {year} {2010})}\BibitemShut {NoStop}%
\bibitem [{\citenamefont {Scazza}\ \emph {et~al.}(2014)\citenamefont {Scazza},
  \citenamefont {Hofrichter}, \citenamefont {H{\"{o}}fer}, \citenamefont {{De
  Groot}}, \citenamefont {Bloch},\ and\ \citenamefont
  {F{\"{o}}lling}}]{Scazza14}%
  \BibitemOpen
  \bibfield  {author} {\bibinfo {author} {\bibfnamefont {F.}~\bibnamefont
  {Scazza}}, \bibinfo {author} {\bibfnamefont {C.}~\bibnamefont {Hofrichter}},
  \bibinfo {author} {\bibfnamefont {M.}~\bibnamefont {H{\"{o}}fer}}, \bibinfo
  {author} {\bibfnamefont {P.~C.}\ \bibnamefont {{De Groot}}}, \bibinfo
  {author} {\bibfnamefont {I.}~\bibnamefont {Bloch}},\ and\ \bibinfo {author}
  {\bibfnamefont {S.}~\bibnamefont {F{\"{o}}lling}},\ }\bibfield  {title}
  {\bibinfo {title} {{Observation of two-orbital spin-exchange interactions
  with ultracold SU(N)-symmetric fermions}},\ }\href
  {https://doi.org/10.1038/nphys3061} {\bibfield  {journal} {\bibinfo
  {journal} {Nature Physics}\ }\textbf {\bibinfo {volume} {10}},\ \bibinfo
  {pages} {779} (\bibinfo {year} {2014})}\BibitemShut {NoStop}%
\bibitem [{\citenamefont {Zhang}\ \emph {et~al.}(2014)\citenamefont {Zhang},
  \citenamefont {Bishof}, \citenamefont {Bromley}, \citenamefont {Kraus},
  \citenamefont {Safronova}, \citenamefont {Zoller}, \citenamefont {Rey},\ and\
  \citenamefont {Ye}}]{zhang_spectroscopic_2014}%
  \BibitemOpen
  \bibfield  {author} {\bibinfo {author} {\bibfnamefont {X.}~\bibnamefont
  {Zhang}}, \bibinfo {author} {\bibfnamefont {M.}~\bibnamefont {Bishof}},
  \bibinfo {author} {\bibfnamefont {S.~L.}\ \bibnamefont {Bromley}}, \bibinfo
  {author} {\bibfnamefont {C.~V.}\ \bibnamefont {Kraus}}, \bibinfo {author}
  {\bibfnamefont {M.~S.}\ \bibnamefont {Safronova}}, \bibinfo {author}
  {\bibfnamefont {P.}~\bibnamefont {Zoller}}, \bibinfo {author} {\bibfnamefont
  {A.~M.}\ \bibnamefont {Rey}},\ and\ \bibinfo {author} {\bibfnamefont
  {J.}~\bibnamefont {Ye}},\ }\bibfield  {title} {\bibinfo {title}
  {Spectroscopic observation of {SU}({N})-symmetric interactions in {Sr}
  orbital magnetism},\ }\href {https://doi.org/10.1126/science.1254978}
  {\bibfield  {journal} {\bibinfo  {journal} {Science}\ }\textbf {\bibinfo
  {volume} {345}},\ \bibinfo {pages} {1467} (\bibinfo {year}
  {2014})}\BibitemShut {NoStop}%
\bibitem [{\citenamefont {Takamoto}\ \emph {et~al.}(2005)\citenamefont
  {Takamoto}, \citenamefont {Hong}, \citenamefont {Higashi},\ and\
  \citenamefont {Katori}}]{takamoto_optical_2005}%
  \BibitemOpen
  \bibfield  {author} {\bibinfo {author} {\bibfnamefont {M.}~\bibnamefont
  {Takamoto}}, \bibinfo {author} {\bibfnamefont {F.-L.}\ \bibnamefont {Hong}},
  \bibinfo {author} {\bibfnamefont {R.}~\bibnamefont {Higashi}},\ and\ \bibinfo
  {author} {\bibfnamefont {H.}~\bibnamefont {Katori}},\ }\bibfield  {title}
  {\bibinfo {title} {An optical lattice clock},\ }\href
  {https://doi.org/10.1038/nature03541} {\bibfield  {journal} {\bibinfo
  {journal} {Nature}\ }\textbf {\bibinfo {volume} {435}},\ \bibinfo {pages}
  {321} (\bibinfo {year} {2005})}\BibitemShut {NoStop}%
\bibitem [{\citenamefont {Ludlow}\ \emph {et~al.}(2006)\citenamefont {Ludlow},
  \citenamefont {Boyd}, \citenamefont {Zelevinsky}, \citenamefont {Foreman},
  \citenamefont {Blatt}, \citenamefont {Notcutt}, \citenamefont {Ido},\ and\
  \citenamefont {Ye}}]{ludlow_systematic_2006}%
  \BibitemOpen
  \bibfield  {author} {\bibinfo {author} {\bibfnamefont {A.~D.}\ \bibnamefont
  {Ludlow}}, \bibinfo {author} {\bibfnamefont {M.~M.}\ \bibnamefont {Boyd}},
  \bibinfo {author} {\bibfnamefont {T.}~\bibnamefont {Zelevinsky}}, \bibinfo
  {author} {\bibfnamefont {S.~M.}\ \bibnamefont {Foreman}}, \bibinfo {author}
  {\bibfnamefont {S.}~\bibnamefont {Blatt}}, \bibinfo {author} {\bibfnamefont
  {M.}~\bibnamefont {Notcutt}}, \bibinfo {author} {\bibfnamefont
  {T.}~\bibnamefont {Ido}},\ and\ \bibinfo {author} {\bibfnamefont
  {J.}~\bibnamefont {Ye}},\ }\bibfield  {title} {\bibinfo {title} {Systematic
  {Study} of the $^{87}${Sr} {Clock} {Transition} in an {Optical} {Lattice}},\
  }\href {https://doi.org/10.1103/PhysRevLett.96.033003} {\bibfield  {journal}
  {\bibinfo  {journal} {Phys. Rev. Lett.}\ }\textbf {\bibinfo {volume} {96}},\
  \bibinfo {pages} {033003} (\bibinfo {year} {2006})}\BibitemShut {NoStop}%
\bibitem [{\citenamefont {Yi}\ \emph {et~al.}(2008)\citenamefont {Yi},
  \citenamefont {Daley}, \citenamefont {Pupillo},\ and\ \citenamefont
  {Zoller}}]{Yi08}%
  \BibitemOpen
  \bibfield  {author} {\bibinfo {author} {\bibfnamefont {W.}~\bibnamefont
  {Yi}}, \bibinfo {author} {\bibfnamefont {A.~J.}\ \bibnamefont {Daley}},
  \bibinfo {author} {\bibfnamefont {G.}~\bibnamefont {Pupillo}},\ and\ \bibinfo
  {author} {\bibfnamefont {P.}~\bibnamefont {Zoller}},\ }\bibfield  {title}
  {\bibinfo {title} {State-dependent, addressable subwavelength lattices with
  cold atoms},\ }\href {https://doi.org/10.1088/1367-2630/10/7/073015}
  {\bibfield  {journal} {\bibinfo  {journal} {New J. Phys.}\ }\textbf {\bibinfo
  {volume} {10}},\ \bibinfo {pages} {073015} (\bibinfo {year}
  {2008})}\BibitemShut {NoStop}%
\bibitem [{\citenamefont {Sebby-Strabley}\ \emph {et~al.}(2006)\citenamefont
  {Sebby-Strabley}, \citenamefont {Anderlini}, \citenamefont {Jessen},\ and\
  \citenamefont {Porto}}]{sebby-strabley_lattice_2006}%
  \BibitemOpen
  \bibfield  {author} {\bibinfo {author} {\bibfnamefont {J.}~\bibnamefont
  {Sebby-Strabley}}, \bibinfo {author} {\bibfnamefont {M.}~\bibnamefont
  {Anderlini}}, \bibinfo {author} {\bibfnamefont {P.~S.}\ \bibnamefont
  {Jessen}},\ and\ \bibinfo {author} {\bibfnamefont {J.~V.}\ \bibnamefont
  {Porto}},\ }\bibfield  {title} {\bibinfo {title} {Lattice of double wells for
  manipulating pairs of cold atoms},\ }\href
  {https://doi.org/10.1103/PhysRevA.73.033605} {\bibfield  {journal} {\bibinfo
  {journal} {Phys. Rev. A}\ }\textbf {\bibinfo {volume} {73}},\ \bibinfo
  {pages} {033605} (\bibinfo {year} {2006})}\BibitemShut {NoStop}%
\bibitem [{\citenamefont {Fölling}\ \emph {et~al.}(2007)\citenamefont
  {Fölling}, \citenamefont {Trotzky}, \citenamefont {Cheinet}, \citenamefont
  {Feld}, \citenamefont {Saers}, \citenamefont {Widera}, \citenamefont
  {Müller},\ and\ \citenamefont {Bloch}}]{folling_direct_2007}%
  \BibitemOpen
  \bibfield  {author} {\bibinfo {author} {\bibfnamefont {S.}~\bibnamefont
  {Fölling}}, \bibinfo {author} {\bibfnamefont {S.}~\bibnamefont {Trotzky}},
  \bibinfo {author} {\bibfnamefont {P.}~\bibnamefont {Cheinet}}, \bibinfo
  {author} {\bibfnamefont {M.}~\bibnamefont {Feld}}, \bibinfo {author}
  {\bibfnamefont {R.}~\bibnamefont {Saers}}, \bibinfo {author} {\bibfnamefont
  {A.}~\bibnamefont {Widera}}, \bibinfo {author} {\bibfnamefont
  {T.}~\bibnamefont {Müller}},\ and\ \bibinfo {author} {\bibfnamefont
  {I.}~\bibnamefont {Bloch}},\ }\bibfield  {title} {\bibinfo {title} {Direct
  observation of second-order atom tunnelling},\ }\href
  {https://doi.org/10.1038/nature06112} {\bibfield  {journal} {\bibinfo
  {journal} {Nature}\ }\textbf {\bibinfo {volume} {448}},\ \bibinfo {pages}
  {1029} (\bibinfo {year} {2007})}\BibitemShut {NoStop}%
\bibitem [{\citenamefont {Spar}\ \emph {et~al.}(2022)\citenamefont {Spar},
  \citenamefont {Guardado-Sanchez}, \citenamefont {Chi}, \citenamefont {Yan},\
  and\ \citenamefont {Bakr}}]{Spar21}%
  \BibitemOpen
  \bibfield  {author} {\bibinfo {author} {\bibfnamefont {B.~M.}\ \bibnamefont
  {Spar}}, \bibinfo {author} {\bibfnamefont {E.}~\bibnamefont
  {Guardado-Sanchez}}, \bibinfo {author} {\bibfnamefont {S.}~\bibnamefont
  {Chi}}, \bibinfo {author} {\bibfnamefont {Z.~Z.}\ \bibnamefont {Yan}},\ and\
  \bibinfo {author} {\bibfnamefont {W.~S.}\ \bibnamefont {Bakr}},\ }\bibfield
  {title} {\bibinfo {title} {Realization of a {Fermi-Hubbard} {Optical}
  {Tweezer} {Array}},\ }\href {https://doi.org/10.1103/PhysRevLett.128.223202}
  {\bibfield  {journal} {\bibinfo  {journal} {Phys. Rev. Lett.}\ }\textbf
  {\bibinfo {volume} {128}},\ \bibinfo {pages} {223202} (\bibinfo {year}
  {2022})}\BibitemShut {NoStop}%
\bibitem [{\citenamefont {Young}\ \emph {et~al.}(2022)\citenamefont {Young},
  \citenamefont {Eckner}, \citenamefont {Schine}, \citenamefont {Childs},\ and\
  \citenamefont {Kaufman}}]{Young22b}%
  \BibitemOpen
  \bibfield  {author} {\bibinfo {author} {\bibfnamefont {A.~W.}\ \bibnamefont
  {Young}}, \bibinfo {author} {\bibfnamefont {W.~J.}\ \bibnamefont {Eckner}},
  \bibinfo {author} {\bibfnamefont {N.}~\bibnamefont {Schine}}, \bibinfo
  {author} {\bibfnamefont {A.~M.}\ \bibnamefont {Childs}},\ and\ \bibinfo
  {author} {\bibfnamefont {A.~M.}\ \bibnamefont {Kaufman}},\ }\bibfield
  {title} {\bibinfo {title} {Tweezer-programmable {2D} quantum walks in a
  {Hubbard}-regime lattice},\ }\href {https://doi.org/10.1126/science.abo0608}
  {\bibfield  {journal} {\bibinfo  {journal} {Science}\ }\textbf {\bibinfo
  {volume} {377}},\ \bibinfo {pages} {885} (\bibinfo {year}
  {2022})}\BibitemShut {NoStop}%
\bibitem [{\citenamefont {H\"ofer}\ \emph {et~al.}(2015)\citenamefont
  {H\"ofer}, \citenamefont {Riegger}, \citenamefont {Scazza}, \citenamefont
  {Hofrichter}, \citenamefont {Fernandes}, \citenamefont {Parish},
  \citenamefont {Levinsen}, \citenamefont {Bloch},\ and\ \citenamefont
  {F\"olling}}]{Hoefer15}%
  \BibitemOpen
  \bibfield  {author} {\bibinfo {author} {\bibfnamefont {M.}~\bibnamefont
  {H\"ofer}}, \bibinfo {author} {\bibfnamefont {L.}~\bibnamefont {Riegger}},
  \bibinfo {author} {\bibfnamefont {F.}~\bibnamefont {Scazza}}, \bibinfo
  {author} {\bibfnamefont {C.}~\bibnamefont {Hofrichter}}, \bibinfo {author}
  {\bibfnamefont {D.~R.}\ \bibnamefont {Fernandes}}, \bibinfo {author}
  {\bibfnamefont {M.~M.}\ \bibnamefont {Parish}}, \bibinfo {author}
  {\bibfnamefont {J.}~\bibnamefont {Levinsen}}, \bibinfo {author}
  {\bibfnamefont {I.}~\bibnamefont {Bloch}},\ and\ \bibinfo {author}
  {\bibfnamefont {S.}~\bibnamefont {F\"olling}},\ }\bibfield  {title} {\bibinfo
  {title} {Observation of an orbital interaction-induced {Feshbach} resonance
  in ${}^\mathrm{173}\mathrm{Yb}$},\ }\href
  {https://doi.org/10.1103/PhysRevLett.115.265302} {\bibfield  {journal}
  {\bibinfo  {journal} {Phys. Rev. Lett.}\ }\textbf {\bibinfo {volume} {115}},\
  \bibinfo {pages} {265302} (\bibinfo {year} {2015})}\BibitemShut {NoStop}%
\bibitem [{\citenamefont {Goban}\ \emph {et~al.}(2018)\citenamefont {Goban},
  \citenamefont {Hutson}, \citenamefont {Marti}, \citenamefont {Campbell},
  \citenamefont {Perlin}, \citenamefont {Julienne}, \citenamefont {D’incao},
  \citenamefont {Rey},\ and\ \citenamefont {Ye}}]{Goban18}%
  \BibitemOpen
  \bibfield  {author} {\bibinfo {author} {\bibfnamefont {A.}~\bibnamefont
  {Goban}}, \bibinfo {author} {\bibfnamefont {R.}~\bibnamefont {Hutson}},
  \bibinfo {author} {\bibfnamefont {G.}~\bibnamefont {Marti}}, \bibinfo
  {author} {\bibfnamefont {S.}~\bibnamefont {Campbell}}, \bibinfo {author}
  {\bibfnamefont {M.}~\bibnamefont {Perlin}}, \bibinfo {author} {\bibfnamefont
  {P.}~\bibnamefont {Julienne}}, \bibinfo {author} {\bibfnamefont
  {J.}~\bibnamefont {D’incao}}, \bibinfo {author} {\bibfnamefont
  {A.}~\bibnamefont {Rey}},\ and\ \bibinfo {author} {\bibfnamefont
  {J.}~\bibnamefont {Ye}},\ }\bibfield  {title} {\bibinfo {title} {Emergence of
  multi-body interactions in a fermionic lattice clock},\ }\href
  {https://doi.org/10.1038/s41586-018-0661-6} {\bibfield  {journal} {\bibinfo
  {journal} {Nature}\ }\textbf {\bibinfo {volume} {563}},\ \bibinfo {pages}
  {369} (\bibinfo {year} {2018})}\BibitemShut {NoStop}%
\bibitem [{\citenamefont {Ono}\ \emph {et~al.}(2019)\citenamefont {Ono},
  \citenamefont {Kobayashi}, \citenamefont {Amano}, \citenamefont {Sato},\ and\
  \citenamefont {Takahashi}}]{Ono19}%
  \BibitemOpen
  \bibfield  {author} {\bibinfo {author} {\bibfnamefont {K.}~\bibnamefont
  {Ono}}, \bibinfo {author} {\bibfnamefont {J.}~\bibnamefont {Kobayashi}},
  \bibinfo {author} {\bibfnamefont {Y.}~\bibnamefont {Amano}}, \bibinfo
  {author} {\bibfnamefont {K.}~\bibnamefont {Sato}},\ and\ \bibinfo {author}
  {\bibfnamefont {Y.}~\bibnamefont {Takahashi}},\ }\bibfield  {title} {\bibinfo
  {title} {{Antiferromagnetic interorbital spin-exchange interaction of
  $^{171}$Yb}},\ }\href {https://doi.org/10.1103/PhysRevA.99.032707} {\bibfield
   {journal} {\bibinfo  {journal} {Phys. Rev. A}\ }\textbf {\bibinfo {volume}
  {99}},\ \bibinfo {pages} {032707} (\bibinfo {year} {2019})}\BibitemShut
  {NoStop}%
\bibitem [{\citenamefont {Lemke}\ \emph {et~al.}(2009)\citenamefont {Lemke},
  \citenamefont {Ludlow}, \citenamefont {Barber}, \citenamefont {Fortier},
  \citenamefont {Diddams}, \citenamefont {Jiang}, \citenamefont {Jefferts},
  \citenamefont {Heavner}, \citenamefont {Parker},\ and\ \citenamefont
  {Oates}}]{Lemke09}%
  \BibitemOpen
  \bibfield  {author} {\bibinfo {author} {\bibfnamefont {N.~D.}\ \bibnamefont
  {Lemke}}, \bibinfo {author} {\bibfnamefont {A.~D.}\ \bibnamefont {Ludlow}},
  \bibinfo {author} {\bibfnamefont {Z.~W.}\ \bibnamefont {Barber}}, \bibinfo
  {author} {\bibfnamefont {T.~M.}\ \bibnamefont {Fortier}}, \bibinfo {author}
  {\bibfnamefont {S.~A.}\ \bibnamefont {Diddams}}, \bibinfo {author}
  {\bibfnamefont {Y.}~\bibnamefont {Jiang}}, \bibinfo {author} {\bibfnamefont
  {S.~R.}\ \bibnamefont {Jefferts}}, \bibinfo {author} {\bibfnamefont {T.~P.}\
  \bibnamefont {Heavner}}, \bibinfo {author} {\bibfnamefont {T.~E.}\
  \bibnamefont {Parker}},\ and\ \bibinfo {author} {\bibfnamefont {C.~W.}\
  \bibnamefont {Oates}},\ }\bibfield  {title} {\bibinfo {title} {Spin-$1/2$
  optical lattice clock},\ }\href
  {https://doi.org/10.1103/PhysRevLett.103.063001} {\bibfield  {journal}
  {\bibinfo  {journal} {Phys. Rev. Lett.}\ }\textbf {\bibinfo {volume} {103}},\
  \bibinfo {pages} {063001} (\bibinfo {year} {2009})}\BibitemShut {NoStop}%
\bibitem [{\citenamefont {Weitenberg}\ \emph {et~al.}(2011)\citenamefont
  {Weitenberg}, \citenamefont {Endres}, \citenamefont {Sherson}, \citenamefont
  {Cheneau}, \citenamefont {Schauß}, \citenamefont {Fukuhara}, \citenamefont
  {Bloch},\ and\ \citenamefont {Kuhr}}]{weitenberg_single-spin_2011}%
  \BibitemOpen
  \bibfield  {author} {\bibinfo {author} {\bibfnamefont {C.}~\bibnamefont
  {Weitenberg}}, \bibinfo {author} {\bibfnamefont {M.}~\bibnamefont {Endres}},
  \bibinfo {author} {\bibfnamefont {J.~F.}\ \bibnamefont {Sherson}}, \bibinfo
  {author} {\bibfnamefont {M.}~\bibnamefont {Cheneau}}, \bibinfo {author}
  {\bibfnamefont {P.}~\bibnamefont {Schauß}}, \bibinfo {author} {\bibfnamefont
  {T.}~\bibnamefont {Fukuhara}}, \bibinfo {author} {\bibfnamefont
  {I.}~\bibnamefont {Bloch}},\ and\ \bibinfo {author} {\bibfnamefont
  {S.}~\bibnamefont {Kuhr}},\ }\bibfield  {title} {\bibinfo {title}
  {Single-spin addressing in an atomic {Mott} insulator},\ }\href
  {https://doi.org/10.1038/nature09827} {\bibfield  {journal} {\bibinfo
  {journal} {Nature}\ }\textbf {\bibinfo {volume} {471}},\ \bibinfo {pages}
  {319} (\bibinfo {year} {2011})}\BibitemShut {NoStop}%
\bibitem [{\citenamefont {Gross}\ and\ \citenamefont
  {Bakr}(2021)}]{gross_quantum_2021}%
  \BibitemOpen
  \bibfield  {author} {\bibinfo {author} {\bibfnamefont {C.}~\bibnamefont
  {Gross}}\ and\ \bibinfo {author} {\bibfnamefont {W.~S.}\ \bibnamefont
  {Bakr}},\ }\bibfield  {title} {\bibinfo {title} {Quantum gas microscopy for
  single atom and spin detection},\ }\href
  {https://doi.org/10.1038/s41567-021-01370-5} {\bibfield  {journal} {\bibinfo
  {journal} {Nature Physics}\ }\textbf {\bibinfo {volume} {17}},\ \bibinfo
  {pages} {1316} (\bibinfo {year} {2021})}\BibitemShut {NoStop}%
\bibitem [{\citenamefont {Riegger}\ \emph
  {et~al.}(2018{\natexlab{b}})\citenamefont {Riegger}, \citenamefont {{Darkwah
  Oppong}}, \citenamefont {H\"ofer}, \citenamefont {Fernandes}, \citenamefont
  {Bloch},\ and\ \citenamefont {F\"olling}}]{Riegger18}%
  \BibitemOpen
  \bibfield  {author} {\bibinfo {author} {\bibfnamefont {L.}~\bibnamefont
  {Riegger}}, \bibinfo {author} {\bibfnamefont {N.}~\bibnamefont {{Darkwah
  Oppong}}}, \bibinfo {author} {\bibfnamefont {M.}~\bibnamefont {H\"ofer}},
  \bibinfo {author} {\bibfnamefont {D.~R.}\ \bibnamefont {Fernandes}}, \bibinfo
  {author} {\bibfnamefont {I.}~\bibnamefont {Bloch}},\ and\ \bibinfo {author}
  {\bibfnamefont {S.}~\bibnamefont {F\"olling}},\ }\bibfield  {title} {\bibinfo
  {title} {Localized magnetic moments with tunable spin exchange in a gas of
  ultracold fermions},\ }\href {https://doi.org/10.1103/PhysRevLett.120.143601}
  {\bibfield  {journal} {\bibinfo  {journal} {Phys. Rev. Lett.}\ }\textbf
  {\bibinfo {volume} {120}},\ \bibinfo {pages} {143601} (\bibinfo {year}
  {2018}{\natexlab{b}})}\BibitemShut {NoStop}%
\bibitem [{\citenamefont {Darkwah~Oppong}\ \emph {et~al.}(2022)\citenamefont
  {Darkwah~Oppong}, \citenamefont {Pasqualetti}, \citenamefont {Bettermann},
  \citenamefont {Zechmann}, \citenamefont {Knap}, \citenamefont {Bloch},\ and\
  \citenamefont {F\"olling}}]{darkwahoppong:2022}%
  \BibitemOpen
  \bibfield  {author} {\bibinfo {author} {\bibfnamefont {N.}~\bibnamefont
  {Darkwah~Oppong}}, \bibinfo {author} {\bibfnamefont {G.}~\bibnamefont
  {Pasqualetti}}, \bibinfo {author} {\bibfnamefont {O.}~\bibnamefont
  {Bettermann}}, \bibinfo {author} {\bibfnamefont {P.}~\bibnamefont
  {Zechmann}}, \bibinfo {author} {\bibfnamefont {M.}~\bibnamefont {Knap}},
  \bibinfo {author} {\bibfnamefont {I.}~\bibnamefont {Bloch}},\ and\ \bibinfo
  {author} {\bibfnamefont {S.}~\bibnamefont {F\"olling}},\ }\bibfield  {title}
  {\bibinfo {title} {Probing transport and slow relaxation in the
  mass-imbalanced {Fermi-Hubbard} model},\ }\href
  {https://doi.org/10.1103/PhysRevX.12.031026} {\bibfield  {journal} {\bibinfo
  {journal} {Phys. Rev. X}\ }\textbf {\bibinfo {volume} {12}},\ \bibinfo
  {pages} {031026} (\bibinfo {year} {2022})}\BibitemShut {NoStop}%
\bibitem [{\citenamefont {Endres}\ \emph {et~al.}(2016)\citenamefont {Endres},
  \citenamefont {Bernien}, \citenamefont {Keesling}, \citenamefont {Levine},
  \citenamefont {Anschuetz}, \citenamefont {Krajenbrink}, \citenamefont
  {Senko}, \citenamefont {Vuletic}, \citenamefont {Greiner},\ and\
  \citenamefont {Lukin}}]{Endres16}%
  \BibitemOpen
  \bibfield  {author} {\bibinfo {author} {\bibfnamefont {M.}~\bibnamefont
  {Endres}}, \bibinfo {author} {\bibfnamefont {H.}~\bibnamefont {Bernien}},
  \bibinfo {author} {\bibfnamefont {A.}~\bibnamefont {Keesling}}, \bibinfo
  {author} {\bibfnamefont {H.}~\bibnamefont {Levine}}, \bibinfo {author}
  {\bibfnamefont {E.~R.}\ \bibnamefont {Anschuetz}}, \bibinfo {author}
  {\bibfnamefont {A.}~\bibnamefont {Krajenbrink}}, \bibinfo {author}
  {\bibfnamefont {C.}~\bibnamefont {Senko}}, \bibinfo {author} {\bibfnamefont
  {V.}~\bibnamefont {Vuletic}}, \bibinfo {author} {\bibfnamefont
  {M.}~\bibnamefont {Greiner}},\ and\ \bibinfo {author} {\bibfnamefont {M.~D.}\
  \bibnamefont {Lukin}},\ }\bibfield  {title} {\bibinfo {title} {Atom-by-atom
  assembly of defect-free one-dimensional cold atom arrays},\ }\href
  {https://doi.org/10.1126/science.aah3752} {\bibfield  {journal} {\bibinfo
  {journal} {Science}\ }\textbf {\bibinfo {volume} {354}},\ \bibinfo {pages}
  {1024} (\bibinfo {year} {2016})}\BibitemShut {NoStop}%
\bibitem [{Note1()}]{Note1}%
  \BibitemOpen
  \bibinfo {note} {Note that here the matter gauge coupling has the same sign
  on every link, so this model differs from the quantum link model studied, for
  example, in~\cite {Banerjee2013}. With this choice, the model has a simpler
  implementation in the experimental setup that we propose. To obtain couplings
  with sign $\protect \tilde s_{r,r+\protect \hat x}=1$, $\protect \tilde
  s_{r,r+\protect \hat y}=(-1)^i$, an approach similar to Ref.~\cite
  {Jaksch_2003} could be applied.}\BibitemShut {Stop}%
\bibitem [{\citenamefont {Banerjee}\ \emph {et~al.}(2013)\citenamefont
  {Banerjee}, \citenamefont {B{\"{o}}gli}, \citenamefont {Dalmonte},
  \citenamefont {Rico}, \citenamefont {Stebler}, \citenamefont {Wiese},\ and\
  \citenamefont {Zoller}}]{Banerjee2013}%
  \BibitemOpen
  \bibfield  {author} {\bibinfo {author} {\bibfnamefont {D.}~\bibnamefont
  {Banerjee}}, \bibinfo {author} {\bibfnamefont {M.}~\bibnamefont
  {B{\"{o}}gli}}, \bibinfo {author} {\bibfnamefont {M.}~\bibnamefont
  {Dalmonte}}, \bibinfo {author} {\bibfnamefont {E.}~\bibnamefont {Rico}},
  \bibinfo {author} {\bibfnamefont {P.}~\bibnamefont {Stebler}}, \bibinfo
  {author} {\bibfnamefont {U.-J.}\ \bibnamefont {Wiese}},\ and\ \bibinfo
  {author} {\bibfnamefont {P.}~\bibnamefont {Zoller}},\ }\bibfield  {title}
  {\bibinfo {title} {{Atomic Quantum Simulation of {U(N)} and {SU(N)}
  {Non-Abelian} {Lattice} {Gauge} {Theories}}},\ }\href
  {https://link.aps.org/doi/10.1103/PhysRevLett.110.125303} {\bibfield
  {journal} {\bibinfo  {journal} {Phys. Rev. Lett.}\ }\textbf {\bibinfo
  {volume} {110}},\ \bibinfo {pages} {125303} (\bibinfo {year}
  {2013})}\BibitemShut {NoStop}%
\bibitem [{\citenamefont {Zohar}\ \emph {et~al.}(2013)\citenamefont {Zohar},
  \citenamefont {Cirac},\ and\ \citenamefont {Reznik}}]{Zohar_2013}%
  \BibitemOpen
  \bibfield  {author} {\bibinfo {author} {\bibfnamefont {E.}~\bibnamefont
  {Zohar}}, \bibinfo {author} {\bibfnamefont {J.~I.}\ \bibnamefont {Cirac}},\
  and\ \bibinfo {author} {\bibfnamefont {B.}~\bibnamefont {Reznik}},\
  }\bibfield  {title} {\bibinfo {title} {{Cold-Atom Quantum Simulator for SU(2)
  Yang-Mills Lattice Gauge Theory}},\ }\href
  {https://doi.org/10.1103/PhysRevLett.110.125304} {\bibfield  {journal}
  {\bibinfo  {journal} {Phys. Rev. Lett.}\ }\textbf {\bibinfo {volume} {110}},\
  \bibinfo {pages} {125304} (\bibinfo {year} {2013})}\BibitemShut {NoStop}%
\bibitem [{\citenamefont {Tagliacozzo}\ \emph {et~al.}(2013)\citenamefont
  {Tagliacozzo}, \citenamefont {Celi}, \citenamefont {Orland}, \citenamefont
  {Mitchell},\ and\ \citenamefont {Lewenstein}}]{Tagliacozzo_2013}%
  \BibitemOpen
  \bibfield  {author} {\bibinfo {author} {\bibfnamefont {L.}~\bibnamefont
  {Tagliacozzo}}, \bibinfo {author} {\bibfnamefont {A.}~\bibnamefont {Celi}},
  \bibinfo {author} {\bibfnamefont {P.}~\bibnamefont {Orland}}, \bibinfo
  {author} {\bibfnamefont {M.}~\bibnamefont {Mitchell}},\ and\ \bibinfo
  {author} {\bibfnamefont {M.}~\bibnamefont {Lewenstein}},\ }\bibfield  {title}
  {\bibinfo {title} {{Simulation of non-Abelian gauge theories with optical
  lattices}},\ }\href {https://doi.org/10.1038/ncomms3615} {\bibfield
  {journal} {\bibinfo  {journal} {Nature Commun.}\ }\textbf {\bibinfo {volume}
  {4}},\ \bibinfo {pages} {1} (\bibinfo {year} {2013})}\BibitemShut {NoStop}%
\bibitem [{\citenamefont {Stannigel}\ \emph {et~al.}(2014)\citenamefont
  {Stannigel}, \citenamefont {Hauke}, \citenamefont {Marcos}, \citenamefont
  {Hafezi}, \citenamefont {Diehl}, \citenamefont {Dalmonte},\ and\
  \citenamefont {Zoller}}]{Stannigel:2014bf}%
  \BibitemOpen
  \bibfield  {author} {\bibinfo {author} {\bibfnamefont {K.}~\bibnamefont
  {Stannigel}}, \bibinfo {author} {\bibfnamefont {P.}~\bibnamefont {Hauke}},
  \bibinfo {author} {\bibfnamefont {D.}~\bibnamefont {Marcos}}, \bibinfo
  {author} {\bibfnamefont {M.}~\bibnamefont {Hafezi}}, \bibinfo {author}
  {\bibfnamefont {S.}~\bibnamefont {Diehl}}, \bibinfo {author} {\bibfnamefont
  {M.}~\bibnamefont {Dalmonte}},\ and\ \bibinfo {author} {\bibfnamefont
  {P.}~\bibnamefont {Zoller}},\ }\bibfield  {title} {\bibinfo {title}
  {{Constrained Dynamics via the Zeno Effect in Quantum Simulation:
  Implementing Non-Abelian Lattice Gauge Theories with Cold Atoms}},\ }\href
  {https://link.aps.org/doi/10.1103/PhysRevLett.112.120406} {\bibfield
  {journal} {\bibinfo  {journal} {Phys. Rev. Lett.}\ }\textbf {\bibinfo
  {volume} {112}},\ \bibinfo {pages} {120406} (\bibinfo {year}
  {2014})}\BibitemShut {NoStop}%
\bibitem [{\citenamefont {Mezzacapo}\ \emph {et~al.}(2015)\citenamefont
  {Mezzacapo}, \citenamefont {Rico}, \citenamefont {Sab\'{\i}n}, \citenamefont
  {Egusquiza}, \citenamefont {Lamata},\ and\ \citenamefont
  {Solano}}]{Mezzacapo2015}%
  \BibitemOpen
  \bibfield  {author} {\bibinfo {author} {\bibfnamefont {A.}~\bibnamefont
  {Mezzacapo}}, \bibinfo {author} {\bibfnamefont {E.}~\bibnamefont {Rico}},
  \bibinfo {author} {\bibfnamefont {C.}~\bibnamefont {Sab\'{\i}n}}, \bibinfo
  {author} {\bibfnamefont {I.~L.}\ \bibnamefont {Egusquiza}}, \bibinfo {author}
  {\bibfnamefont {L.}~\bibnamefont {Lamata}},\ and\ \bibinfo {author}
  {\bibfnamefont {E.}~\bibnamefont {Solano}},\ }\bibfield  {title} {\bibinfo
  {title} {{Non-Abelian SU(2) Lattice Gauge Theories in Superconducting
  Circuits}},\ }\href {https://link.aps.org/doi/10.1103/PhysRevLett.115.240502}
  {\bibfield  {journal} {\bibinfo  {journal} {Phys. Rev. Lett.}\ }\textbf
  {\bibinfo {volume} {115}},\ \bibinfo {pages} {240502} (\bibinfo {year}
  {2015})}\BibitemShut {NoStop}%
\bibitem [{\citenamefont {Rico}\ \emph {et~al.}(2018)\citenamefont {Rico},
  \citenamefont {Dalmonte}, \citenamefont {Zoller}, \citenamefont {Banerjee},
  \citenamefont {Bögli}, \citenamefont {Stebler},\ and\ \citenamefont
  {Wiese}}]{Rico_2018}%
  \BibitemOpen
  \bibfield  {author} {\bibinfo {author} {\bibfnamefont {E.}~\bibnamefont
  {Rico}}, \bibinfo {author} {\bibfnamefont {M.}~\bibnamefont {Dalmonte}},
  \bibinfo {author} {\bibfnamefont {P.}~\bibnamefont {Zoller}}, \bibinfo
  {author} {\bibfnamefont {D.}~\bibnamefont {Banerjee}}, \bibinfo {author}
  {\bibfnamefont {M.}~\bibnamefont {Bögli}}, \bibinfo {author} {\bibfnamefont
  {P.}~\bibnamefont {Stebler}},\ and\ \bibinfo {author} {\bibfnamefont {U.-J.}\
  \bibnamefont {Wiese}},\ }\bibfield  {title} {\bibinfo {title} {{SO(3)
  “Nuclear Physics” with ultracold Gases}},\ }\href
  {https://doi.org/10.1016/j.aop.2018.03.020} {\bibfield  {journal} {\bibinfo
  {journal} {Annals of Physics}\ }\textbf {\bibinfo {volume} {393}},\ \bibinfo
  {pages} {466} (\bibinfo {year} {2018})}\BibitemShut {NoStop}%
\bibitem [{\citenamefont {Kasper}\ \emph {et~al.}(2020)\citenamefont {Kasper},
  \citenamefont {Zache}, \citenamefont {Jendrzejewski}, \citenamefont
  {Lewenstein},\ and\ \citenamefont {Zohar}}]{Kasper2020}%
  \BibitemOpen
  \bibfield  {author} {\bibinfo {author} {\bibfnamefont {V.}~\bibnamefont
  {Kasper}}, \bibinfo {author} {\bibfnamefont {T.~V.}\ \bibnamefont {Zache}},
  \bibinfo {author} {\bibfnamefont {F.}~\bibnamefont {Jendrzejewski}}, \bibinfo
  {author} {\bibfnamefont {M.}~\bibnamefont {Lewenstein}},\ and\ \bibinfo
  {author} {\bibfnamefont {E.}~\bibnamefont {Zohar}},\ }\bibfield  {title}
  {\bibinfo {title} {Non-abelian gauge invariance from dynamical decoupling},\
  }\href {https://arxiv.org/abs/2012.08620} {\bibfield  {journal} {\bibinfo
  {journal} {arXiv:2012.08620}\ } (\bibinfo {year} {2020})}\BibitemShut
  {NoStop}%
\bibitem [{\citenamefont {Davoudi}\ \emph {et~al.}(2021)\citenamefont
  {Davoudi}, \citenamefont {Raychowdhury},\ and\ \citenamefont
  {Shaw}}]{Davoudi20}%
  \BibitemOpen
  \bibfield  {author} {\bibinfo {author} {\bibfnamefont {Z.}~\bibnamefont
  {Davoudi}}, \bibinfo {author} {\bibfnamefont {I.}~\bibnamefont
  {Raychowdhury}},\ and\ \bibinfo {author} {\bibfnamefont {A.}~\bibnamefont
  {Shaw}},\ }\bibfield  {title} {\bibinfo {title} {{Search for efficient
  formulations for Hamiltonian simulation of non-Abelian lattice gauge
  theories}},\ }\href {https://doi.org/10.1103/PhysRevD.104.074505} {\bibfield
  {journal} {\bibinfo  {journal} {Phys. Rev. D}\ }\textbf {\bibinfo {volume}
  {104}},\ \bibinfo {pages} {074505} (\bibinfo {year} {2021})}\BibitemShut
  {NoStop}%
\bibitem [{\citenamefont {Gonz{\'a}lez-Cuadra}\ \emph
  {et~al.}(2022)\citenamefont {Gonz{\'a}lez-Cuadra}, \citenamefont {Zache},
  \citenamefont {Carrasco}, \citenamefont {Kraus},\ and\ \citenamefont
  {Zoller}}]{gonzalez2022hardware}%
  \BibitemOpen
  \bibfield  {author} {\bibinfo {author} {\bibfnamefont {D.}~\bibnamefont
  {Gonz{\'a}lez-Cuadra}}, \bibinfo {author} {\bibfnamefont {T.~V.}\
  \bibnamefont {Zache}}, \bibinfo {author} {\bibfnamefont {J.}~\bibnamefont
  {Carrasco}}, \bibinfo {author} {\bibfnamefont {B.}~\bibnamefont {Kraus}},\
  and\ \bibinfo {author} {\bibfnamefont {P.}~\bibnamefont {Zoller}},\
  }\bibfield  {title} {\bibinfo {title} {Hardware efficient quantum simulation
  of non-abelian gauge theories with qudits on {Rydberg} platforms},\ }\href
  {https://arxiv.org/abs/2203.15541} {\bibfield  {journal} {\bibinfo  {journal}
  {arXiv:2203.15541}\ } (\bibinfo {year} {2022})}\BibitemShut {NoStop}%
\bibitem [{\citenamefont {Jaksch}\ and\ \citenamefont
  {Zoller}(2003)}]{Jaksch_2003}%
  \BibitemOpen
  \bibfield  {author} {\bibinfo {author} {\bibfnamefont {D.}~\bibnamefont
  {Jaksch}}\ and\ \bibinfo {author} {\bibfnamefont {P.}~\bibnamefont
  {Zoller}},\ }\bibfield  {title} {\bibinfo {title} {Creation of effective
  magnetic fields in optical lattices: the hofstadter butterfly for cold
  neutral atoms},\ }\href {https://doi.org/10.1088/1367-2630/5/1/356}
  {\bibfield  {journal} {\bibinfo  {journal} {New J. Phys.}\ }\textbf {\bibinfo
  {volume} {5}},\ \bibinfo {pages} {56} (\bibinfo {year} {2003})}\BibitemShut
  {NoStop}%
\bibitem [{\citenamefont {Kivelson}(1982)}]{Kivelson1982}%
  \BibitemOpen
  \bibfield  {author} {\bibinfo {author} {\bibfnamefont {S.}~\bibnamefont
  {Kivelson}},\ }\bibfield  {title} {\bibinfo {title} {Wannier functions in
  one-dimensional disordered systems: Application to fractionally charged
  solitons},\ }\href {https://doi.org/10.1103/PhysRevB.26.4269} {\bibfield
  {journal} {\bibinfo  {journal} {Phys. Rev. B}\ }\textbf {\bibinfo {volume}
  {26}},\ \bibinfo {pages} {4269} (\bibinfo {year} {1982})}\BibitemShut
  {NoStop}%
\bibitem [{\citenamefont {Uehlinger}\ \emph {et~al.}(2013)\citenamefont
  {Uehlinger}, \citenamefont {Jotzu}, \citenamefont {Messer}, \citenamefont
  {Greif}, \citenamefont {Hofstetter}, \citenamefont {Bissbort},\ and\
  \citenamefont {Esslinger}}]{Uehlinger2013}%
  \BibitemOpen
  \bibfield  {author} {\bibinfo {author} {\bibfnamefont {T.}~\bibnamefont
  {Uehlinger}}, \bibinfo {author} {\bibfnamefont {G.}~\bibnamefont {Jotzu}},
  \bibinfo {author} {\bibfnamefont {M.}~\bibnamefont {Messer}}, \bibinfo
  {author} {\bibfnamefont {D.}~\bibnamefont {Greif}}, \bibinfo {author}
  {\bibfnamefont {W.}~\bibnamefont {Hofstetter}}, \bibinfo {author}
  {\bibfnamefont {U.}~\bibnamefont {Bissbort}},\ and\ \bibinfo {author}
  {\bibfnamefont {T.}~\bibnamefont {Esslinger}},\ }\bibfield  {title} {\bibinfo
  {title} {Artificial graphene with tunable interactions},\ }\href
  {https://doi.org/10.1103/PhysRevLett.111.185307} {\bibfield  {journal}
  {\bibinfo  {journal} {Phys. Rev. Lett.}\ }\textbf {\bibinfo {volume} {111}},\
  \bibinfo {pages} {185307} (\bibinfo {year} {2013})}\BibitemShut {NoStop}%
\bibitem [{\citenamefont {Marzari}\ and\ \citenamefont
  {Vanderbilt}(1997)}]{Marzari1997}%
  \BibitemOpen
  \bibfield  {author} {\bibinfo {author} {\bibfnamefont {N.}~\bibnamefont
  {Marzari}}\ and\ \bibinfo {author} {\bibfnamefont {D.}~\bibnamefont
  {Vanderbilt}},\ }\bibfield  {title} {\bibinfo {title} {Maximally localized
  generalized wannier functions for composite energy bands},\ }\href
  {https://doi.org/10.1103/PhysRevB.56.12847} {\bibfield  {journal} {\bibinfo
  {journal} {Phys. Rev. B}\ }\textbf {\bibinfo {volume} {56}},\ \bibinfo
  {pages} {12847} (\bibinfo {year} {1997})}\BibitemShut {NoStop}%
\bibitem [{\citenamefont {Marzari}\ \emph {et~al.}(2012)\citenamefont
  {Marzari}, \citenamefont {Mostofi}, \citenamefont {Yates}, \citenamefont
  {Souza},\ and\ \citenamefont {Vanderbilt}}]{Marzari2012}%
  \BibitemOpen
  \bibfield  {author} {\bibinfo {author} {\bibfnamefont {N.}~\bibnamefont
  {Marzari}}, \bibinfo {author} {\bibfnamefont {A.~A.}\ \bibnamefont
  {Mostofi}}, \bibinfo {author} {\bibfnamefont {J.~R.}\ \bibnamefont {Yates}},
  \bibinfo {author} {\bibfnamefont {I.}~\bibnamefont {Souza}},\ and\ \bibinfo
  {author} {\bibfnamefont {D.}~\bibnamefont {Vanderbilt}},\ }\bibfield  {title}
  {\bibinfo {title} {Maximally localized wannier functions: Theory and
  applications},\ }\href {https://doi.org/10.1103/RevModPhys.84.1419}
  {\bibfield  {journal} {\bibinfo  {journal} {Rev. Mod. Phys.}\ }\textbf
  {\bibinfo {volume} {84}},\ \bibinfo {pages} {1419} (\bibinfo {year}
  {2012})}\BibitemShut {NoStop}%
\end{thebibliography}%

\appendix

\section{Wannier functions and lattice Hamiltonian}
\label{app:wannier}
\subsection{One-dimensional case}
In this appendix, we discuss the ab-initio derivation of the lattice Hamiltonian from the Wannier functions for the case of the one-dimensional model. First, we solve the non interacting Hamiltonian $H_\text{non-int}$. We diagonalize the single-particle Hamiltonian $h_\alpha(\mathbf{r})$
\begin{equation}\label{eq:sgham}
    h_\alpha(\mathbf{r})=- \frac{\hbar^2}{2M}\nabla^2+V_{\alpha} \left(\mathbf{r}\right).
\end{equation}
with $\alpha=g,e$.
For the quantum simulation of the one-dimensional model discussed in Section \ref{sec:onedim}, the potential is
\begin{equation}\label{eq:factpot}
    V_\alpha(\mathbf{r})=V_\alpha^x(x)+V_\alpha^y(y)+V_\alpha^z(z).
\end{equation}

\begin{figure*}
    \centering
    \includegraphics[width=\linewidth]{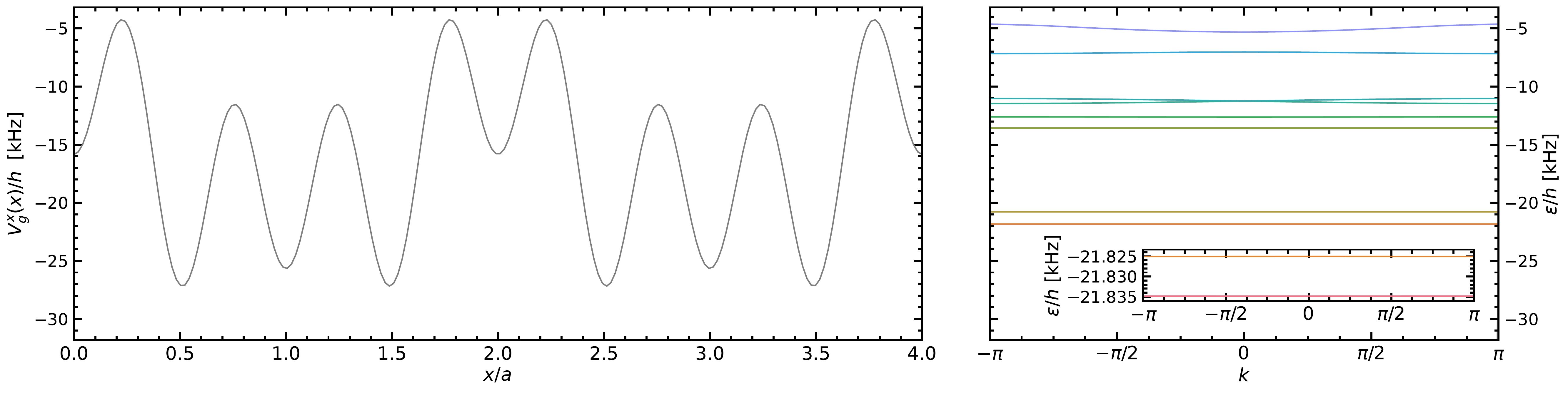}
    \caption{\textbf{Lattice potential and band structure along $x$.} Left panel: $x$ component of the optical lattice potential $V_g^x(x)$. The defining parameters are reported in Table~\ref{tab:params}. Right panel: corresponding band structure using the same scale as the panel on the left. The two lowest bands are almost degenerate (they correspond to symmetric and antisymmetric superpositions of the $s=\{+,-\}$ sites in the triple well), and are separated from the third one (corresponding to the central site $0$ of the triple well) by an energy $\sim \delta_g$. Higher bands are separated from the first three by an energy gap $\Delta_g$.}
    \label{fig:bands1d}
\end{figure*}
We then look for a factorized and localized three-dimensional complete basis of wavefunctions for all states involved in the dynamic evolution of the system. The potential is periodic in space for each component $x,y,z$ such that the Bloch theorem applies to Eq.~(\ref{eq:sgham}). In Fig.~\ref{fig:bands1d}, we plot the potential $V_g^x$ (with the parameters of Table~\ref{tab:params}) and the lowest bands obtained by solving the corresponding periodic Hamiltonian. Using a unitary transformation (similar to the reverse Fourier transformation) of the Bloch eigenfunctions, we obtain a set of localized orthonormal wavefunctions $w_{\alpha,s}$ called the Wannier functions. Because of the form of the potential Eq.~(\ref{eq:factpot}), the Bloch functions are factorizable. Likewise, the Wannier functions factorize along $x,y,z$:
\begin{equation}
    w_{\alpha,s}(\mathbf{r}-\mathbf{r}_j)=\phi_{\alpha,s}^x(x-ja)\phi_\alpha^y(y) \phi_\alpha^z(z),
\end{equation}
where $w$ is the three-dimensional Wannier function, and the $\phi^i$ are the one dimensional Wannier functions for each direction $i=x,y,z$. $\alpha=g,e$ is the electronic state, $s=\{0,+,-\}$ denotes the three functions corresponding to the three sites of a triple well, and $\mathbf{r}_j=ja\hat x$ is the Wannier center. As we target a one dimensional system, we consider only Wannier functions in the transverse direction $y,z$ centered around $y=z=0$ by convention. They involve only the lowest Bloch band. The $x$ component is instead obtained from the three lowest bands such that we have three Wannier centers per unit cell. 

To derive an expression for the localized Wannier functions useful to estimate the overlap, we compute the eigenstates of the projection of the position operator onto a given set of Bloch states~\cite{Kivelson1982,Uehlinger2013}. In one dimension, this method always gives the maximally localized Wannier functions~\cite{Marzari1997,Marzari2012}. In many other cases, this derivation still gives a good approximation of the maximally localized Wannier functions.

We can then define discrete operators for the discrete Hamiltonian. We expand the fermionic operators in the basis of the Wannier functions
\begin{align}
\label{eq:wfg}
\begin{split}
    \Psi_g(\mathbf{r})=&\sum_{j \text{ odd}} \left[w_{g,+}(\mathbf{r}-\mathbf{r}_{j})c_{j+1/2}\right.\\
    &\left.+w_{g,0}(\mathbf{r}-\mathbf{r}_j)c_j +w_{g,-}(\mathbf{r}-\mathbf{r}_{j})c_{j-1/2}\right],
\end{split}\\\label{eq:wfe}
\begin{split}
    \Psi_e(\mathbf{r})=&\sum_{j \text{ even}} \left[w_{e,+}(\mathbf{r}-\mathbf{r}_{j})d_{j+1/2}\right.\\
    &\left.+w_{e,0}(\mathbf{r}-\mathbf{r}_j)d_j +w_{e,-}(\mathbf{r}-\mathbf{r}_{j})d_{j-1/2}\right].
    \end{split}
\end{align}
We stress that the only approximations performed so far are (i) neglecting the higher bands and (ii) only considering the chain localized at $y,z=0$. (i) is justified when the gaps $\Delta_g, \Delta_e$ between the third and the closest higher bands (see Fig.~\ref{fig:bands1d}) are much larger than the energy scales of the dynamics that we are interested in. (ii) is justified if the transverse hoppings $t_\alpha^{y},t_\alpha^{z}$ are small compared to the energy scales of our interest. For the gaps and the transverse hoppings of Section~\ref{sec:param}, both approximations are appropriate.

The discrete parameters emerge when substituting Eqs.~(\ref{eq:wfg}) and (\ref{eq:wfe}) in the Hamiltonian $H=H_\text{non-int}+H_\text{int}$ in Eq.~(\ref{eq:opt}). These parameters include the chemical potentials, the hoppings (both from $H_\text{non-int}$), the on-site and off-site density-density interactions, the density mediated hoppings, and the correlated hoppings between $g$ and $e$ atoms. We included all these terms in the numerical simulations of the real-time dynamics in Section~\ref{sec:dyn}. For clarity, we give the lattice Hamiltonian with the terms of highest amplitude. We define the chemical potentials such that
\begin{align}
\begin{split}
    \mu_{\alpha,s}&=\int d^3 \mathbf{r} \,w^*_{\alpha,s}(\mathbf{r}-\mathbf{r}_\alpha)h_\alpha(\mathbf{r}) w_{\alpha,s}(\mathbf{r}-\mathbf{r}_\alpha)
\\
&=\mu_{\alpha,s}^x+\mu_{\alpha}^{y}+\mu_{\alpha}^{z}
\end{split}
\end{align}
where $\alpha=g,e$, $s=\{0,+,-\}$, and the Wannier centers are $\mathbf{r}_g=a\hat x$, $\mathbf{r}_e=0$. We further define the difference between the chemical potentials in each triple well
\begin{equation}
    \delta_{\alpha,\pm}=\mu_{\alpha,0}-\mu_{\alpha,\pm}=\mu_{\alpha,0}^x-\mu_{\alpha,\pm}^x,
\end{equation}
and the nearest-neighbour hoppings within a triple well
\begin{equation}
\begin{split}
t_{\alpha,\pm}&=-\int d^3 \mathbf{r} \,w^*_{\alpha,0}(\mathbf{r}-\mathbf{r}_\alpha)h_\alpha(\mathbf{r}) w_{\alpha,\pm}(\mathbf{r}-\mathbf{r}_\alpha)\\
 &= -\int \mathrm d x\, [\phi_{\alpha,0}^x(x-ja)]^*h^x_\alpha(x)\phi_{\alpha,\pm}^x(x-ja).
\end{split}
\end{equation}
For $\varphi=0$ the triple well is designed to be symmetrical, such that $\delta_{\alpha, +}=\delta_{\alpha, -}=\delta_{\alpha}$ and $t_{\alpha, +}=t_{\alpha, -}=t_\alpha$. These two terms result in the Hamiltonian terms $H_g$ and $H_e$ in Eq.~(\ref{eq:Hlatt}).

The term with largest amplitude obtained from $H_\text{int}$ is the on-site interaction on the sites where $g$ and $e$ triple wells overlap. This amplitude is given by
\begin{equation}
\begin{split}
    U&=g_{eg}^{-}\int d^3 \mathbf{r}\, |w_{g,-}(\mathbf{r}-a\hat x)|^2 |w_{e,+}(\mathbf{r})|^2\\
    &=g_{eg}^{-}J_{yz}\int dx \, |\phi_{g,-}^x(x-ja)|^2 |\phi_{e,+}^x(x)|^2,
\end{split}
\end{equation}
with
\begin{equation}
    J_{yz}=\int \mathrm d y \, \mathrm d z \, |\phi_g^y(y)|^2 |\phi_e^y(y)|^2 |\phi_g^z(z)|^2 |\phi_e^z(z)|^2,
\end{equation}
and yields the Hamiltonian term $H_U$ in Eq.~(\ref{eq:Hlatt}). The terms with the next largest amplitude with our choice of parameters are density-assisted hoppings. Specifically, where a $g$ or $e$ atom hops between two sites of a triple well, provided that an atom of the opposite electronic state sits in either the initial or the final site (see Fig.~\ref{fig:UD}). The amplitude has the form
\begin{align}
\begin{split}
    D_g&=g_{eg}^{-}\int d^3 \mathbf{r}\, w_{g,0}^*(\mathbf{r}-a\hat x)w_{g,-}(\mathbf{r}-a) |w_{e,+}(\mathbf{r})|^2\\
    &=g_{eg}^{-}J_{yz}\int dx \, [\phi_{g,0}^x(x-ja)]^*\phi_{g,-}^x(x-ja)|\phi_{e,+}^x(x)|^2,
\end{split}\\
\begin{split}
    D_e&=g_{eg}^{-}\int d^3 \mathbf{r}\, |w_{g,-}(\mathbf{r}-a\hat x)|^2 w_{e,0}^*(\mathbf{r})w_{e,+}(\mathbf{r}),\\
    &=g_{eg}^{-}J_{yz}\int dx \, |\phi_{g,-}^x(x-ja)|^2 [\phi_{e,0}^x(x)]^*\phi_{e,+}^x(x),
\end{split}
\end{align}
and results in the Hamiltonian $H_D$ in Eq.~(\ref{eq:Hlatt}).
For our choice of parameters, all the other terms (that we generically include in $H_{\textrm{lr}}$) have sufficiently small amplitudes to be negligible  according to Sec.~\ref{sec:gauge}.

\subsection{Two-dimensional case}
\begin{figure}
    \centering
    \includegraphics[width=\linewidth]{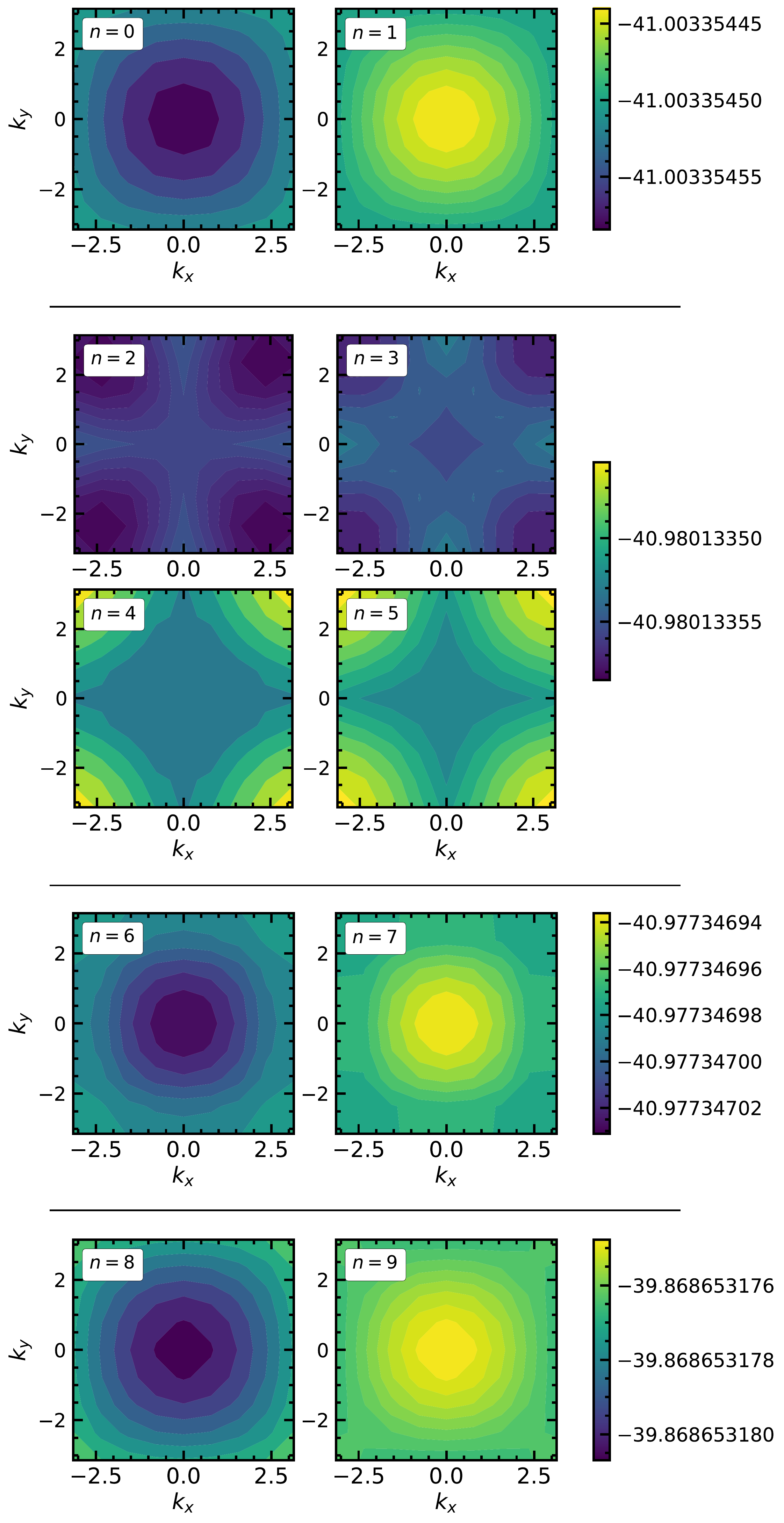}
    \caption{\textbf{Bandstructure for the 2d lattice.} The 10 lowest bands corresponding the two-dimensional lattice $V_e^{x,y}(x,y)$ in Fig.~\ref{fig:2D}.}
    \label{fig:bands2d}
\end{figure}

The steps of the derivation of the lattice Hamiltonian in the two-dimensional system are very similar to the one-dimensional case. The main difference lies in the potential that is now
\begin{equation}
    V_\alpha(\mathbf{r})=V_\alpha^{x,y}(x,y)+V_\alpha^z(z).
\end{equation}
As a consequence, only the $z$ component of the Bloch (and Wannier) functions can be factorized out, while for the $x-y$ plane we have to solve a two-dimensional single particle Hamiltonian. The first 10 two-dimensional Bloch bands for the parameters in Table~\ref{tab:params2D} are plotted in Fig.~\ref{fig:bands2d}. 

\begin{figure}
    \centering
    \includegraphics[width=\linewidth]{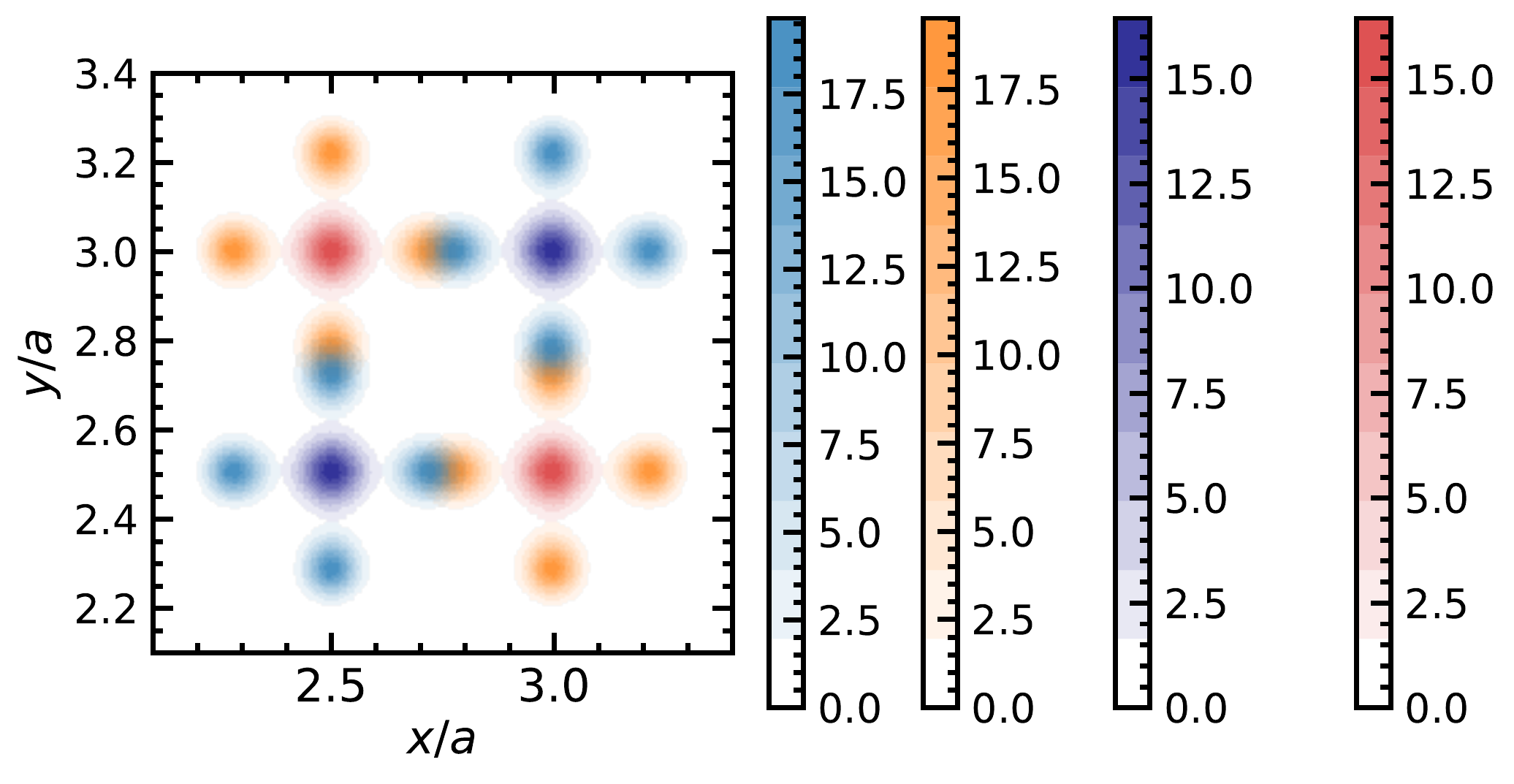}
    \caption{\textbf{Wannier functions for the 2d lattice.} Two-dimensional Wannier functions of two $g$ and two $e$ blocks. In orange/light blue, we plot the 4 Wannier functions localized on the links of each block for the $e/g$ state. In red/dark blue, we plot the Wannier function localized on the center of each block for the $e/g$ state.}
    \label{fig:wannier2d}
\end{figure}

The unit cell we consider is defined by the lattice vectors $(2a,0)$ and $(0,2a)$ and contains two blocks for each electronic state. Because each block contains five sites, we need 10 Wannier functions per unit cell for each state. To find the Wannier functions we first find the eigenstates of the projection of the $x$ position operator on the 10 lowest bands: we collect the groups of eigenstates with (almost) degenerate eigenvalue and diagonalize the $y$ position operator projected on each group. The Wannier functions obtained with this procedure are plotted in Fig.~\ref{fig:wannier2d}. The coefficients of the lattice Hamiltonian are then obtained as in the one-dimensional case.

\section{Perturbative theory of the lattice Hamiltonian}\label{app:A}

The Hamiltonian Eq.~\eqref{eq:opt} describing the cold atoms in the optical lattice and its lattice formulation Eq.~\eqref{eq:Hlatt} can be mapped to the QLM Hamiltonian Eq.~\eqref{eq:QLM}, when considering the coupling between the targeted gauge-invariant Hilbert subspace and the rest of the Hilbert space as a perturbation. Such a regime occurs when $\epsilon, t_\alpha, D_\alpha \ll \delta, U-\delta$ and $H_{\textrm{lr}}$ is negligible. To second order in perturbation, we obtain the correction Eq.~\eqref{eq:correction}. We present here the computation in more detail.

The resonant targeted subspace verifies:
\begin{align}
    \forall j\text{ even}, \qquad &  n^g_{j-1/2}+ n^g_{j}+ n^g_{j+1/2}=2,\\
    \forall j\text{ odd}, \qquad &  n^e_{j-1/2}+ n^e_{j}+ n^e_{j+1/2}=1,\\
    \forall j, \qquad &  n^e_{j+1/2}+ n^g_{j+1/2}=1,
\end{align}
which satisfy the gauge-invariant condition $G_i\lvert \psi \rangle = 0$ for all sites $i$ and $\lvert \psi \rangle$ in the targeted subspace. All states within this targeted subspace have the same energy relatively to the Hamiltonian $H_0$ in Eq.~\eqref{eq:H0}, although they are not its ground states. This subspace is separated from all other orthogonal states coupled by $H_1$ in Eq.~\eqref{eq:H1} by an energy proportional to $\delta$ and $U$. It is thus possible to apply standard quantum perturbation theory by treating $H_1$  as a perturbation to $H_0$ with ratios of $t_\alpha$ and $D_\alpha$ with $1/\delta$ or $1/(U-\delta)$ as the small parameters. To first order, the terms in $\epsilon$ of $H_1$ generate a contribution in the staggered mass $m$. Further corrections due to these terms are negligible and neglected. The eigenfunctions of the resonant subspace in the Fock basis are not modified to first order.

To find Eq.~\eqref{eq:correction}, we continue the perturbation to second order. We use the perturbation formula:
\begin{equation}\label{eq:perturb2}
    H^{(2)}_{\mathrm{eff}}=\sum_{i,j} \sum_\phi \frac{\lvert \psi_i \rangle \langle \psi_i \rvert H_1 \lvert \phi \rangle \langle \phi \rvert H_1 \lvert \psi_j \rangle \langle \psi_j \rvert}{E_\phi - E_\psi},
\end{equation}
with $H_0 \lvert \psi_i \rangle = E_\psi \lvert \psi_i \rangle$ for all $i$ where the $\psi_i$ generate the resonant subspace, and the $\phi$ are all states orthogonal to the $\psi_i$ such that $\langle \psi_i \rvert H_1 \lvert \phi \rangle \neq 0$. By definition, we take $H_0 \lvert \phi \rangle = E_\phi \lvert \phi \rangle$. Both set of states $\lbrace \lvert \psi_i \rangle \rbrace$ and $\lbrace \lvert \phi \rangle \rbrace$ are separable in the local Fock basis and $H_1$ is short-ranged such that, in Eq.~\eqref{eq:perturb2}, we may only consider a couple of processes (i.e., matrix elements of $\langle \psi_i \rvert H_1 \lvert \phi \rangle \langle \phi \rvert H_1 \lvert \psi_j \rangle$) involving one link or one site. All of these processes, their amplitude, and the amplitude they contribute to are listed in Table~\ref{tab:perturbation}. 
\vspace{5em}
 \begin{longtable}{p{.5\linewidth}  p{.30\linewidth} p{.15\linewidth}} 
         Processes & Amplitude & Add to \\ \hline
         \parbox[c]{\linewidth}{\includegraphics[width=0.8\linewidth]{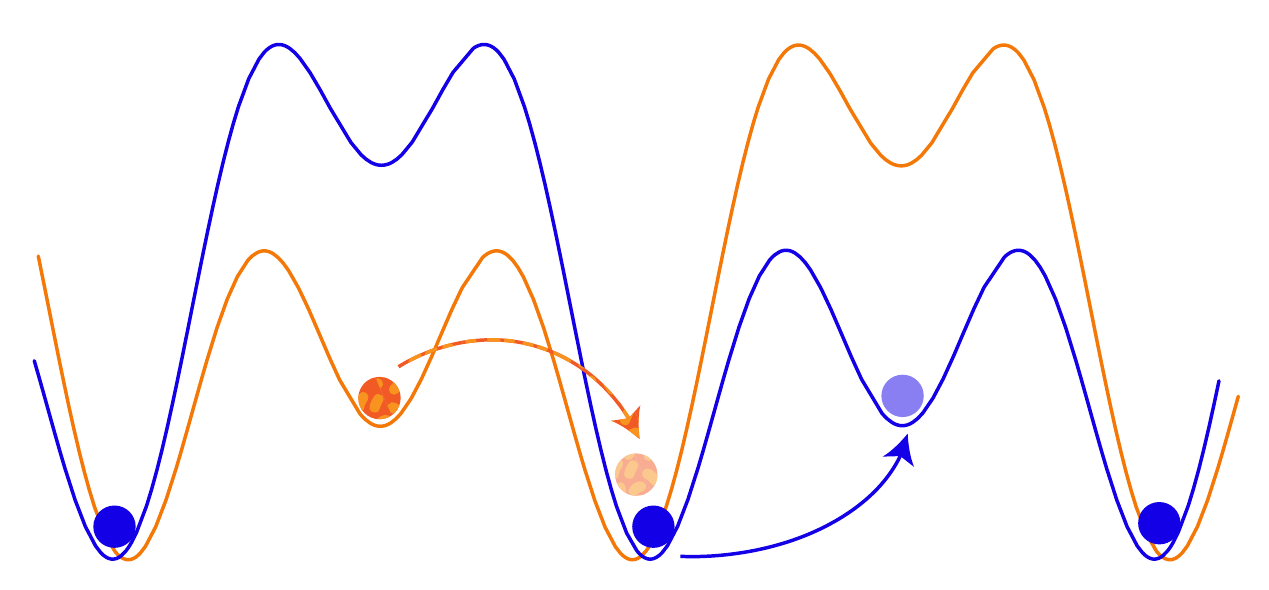}}& $-\frac{t_gt_e U}{\delta(U-\delta)} -\frac{D_e t_g}{U-\delta}$ \newline$ -\frac{D_g t_e}{U-\delta} -\frac{D_e D_g}{U-\delta} $ & $w$ \\\hline %%%%%%%%%%%
         \parbox[c]{0.55\linewidth}{\includegraphics[width=0.8\linewidth]{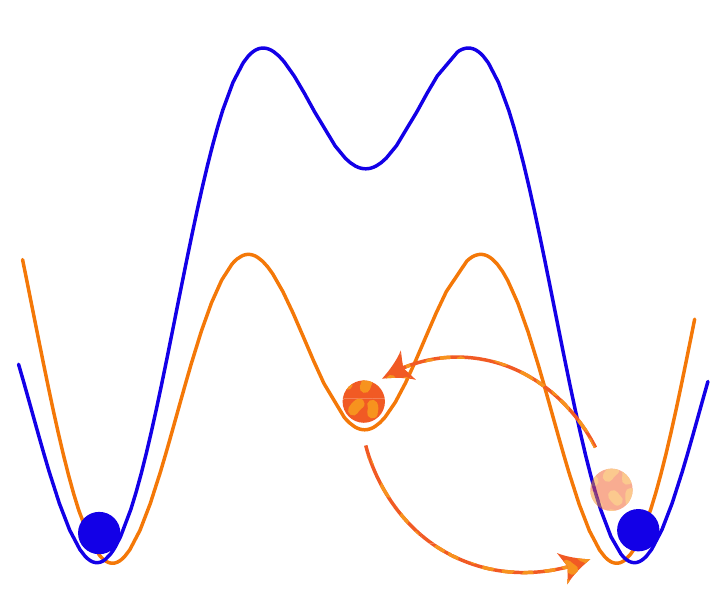}}& $-\frac{t_e^2}{U-\delta}$ \newline$ -2\frac{t_e D_e}{U-\delta}-\frac{D_e^2}{U-\delta}$ & $\delta_e$ (x2)  \\ \hline %%%%%%%%%%%%%%%%%%%%%%%%
         \parbox[c]{0.55\linewidth}{\includegraphics[width=0.8\linewidth]{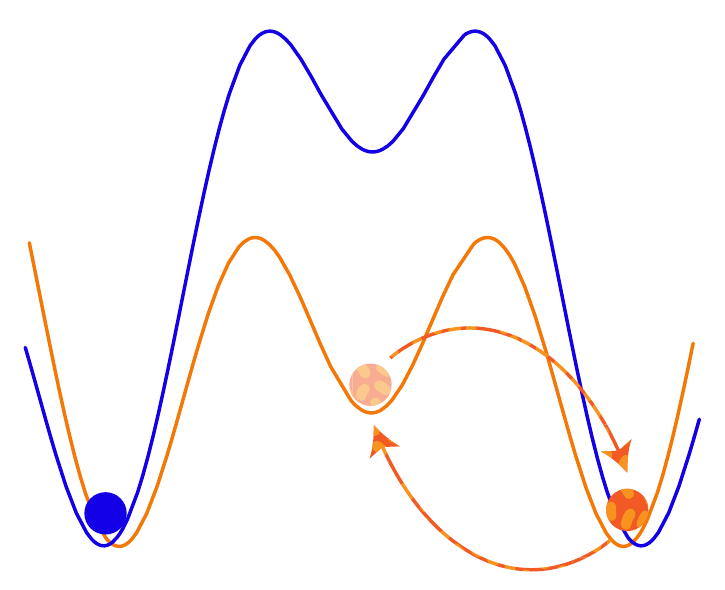}}& $-\frac{t_e^2}{\delta}$ & $\delta_e$ \\ \hline
         \parbox[c]{0.55\linewidth}{\includegraphics[width=0.8\linewidth]{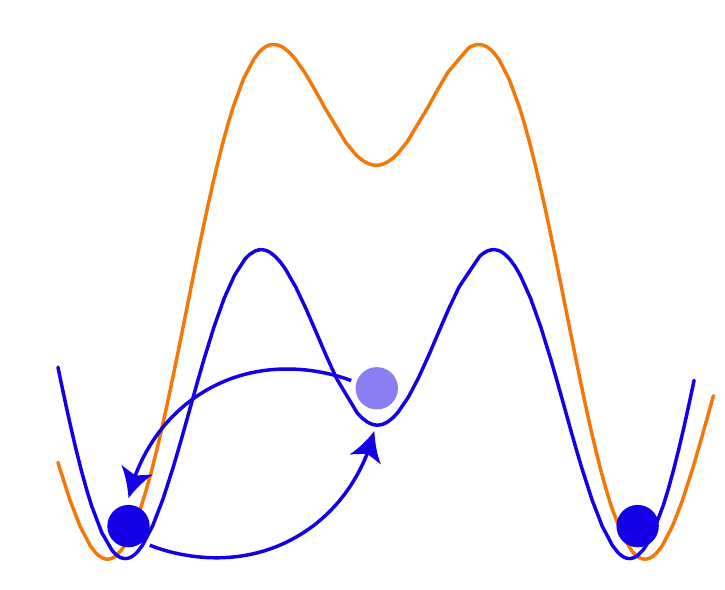}}& $-\frac{t_g^2}{\delta}$ & $\delta_g$ (x2) \\ \hline
         \parbox[c]{0.55\linewidth}{\includegraphics[width=0.8\linewidth]{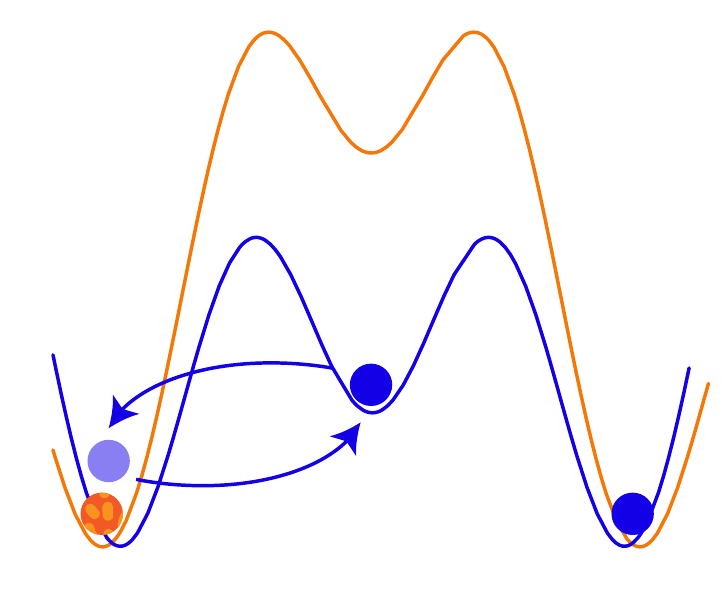}}& $-\frac{t_g^2}{U-\delta} $\newline $-2\frac{t_g D_g}{U-\delta}-\frac{D_g^2}{U-\delta}$ & $\delta_g$ \\ \hline \\%%%%%%%%%%%%%%
         \caption{A couple of processes (arrows) starting and ending with a gauge-invariant state (full color) are illustrated. Transparent dots correspond to the intermediate state. First order corrections to the energy are neglected in the amplitude of each processes. $m=(\delta_e-\delta_g)/2$.}
    \label{tab:perturbation}
    \end{longtable}

\end{document}